\documentclass[twocolumn,twocolappendix]{aastex631} 

\usepackage{graphicx}
\usepackage{epstopdf}
\usepackage[colorinlistoftodos]{todonotes}
\usepackage{natbib}
\usepackage{morefloats}
\usepackage{amsmath}
\usepackage{dirtytalk}
\usepackage{subfigure}
\usepackage{verbatim}
\usepackage{float}
\usepackage{supertabular}

\newcommand{\iMpc}{\mbox{ Mpc$^{-1}$}}
\newcommand{\hi}{H~{\sc i}}
\newcommand{\hii}{H~{\sc ii}}

\defcitealias{2018MNRAS.476L..60A}{AW18}
\defcitealias{2018MNRAS.477.2874A}{AWF18}
\defcitealias{2019MNRAS.488.4440A}{A19}
\defcitealias{2022ApJ...926..151Z}{Z22}

\submitjournal{ApJ}

\shorttitle{Model Selection with \textsc{Delfi-Nest}}
\shortauthors{Binnie et al.}

\graphicspath{{./}{figures/}}

\begin{document}

\title{Likelihood-free Model Selection in Cosmic Reionization with Three-dimensional Tomographic 21~cm Lightcone Images}

\correspondingauthor{T. Binnie, Yi Mao}
\email{binnietom@tsinghua.edu.cn (TB), ymao@tsinghua.edu.cn (YM)}

\author{T. Binnie}
\affiliation{Department of Astronomy, Tsinghua University, Beijing 100084, China}
\affiliation{Blackett Laboratory, Imperial College, London, SW7 2AZ, UK}

\author[0000-0002-8328-1447]{Xiaosheng Zhao}
\affiliation{Department of Astronomy, Tsinghua University, Beijing 100084, China}
\affiliation{Sorbonne Universit\'e, CNRS, UMR 7095, Institut d'Astrophysique de Paris (IAP), 98 bis bd Arago, 75014 Paris, France}

\author{J. R. Pritchard}
\affiliation{Blackett Laboratory, Imperial College, London, SW7 2AZ, UK}

\author[0000-0002-1301-3893]{Yi Mao}
\affiliation{Department of Astronomy, Tsinghua University, Beijing 100084, China}

\begin{abstract}
We explore likelihood-free (aka simulation-based) Bayesian model selection to quantify model comparison analyses of reionisation scenarios. 
We iteratively train the 3D Convolutional Neural Network (CNN) on four toy EoR models based on \textsc{21cmFAST} simulations with contrasting morphology to obtain summaries of the 21~cm lightcone. 
Within the \textsc{pyDelfi} framework, we replaced the \textsc{Emcee} sampler with \textsc{MultiNest} to integrate learnt posteriors and produce the Bayesian Evidence. 
We comfortably distinguish the model used to produce the mock data set in all cases. 
However, we struggle to produce accurate posterior distributions for \textit{outside-in} reionisation models. 
After a variety of cross-checks and alternate analyses we discuss the flexibility of summarising models that differ from precisely the intended network training conditions as this should be more widely scrutinised before CNN can reliably analyse observed data. 
\end{abstract}



\section{Introduction}
The epoch of reionisation (EoR) is the period in cosmic history where our universe's intergalactic medium (IGM) changes phase from neutral hydrogen (\hi) gas to ionised hydrogen (\hii). 
After cosmic recombination ($z\sim 1100$), the first stars formed, the first galaxies developed, and the IGM was reionised. 
These events depend on the timing of the first star formation in the universe, known as the cosmic dawn (CD), as well as the nature of the first galaxies. 
Some examples of unknown galactic properties throughout this epoch include star formation rates, supernovae density, how star formation rates feedback through interstellar dust, and how dark matter species, including primordial black holes, can produce detectable signatures. 
Interested readers are referred to, e.g.\ \citet{FGU}, \citet{2014arXiv1411.3330E}, \citet{2014MNRAS.445.2545D} and \citet{2019arXiv190706653W} for more detail about the vast amount of astrophysical information that can be learned from studying the EoR and beyond. 

The CD and EoR are two of the least observed periods in cosmic history, but the most promising observational analysis of these epochs comes from the hyperfine transition of \hi\ gas known as the 21~cm line. 
Observations of the 21~cm line will be made in difference to the cosmic microwave background (CMB) \citep{1958PIRE...46..240F}, which is absorbed or stimulated depending on the temperature of the IGM. 
The \hi\ gas will span throughout the IGM from recombination until the end of the EoR, meaning the 21~cm line has the potential to provide tomographic maps that span most of the cosmos \citep{2012RPPh...75h6901P}. Since all species of astrophysical sources live or form within this window of cosmic history, their imprints are left on the IGM through the high-redshift 21~cm signal for us to decipher. 

Efforts to observe the 21~cm signal globally and with radio interferometers are in progress at various sites. 
The EDGES (Experiment to Detect the Global EoR Signature\footnote{\url{https://www.haystack.mit.edu/ast/arrays/Edges/}}) team claimed the first detection of an absorption signal in the sky-averaged radio spectrum, which could correspond to a 21~cm absorption signal at $z\approx17$ \citep{2018Natur.555...67B}.
This measurement is yet to be widely accepted in the community \citep{2018Natur.564E..32H, 2018Natur.564E..35B}. 
The precision required to settle this depute is inspiring novel insights into the precise modelling of radio telescope surroundings \citep{2023arXiv230702908P, 2023arXiv231007741M}. 
REACH (The Radio Experiment for the Analysis of Cosmic Hydrogen\footnote{\url{https://www.kicc.cam.ac.uk/projects/reach}}), LEDA (Large-aperture Experiment to detect the Dark Ages\footnote{\url{http://www.tauceti.caltech.edu/leda/}}), and SARAS (Shaped Antenna measurement of the background RAdio Spectrum\footnote{\url{http://www.rri.res.in/DISTORTION/saras.html}}; \citealt{2018ApJ...858...54S}) plan to cross-check this measurement in the near future. 
Current upper limits on the 21~cm brightness temperature power spectrum (PS) have been made by the interferometer HERA (Hydrogen Epoch of Reionisation Array\footnote{\url{https://reionisation.org}}).  
The lowest 2$\sigma$ upper limit for the PS square is $946.18 ~{\rm mK^2}$ for $k = 0.192 \,h {\rm \iMpc}$ at $z \approx 7.9$ \citep{2022ApJ...925..221A}. 
Competing efforts include measurements made by MWA (Murchison Widefield Array\footnote{\url{http://www.mwatelescope.org}}) and LOFAR (The LOw Frequency ARray\footnote{\url{http://www.lofar.org}}) \citep{2020MNRAS.493.1662M, 2021MNRAS.505.4775Y}. 
The amalgamation of instrumental development has been unified in the development of the SKA (Square Kilometre Array\footnote{\url{https://www.skatelescope.org}}; \citealt{2013ExA....36..235M, Koopmans:2015K0}). 
SKA has the potential to tomographically map the \hi\ gas between redshifts of 6 and 27.
Revolutionary insights into early universe cosmology are on the horizon \citep{2004NewAR..48.1039F}. 

The upcoming 21~cm data will vast due to the sheer volume of IGM which contains \hi. 
Statistical techniques must summarise the signal before it can be stored and to facilitate interpretation. 
Typically, the spherically averaged 21~cm PS is measured across discrete chunks of redshift space from a small field of view in the sky. 
This technique can provide valuable insight into the mean properties of early galaxies and can be used to probe the onset of structure evolution throughout the EoR and CD. 
Unfortunately, power spectra cannot capture non-Gaussian information on their own \citep{2015MNRAS.449L..41M}. 
Higher-order statistics, such as the bispectrum, can capture non-Gaussianity \citep{2017MNRAS.472.2436W, 2020arXiv200205992W}; however, they are too expensive to calculate quickly and in statistically significant quantities. 
The 21~cm PS also becomes biased along the line of sight due to structure evolution within each chunk of redshift space \citep{2012MNRAS.424.1877D}. 
This is known as the lightcone effect and can be reduced within the 21~cm PS by careful choices of instrument bandwidth \citep{2014MNRAS.442.1491D} or evolving a transform ergodically with the signal along the line of sight \citep{2016MNRAS.461..126T}. 
A modern alternative to tackling these issues is summarising the entire lightcone with a neural network. Neural networks can be used in a variety of ways; the most popular choices are either emulating the 21~cm PS \citep{2017MNRAS.468.3869S, 2018MNRAS.475.1213S} or extracting summaries from the lightcone directly (\citealt{2019MNRAS.484..282G}; \citealt{2022ApJ...926..151Z}, hereafter referred to as \citetalias{2022ApJ...926..151Z}; \citealt{2022MNRAS.509.3852P}). 
For an in-depth, comprehensive background on machine learning and neural networks, please see, e.g.\ \citet{2007JEI....16d9901B}. 

Simulations help predict the variety in plausible signals that telescopes will observe. 
It is important to understand how changes in physics alter the signal and how these changes in signal alter different data compression techniques and summary statistics. 
Hydrodynamic simulations that capture near-full physical detail are too computationally heavy to be performed in the number and scales necessary to be helpful in EoR interpretation \citep{2006NewA...11..374M, 2011MNRAS.414..727Z, 2012ApJ...745...50W, 2012MNRAS.427..311W, 2014MNRAS.442.2560W, 2016MNRAS.455.2778F, 2016MNRAS.463.1462O, 2018MNRAS.480.2628F, 2016MNRAS.463.1462O, 2018arXiv181111192O,2023MNRAS.tmp..220Y}. 
Fortunately, semi-numerical simulations are faster and therefore better suited for statistical analyses across the required scales \citep{2007ApJ...669..663M, 2010MNRAS.406.2421S, 2011MNRAS.411..955M, 2014Natur.506..197F, 2020JOSS....5.2582M}. 
Recently, neural networks have also been adapted and tuned to cater for accelerated simulations. \citep{2017ApJ...848...23K, 2019arXiv191006274C, 2019MNRAS.483.2524H, 2018MNRAS.475.1213S, 2022ApJ...933..236Z}. 

It is widely known that approximate Bayesian computation (ABC) methods are helpful in the face of intractable likelihoods because true posteriors can be found within an error-compensated approximation (e.g.\ \citealt{Wilkinson+2013+129+141}). 
An excellent example is when measuring the post-reionisation ionising background \citep{2018ApJ...855..106D}. 
In cosmology, the statistical tool of choice for fitting observed or mock (simulated) data to models are Markov Chain Monte-Carlo (MCMC) algorithms, a subgroup of Bayesian analyses. 
Within the EoR, the application of MCMC analyses using likelihood based on 21~cm PS for Bayesian parameter estimation \citep{2015MNRAS.449.4246G, 2018MNRAS.477.3217G, 2019MNRAS.484..933P} and Bayesian model selection \citep{2019MNRAS.487.1160B} have both been established. 
{\it Likelihood-free inference} (aka {\it simulation-based inference}) also removes the need to define a tractable likelihood, avoiding assumptions that influence the final parameter posterior distribution (\citealt{ 2017arXiv171101861L, 2018arXiv180509294L}; \citealt{2019MNRAS.488.4440A}, hereafter referred to as \citetalias{2019MNRAS.488.4440A}; \citetalias{2022ApJ...926..151Z}). 
Please refer to, e.g.\ \citet{10.1093/sysbio/syw077} for a review that contains both ABC and likelihood-free methods. 
The performance of either technique depends on the quality of summary used to compress the data.
How to efficiently compress data into summaries whilst preserving useful information content is an ongoing and active field of research across many fields of science (see, e.g.\ \citealt{10.1214/12-STS406} for a review). 

In this work, we look to expand recent developments in simulation-based parameter estimation with the Density Estimation Likelihood-Free Inference (DELFI) code 
(\citealt{2018MNRAS.477.2874A}, hereafter referred to as \citetalias{2018MNRAS.477.2874A}; \citetalias{2019MNRAS.488.4440A}) by adapting \textsc{pyDelfi} for {\it simulation-based model selection}. 
We then use the 3D Convolutional Neural Network (CNN) developed in \citetalias{2022ApJ...926..151Z}; \citealt{2022ApJ...933..236Z} to summarise the 21~cm lightcone and use the adapted \textsc{pyDelfi} to perform repeat analyses of EoR morphologies with Bayesian model selection as in \citet{2019MNRAS.487.1160B}. 
We aim to build on our previous work by using lightcones rather than coeval cubes and by using likelihood-free inference. 
We do not precisely follow the depth of our previous analyses since more attention is aimed towards confirming the validity of Bayesian model selection analysis in a likelihood-free setting. 

The remainder of this paper is organized as follows. Section \ref{sec: 21cm} contains the necessary background details of the 21~cm signal in the EoR which is largely based on \textsc{21cmFAST} \citep{2011MNRAS.411..955M, 2020JOSS....5.2582M}, including the differing toy scenarios of EoR morphology \citep{2000ApJ...530....1M, 2004ApJ...613....1F, 2014MNRAS.443.3090W}.  
Section \ref{sec: 3D-CNN} contains a summary of the 3D CNN that we use to summarise the 21~cm brightness-temperature lightcones (\citetalias{2022ApJ...926..151Z}; \citealt{2022ApJ...933..236Z}).  
Section \ref{sec: pyDelfi} outlines of the \textsc{pyDelfi} algorithm (\citealt{2018MNRAS.476L..60A}, hereafter referred to as \citetalias{2018MNRAS.476L..60A}; \citetalias{2018MNRAS.477.2874A}; \citetalias{2019MNRAS.488.4440A}). Section \ref{sec: stats} contains the details surrounding Bayesian statistical analysis,
including an introduction to \textsc{MultiNest} \citep{2008MNRAS.384..449F, 2009MNRAS.398.1601F}. 
We discuss the validity of integrating learnt posteriors and detail the changes made in our method in Section \ref{ssec: delfi_nest} before presenting the cross-checks involving cosmic shear in Sections \ref{sec: Delfi-Nest crosscheck} \& \ref{sec: cosmic shear Model selection}. 
The 21~cm results are in Section \ref{sec: results}, starting in \ref{sec: XZcrosscheck} by reproducing the EoR model parameters posteriors shown in \citetalias{2022ApJ...926..151Z}. 
We show the main EoR morphology analyses in Section \ref{sec: EoRresults} where we reproduce the main results from our previous work \citep{2019MNRAS.487.1160B}. 
A discussion of the results and future outlook is included in Section \ref{sec: discussion} before we conclude and summarise in Section \ref{sec: conc+summ}. 
We leave some technical details to Appendix \ref{sec: ABC} (on the validity of ABC techniques), Appendix \ref{sec: remedies} (on alternate analyses involving a 2D-CNN, a no-recurring seed variant of \textsc{21cmFAST}, and a variety of priors),  Appendix \ref{sec: NDEs} (on details of the neural density estimators), and Appendix \ref{sec: Cosmic_shear} (on the details of the cosmic-shear model along with further cross-checks involving importance nested sampling). 

Throughout this work, we use the $\Lambda{\rm CDM}$ cosmology  \citep{2016A&A...594A..13P} with parameters: ($\Omega_{\Lambda}$, $\Omega_{\rm M}$, $\Omega_b$, $\sigma_8$, $h$, $n_s$) = (0.73, 0.27, 0.046, 0.82, 0.7, 0.96) unless specified otherwise. 

\section{21~cm mock observations}\label{sec: 21cm}

For simulating mock observations of the 21~cm brightness temperature, we build upon \textsc{21cmFAST} \citep{2011MNRAS.411..955M, 2020JOSS....5.2582M}. 
Alterations are detailed in the following subsections. 
The toy EoR models used to test the model selection statistics are similar to \citet{2014MNRAS.443.3090W, 2015MNRAS.449.4246G, 2019MNRAS.487.1160B}, but now all toy EoR models use lightcones rather than coeval cubes as in \citet{2018MNRAS.477.3217G}. 
More information on how lightcones are created from coeval simulations is presented in Appendix \ref{sec: no_recurring_seeds}. 

Figure \ref{fig: Toy_model_light-cones} shows a 21~cm brightness temperature lightcone calculated between redshifts $\sim [7.5,12]$ for each of the four EoR models. 
Namely, these models are called FZH, InvFZH, MHR \& InvMHR and are detailed in Sections \ref{sec: FZH&FZHinv} and \ref{sec: MHR&invMHR} respectively. 
Each is clearly defined in EoR morphology with a \textit{global} EoR scale. 
The morphologies are either \textit{outside-in}, based on \citet{2000ApJ...530....1M}, or \textit{inside-out} based on \citet{2004ApJ...613....1F} (as in \textsc{21cmFAST}). 
\textit{Inside-out} (\textit{outside-in}) is when over (under) dense regions re-ionise first, i.e.\ the ionization fraction field is in correlation (anti-correlation) with the underlying density field. 
Previously, the MHR based models followed a \textit{local}  (pixel-by-pixel) ionisation implementation. 
The MHR-based models have been updated so that all models now follow a \textit{global} reionisation similar to \textsc{21cmFAST} \citep{2011MNRAS.411..955M}. 
This addition gives the MHR based models an adaptive density smoothing scale based on the excursion set formalism. 

\begin{figure}
    \centering
    {\includegraphics[scale=0.35]{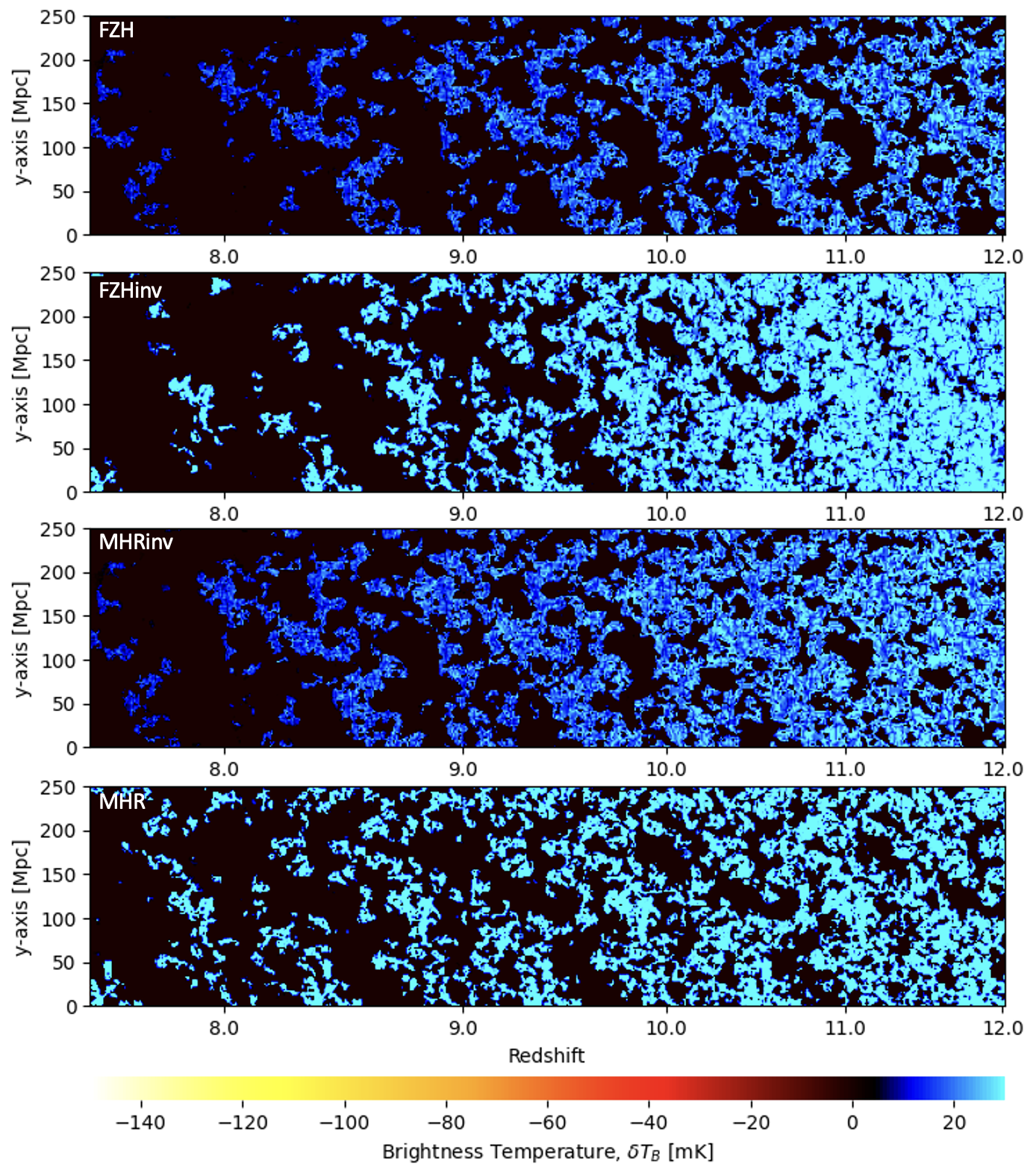}\hfill}
    \caption{\label{fig: LCs}
    Brightness temperatures for the four toy EoR scenarios plotted $z \sim [7.5,12]$ from the same density field. 
    Namely, these are FZH, FZHinv, MHRinv, and MHR, in descending order. 
    FZH and MHRinv have an \textit{inside-out} morphology while FZHinv and MHR contrast with an \textit{outside-in} morphology. 
    A summary of these models is provided in Section \ref{sec: 21cm}. 
    }
    \label{fig: Toy_model_light-cones}
\end{figure}

\begin{table}
    \centering
    \caption{Uniform Prior Distributions of Each Parameter for the Four EoR Morphologies\,\tablenotemark{\scriptsize{a}}}
    \label{tab: priorranges}
    \begin{tabular}{cccc}
    \hline\hline
        Model &  $\zeta$ & ${\rm log_{10}{T_{\rm vir}}}$ & Fiducial\,\tablenotemark{\scriptsize{b}}  \\ \hline
         FZH & [10,~250] & [4.0,~6.0] & [30,~4.7] \\
         FZHInv& [10,~5000] & [4.0,~7.0] & [72,~4.72] \\
         MHR & [10,~4000] & [4.0,~7.0] & [413,~5.80] \\
         MHRinv & [10,~1200] & [4.0,~6.0] & [268,~5.61] \\
         \hline
    \end{tabular}
    \flushleft
\tablenotetext{\scriptsize{a}}{The uniform prior distributions of each parameter follow \citet{2019MNRAS.487.1160B}. For each model, 10,000 lightcones are used to train each 3D CNN. The training, validation, and prediction data sets are divided in the ratio 8:1:1. Training data is generated throughout these ranges. 
} 
\tablenotetext{\scriptsize{b}}{The fourth column shows the fiducial parameters, $[\zeta,{\rm log_{10}{T_{\rm vir}}}]$, for the mock data set of each model.} 
\end{table}

The parameter space used to generate the training set for the 3D CNN is selected randomly with Latin hypercube sampling \citep{ef76b040-2f28-37ba-b0c4-02ed99573416} in the ranges shown in Table \ref{tab: priorranges}. 
These values imply similar prior considerations as the Bayesian analysis in our previous work. 
We consider boxes that are 250 {\rm cMpc} (128 pixels) across; these are larger box sizes than in \citetalias{2022ApJ...926..151Z} so that we improve the robustness of the analysis against cosmic variance \citep{2023A&A...669A...6G}. 

All four of the following toy scenarios follow the same underlying procedure as \textsc{21cmFAST}. 
The 21~cm brightness temperature is calculated as \citep{2006PhR...433..181F},
\begin{equation}
\begin{aligned}
	\delta T_{\rm b} & \approx 27 x_{\rm HI} \left( 1+\delta \right) \left( \frac{H}{ \frac{dv_{ \rm r} }{dr}+H}\right) \left( 1-\frac{T_{\rm CMB}}{T_{\rm s}}\right) \\ 
    & \times \left(\frac{1+z}{10} \frac{0.15}{\Omega_{\rm M}h^2} \right)^{\frac{1}{2}} \left( \frac{\Omega_{\rm b}h^2}{0.023} \right) \rm mK  ,
\end{aligned}    \label{Tbright} 
\end{equation}
where the evolved matter density ($\delta$) and velocity ($dv_{ \rm r}/dr$) fields in the IGM are pre-calculated, $x_{\rm HI}$ is the hydrogen neutral fraction, $T_{\rm CMB}$ is the CMB temperature, $T_{\rm s}$ is the spin temperature.  However, all models and redshifts analysed here assume the IGM to be `post-heated' (i.e. $T_{\rm s} \gg T_{\rm CMB}$). 
All other parameters relate to a conventional $\Lambda$-CDM cosmology. 
We use four 21~cm lightcone simulations as mock observations, one for each toy EoR model using the fiducial parameters in Table \ref{tab: priorranges}. 
Please see \citet{2014MNRAS.443.3090W} and \citet{2019MNRAS.487.1160B} for more details of these models. 

\subsection{FZH (\textsc{21cmFAST}) \& FZHinv}\label{sec: FZH&FZHinv}

Furlanetto, Zaldarriaga,  and Hernquist (FZH) \citep{2004ApJ...613....1F} implement the excursion set formalism as a way of identifying \hii\ regions in the IGM as the EoR progresses. 
This model has been developed into the base of the \textsc{21cmFAST} \citep{2011MNRAS.411..955M, 2020JOSS....5.2582M} --- a state-of-the-art EoR semi-numerical simulation. 
The FZH parameterisation implies the simplest scenario \citep{2001PhR...349..125B} where the mass of collapsed objects relates to the mass of ionised regions via the UV ionising efficiency $\zeta$ as, 
\begin{equation}
\label{eq: m_ion}
m_{\rm ion} = \zeta m_{\rm gal} . 
\end{equation}
To implement this, we integrate the density field from the virial mass, $M_{\rm vir}$, as,
\begin{equation}\label{eq: fcollint}
f_{\rm coll} = \frac{1}{\rho_M} \int^{\infty}_{M_{\rm vir}} m~\frac{dn}{dm}~dm,
\end{equation}
to define the collapse fraction, $f_{\rm coll}$.
With the use of a Press-Schechter collapse fraction, we can expand Equation~(\ref{eq: fcollint}) to become
\begin{equation}\label{eq: fcollerf}
    f_{\rm coll}   =  \text{erfc} \left\{ \frac{ \delta_{\rm crit}  -  \delta }  {  \sqrt[]{   2[  \sigma^2 (m_{\rm min})-\sigma^2  (m) ]  }  }    \right\} .
 \end{equation}
The two parameters of this (and for every) model are $\zeta$ and ${\rm log}_{10}{T_{\rm vir}}$, where $T_{\rm vir} \propto M_{\rm vir}^{2/3}$. 
FZH requires smoothing from large to small scale, naturally incorporating the surrounding behaviour within an approximate photon mean free path. 
This behaviour is adaptive and referred to as global because each pixel's ionisation label is inferred from the  surrounding area. 
Central pixels within each iteration are flagged as fully ionised if 
\begin{equation}\label{eq: threshold}
    \zeta f_{\rm coll} \geq (1 + \bar{n}_{\rm rec}),
\end{equation}
where $\bar{n}_{\rm rec}$ represents inhomogeneous recombination rate density as in \citet{2014MNRAS.440.1662S}. 
For simplicity, we implement $\bar{n}_{\rm rec} = 0$ throughout. 
When ionisation is not achieved before the filter iteration reaches the pixel size, sub-grid regions are assigned a neutral fraction value via $x_{\rm HI} = 1- \zeta f_{\rm coll-per-pixel}$. 

For the inverted threshold equivalent, FZHinv, we change the integration limits in Equation~(\ref{eq: fcollint}) to $[0, M_{\rm vir}]$ to obtain $f'_{coll}$ which accounts for underdense IGM  structures. 
Sparse enough regions will be ionised by a background ionising UV efficiency $\zeta'$. 
We also impose the transformation $\delta \rightarrow -\delta$ in Equation~(\ref{eq: fcollerf}) so that the critical barrier over-density has been inverted. 
Ionisation is now defined as before but by replacing $\zeta$ and $f_{coll}$ with their primed counterparts. 
This change alters the morphology of the \textit{inside-out} FZH model to become \textit{outside-in} for FZHinv.
We assign $x_{\rm HI} = 1 - \zeta'f_{\rm coll}'$ to any partially ionised pixels, and we now have an inverted ionisation threshold, $\zeta ' f_{\rm coll}' \geq 1$, that can be implemented within the \textsc{21cmFAST} framework. 
Both FZH and FZHinv implement reionisation by estimating the total UV photons inside pixels within the scale of each excursion set iteration. 
Neutral fraction is first counted globally and added to sub-grid estimates so that the FZH and FZHinv sub-grid neutral fraction estimate is analogous to the recombination rate in Equation~(\ref{eq: threshold}).  

\subsection{MHR \& MHRinv}\label{sec: MHR&invMHR}

For a contrasting reionisation model, we use a version of the Miralda-Escud\'e, Haenelt and Rees (MHR) model \citep{2000ApJ...530....1M}. 
\textit{Outside-in} reionisation describes how underdense IGM patches ionise first since the recombination rate is correlated with the underlying density field. 
Denser regions of IGM gas recombine quickly and remain neutral since they have a higher opacity to ionising radiation. 
Circumgalactic gas (within each halo) must first be ionised before photons infiltrate into the IGM.
Reionization progresses due to a steadily increasing amount of background UV radiation. 
\hii\ regions percolate in the direction of low gas density, and the EoR is defined to end when these regions overlap. 
The radiation background is driven by a combination of UV photons produced directly by galaxies as well as high redshift objects whose X-ray photons heat the gas due to their large mean free paths \citep{2006MNRAS.367.1057P,2007MNRAS.376.1680P}. 
Since we are not directly calculating the X-ray background here \citep{2017MNRAS.472.2651G}, we summarise the background radiation with the UV ionising efficiency, $\zeta$. 
In both MHR and MHRinv, the background radiation is not directly related to the spatial distribution of the galaxies. 
We implement the ionisation thresholds by initially running the excursion set formalism as before. 
However, within each selected subset of the box, we implement a pixel-by-pixel check to see whether or not the density field at that position qualifies for ionisation. 
We then update $f_{\rm coll}$ and iterate as before via the excursion set formalism. 
Sub-grid neutral fractions are then assigned partial ionisation similarly to the FZHinv and FZH models respectively. 

Within the MHR model, the ionisation threshold is defined via the neutral fraction as, 
\begin{equation}\label{eq: xHI consistensy}
    \bar{x}_{\rm HI} = 1 - \zeta f_{\rm coll} . 
\end{equation}
First, we order pixels by \hi\ density. 
Then we select the $i$-th pixel (of total $N_{\rm p}$) so that the ionisation threshold, $\delta < \delta_i$, selects underdense pixels that define $\bar{x}_{\rm HI}$ to be consistently calculated across all our models with the parameters ${\rm log}_{10}[T_{\rm vir}]$ and $\zeta$. 

For MHRinv, the ionisation threshold is inverted. 
Ionisation is classified via an over-density threshold as $\delta > \delta_{j}$, where the $j$-th pixel is chosen so that the reionisation history is consistent with Equation~(\ref{eq: xHI consistensy}). 
Both MHR and MHRinv will have the same $\bar{x}_{\rm HI}$ for the same set of ${\rm log}_{10}[T_{\rm vir}]$ and $\zeta$ because of this consistency. 
The resulting model provides an \textit{inside-out} reionisation. 

The only deviations from the FZH models are that the definition of ionisation comes from a density field threshold evaluated within each excursion set iteration. 
In summary, all models share very similar reionisation histories by construction because all models have both global and sub-grid neutral fractions prescribed by Equation~(\ref{eq: xHI consistensy}).

\section{Compression of the 21~cm lightcone to data summaries}\label{sec: 3D-CNN}

Most reionization analyses use a summary statistic to compress the information of the brightness temperature lightcone. 
In this work, we primarily use the 3D CNN described in \citetalias{2022ApJ...926..151Z} to create concise summaries from the lightcone directly. 
\citetalias{2022ApJ...926..151Z} suggested that most of the information extracted by the 3D CNN from the lightcone images likely comes from the shape and boundaries of \hii\ regions. 
It is implied that sharp changes in space or frequency contribute the most to the 3D CNN summaries. 

\begin{figure}
    \centering
    \includegraphics[scale=0.55]{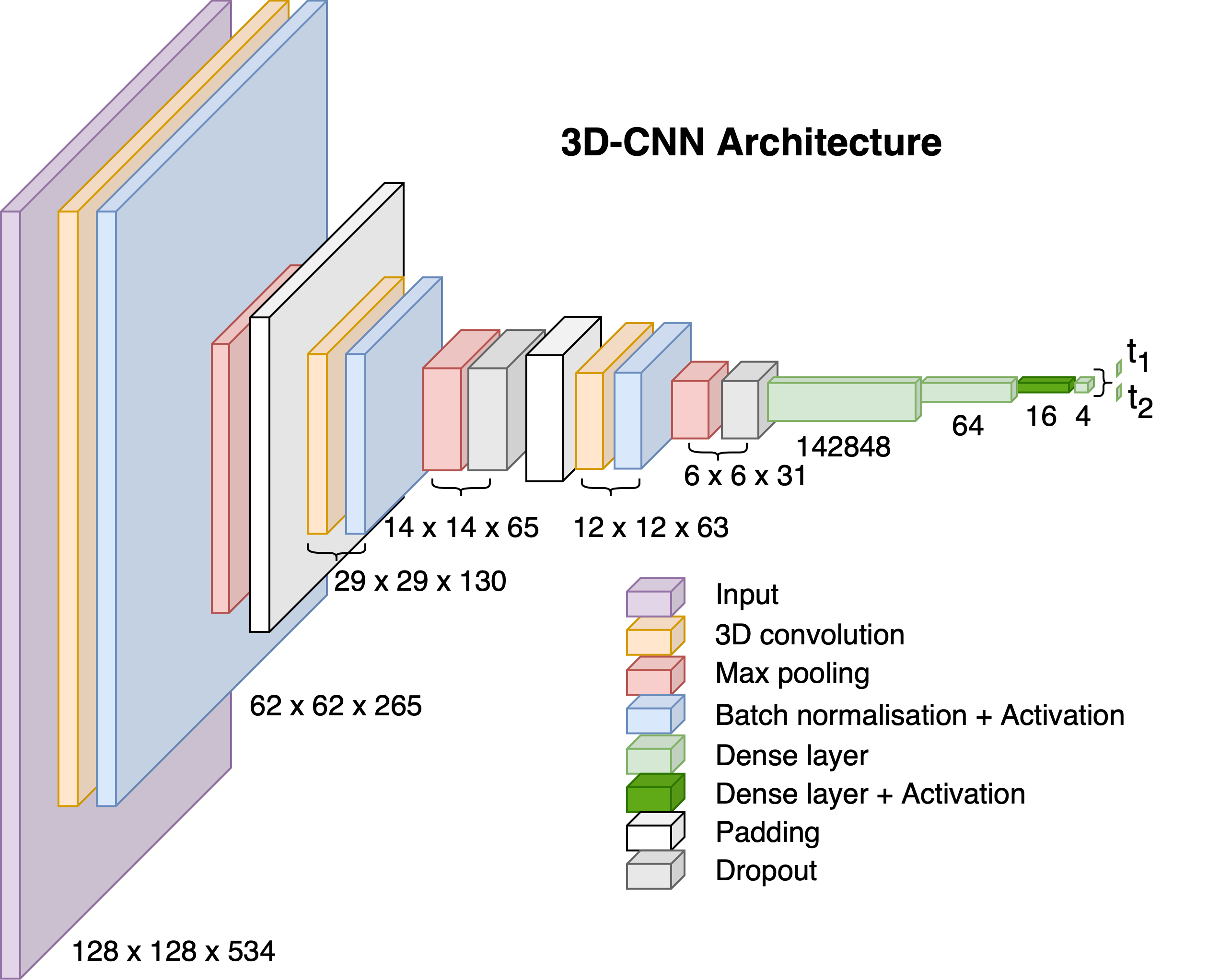}
    \caption{
    A summary of the 3D CNN architecture used in this work. 
    For more detail about the purpose of each layer, please see Section \ref{sec: 3D-CNN}. 
    For a fuller description of the network architecture please see Table \ref{tab: NN architectures} (in Appendix \ref{sec: alt_cnn}). 
    }
    \label{fig: 3D-CNN}
\end{figure}

The 3D CNN summarises the 21~cm brightness temperature lightcone into a concise set of summaries $t(\theta)$ that preserves a dependency of the reionisation parameters. 
There is a 1:1 mapping between the number of parameters $\theta$ and the number of summaries, $t$. 
A summary of the network architecture is presented in Figure \ref{fig: 3D-CNN}, where the simulated data (dependent on $\zeta$ \& ${\rm log}[T_{\rm vir}]$) is input on the left-hand side and passes through the layers until the two summary values ($t_1$ \& $t_2$) are obtained on the right-hand side. 
Please see Table \ref{tab: NN architectures} for a more detailed description of the layer shapes. 
We use the Keras functional API\footnote{\url{https://keras.io/guides/functional\_api/}} for constructing the network, where each type of layer is used for a specific purpose.

Convolution layers reduce the data vector by running a 3D kernel across the data set. 
The batch normalisation constrains the mini-batches of the data vector to have mean and variance of $[\mu=0,~\sigma^2=1]$ \citep{conf/nips/SanturkarTIM18}. 
A linear transform is then applied to ensure the batch normalisation process is an identity transform of the previous layer \citep{2015arXiv150203167I}. 
Max pooling layers down-sample the previous layer to reduce the number of parameters and learn position invariant features. 

Batch normalisation layers follow convolutional layers to increase speed and stability during training. 
Each batch normalisation layer is paired with an Activation function to assess which parts of the data vector are worth progressing into the next layer. 
We use a Rectified Linear Unit (ReLU) activation function \citep{6639346} because it enables complex mappings between the inputs and outputs. 
Dense layers are fully connected, i.e. every activation in the previous layer is connected to each data vector element in the proceeding layer. 
Zero-padding layers fill the edges of the data vector with zeros to offset the decrease in vector size between layers. 
Dropout layers are necessary to compensate for the over-fitting that can occur with so many parameters \citep{NIPS2013_71f6278d, JMLR:v15:srivastava14a}. 
We use a dropout ratio of 0.4. 
The data is split into 16 mini-batches, and the process is separated into a number of channels depending on which function is in progress. 

The 3D CNN architecture presented in this work contains some modifications developed by the author since the original publication in \citetalias{2022ApJ...926..151Z}. 
The network has 10,428,044 trainable parameters (and 366 fixed, untrainable parameters) that are fit by comparing a validation loss measure against the validation set. 
The network is trained using \textsc{RMSprop} \citep{RMSProp}, which includes a variable learning rate by means of a weighting function. 
Initially, learning is fast so that local minima do not stagnate early parameter exploration accross the network. 
As the training run progresses learning slows allowing precise convergence of the network parameters. 
In an example training run, after 37 epochs of training ($\sim6$ hours on a GPU), the validation loss began to increase, triggering the stopping criteria and mitigating over-fitting. 
The prediction set was then used to measure the performance of the network using a coefficient of determination based on regression, $\sim(x_{\rm True}-x_{\rm Predicted})^2$. 
For both $\zeta$ and $\rm log_{10}{T_{\rm vir}}$, agreement was achieved when predicting the parameters from their summaries with $>99.1\%$ agreement. 

To emulate the measurements of an interferometer, the mean brightness temperature is subtracted from each lightcone with \textsc{tools21cm} \citep{2020JOSS....5.2363G} before being summarised by the network. 
We defer a more complete treatment of telescope noise and foreground cleaning to future work. 
In \citetalias{2022ApJ...926..151Z}, ten similarly parameterised lightcone boxes with different initial conditions are concatenated into a single data cube to avoid biasing from periodic boundary conditions. 
We see this as slightly unconventional, and we therefore leave the lightcone data objects untouched in our primary analysis. 
Appendix \ref{sec: no_recurring_seeds} includes an alternate analysis that introduces a version of \textsc{21cmFAST} that draws different random seeds for each interpolated coeval cube in the lightcone. 
Unfortunately this only worsens the boundary conditions, making them observable in the lightcone (Figure \ref{fig: LCs_nowrapping}). 
Despite this, the results produced are indistinguishable (at the level of statistical fluctuation). 
Since this alternative analysis is a more extreme approach to \citetalias{2022ApJ...926..151Z}'s, we therefore conclude that tweaks to these boundary conditions are insignificant. 
In Appendix \ref{sec: alt_cnn}, we include an alternate network analysis inspired by the 2D-CNN in \citet{2019MNRAS.484..282G} that also produces statistically indistinguishable results to our primary analysis. 

\section{Density estimation with likelihood free inference}\label{sec: pyDelfi}

Posterior density estimation with likelihood-free inference works in two halves. \newline
$\bullet$ The first half is model-dependent and requires data to be compressed into a summary, which typically contains information loss depending on the algorithm. 
For example, this can be the 21~cm PS or the 3D CNN summaries, $t$, introduced in Section \ref{sec: 3D-CNN}. \newline
$\bullet$ The second half is model-independent, \textsc{pyDelfi} learns a posterior by fitting a joint density (described in \ref{sec: pydelfi-nittygritty}) to the above data summaries after compressing the entire distribution of, $t(\theta)$, to create new summaries, $s$. In \citetalias{2018MNRAS.476L..60A} the relationship between the two sets of summaries is lossless because the score function is used for the compression (detailed in Section \ref{sec: score_comp}). 

Throughout this section we adopt the terms pseudo-Evidence, $\hat{\mathcal{Z}}$, and pseudo-posterior, $\hat{\mathcal{P}}$, to emphasise that these are defined in relation to the joint-density distribution, differing from the Evidence and posterior which relate to the likelihood as in Section \ref{sec: stats}. 
Since these should converge, we drop this terminology outside of Section \ref{sec: pyDelfi}. 

\subsection{A summary of \textsc{pyDelfi}}\label{sec: pydelfi-nittygritty}

The \textsc{pyDelfi} code is a python implementation\footnote{\url{https://github.com/justinalsing/pydelfi}} of the Density Estimation Likelihood Free Inference (DELFI) algorithm (\citetalias{2018MNRAS.476L..60A}, \citetalias{2018MNRAS.477.2874A}, \citetalias{2019MNRAS.488.4440A}). 
We can estimate a posterior by learning a joint density, $\rho[\theta, D(\theta)]$, for the parameters and data across the prior space. 
We refer to this as a pseudo-posterior, 
\begin{equation}\label{eq: pseudoP}
    \hat{\mathcal{P}}[\theta ~| D(\theta_{f}), M] \propto \rho[\theta,D = D(\theta_{f}); \eta] ,
\end{equation}
where $\eta$ are the model parameters we require when writing down an analytic model for $\rho$, the hat is used to distinguish from the theoretical $\mathcal{P}$ defined in Section \ref{sec: stats}. 
There are a variety of possible models for a tractable joint density space, and recent developments in machine learning make neural density estimators (NDEs) a sensible choice, e.g. \citet{NIPS1998_93279690, 2019arXiv191013233P}. 
We can analytically integrate, $\hat{\mathcal{P}}$ , to obtain a pseudo-Evidence,
\begin{equation}\label{eq: pseudoZ}
    \hat{\mathcal{Z}} = \int_{\Pi} \hat{\mathcal{P}} d\theta ,
\end{equation}
which will be equal to the $\mathcal{Z}$ assuming $\rho$ can accurately fit $D$ and assuming the proportionality in Equation \ref{eq: pseudoP} is a valid assumption. 
Because the Evidence (or pseudo-Evidence) is the normalisation of the posterior (or join-density). 
If the proportionality constant is incorrect or contains parameter dependence, then  $\hat{\mathcal{Z}}\neq \mathcal{Z}$,  while $\hat{\mathcal{P}}(\theta_{\rm  MAP}) = \mathcal{P}(\theta_{\rm  MAP})$ will still likely be true. 
Within \textsc{pyDelfi}, the uncertainties of the parameter fit to the data can be propagated through to an uncertainty on $\hat{\mathcal{Z}}$. 
The validity of the proportionality between the joint-density and posterior however, is less easy to describe and we return to this in section \ref{sec: discussion}. 

The full NDE we use is constructed from a set of stacked Gaussian Mixture Models (GMMs), which form a Mixture Density Network (MDN) in combination with Masked Auto-regressive Flows (MAFs). 
For simplicity, we illustrate a simple NDE constructed from a single type of GMMs in the MDN and encourage the reader to refer to Appendix \ref{sec: NDEs}, \citetalias{2018MNRAS.476L..60A}, \citetalias{2018MNRAS.477.2874A}, and \citetalias{2019MNRAS.488.4440A} for more detail on the full NDE architecture. 
By definition, we can write, 
\begin{equation}\label{eq: GMMpseudoZ} 
    \hat{\mathcal{P}} \left [ \theta|D(\theta_f) \right ] = \sum^K_{i=1} w_i N(\mu_i, \mathcal{C}),
\end{equation}
where N is the Gaussian density with mean, covariance and network weights parameterising and the joint density as, $\rho(\mu, \mathcal{C}, w)$. 
Therefore, the $\hat{\mathcal{Z}}$ of an MDN made solely from GMMs can be obtained via Equation \ref{eq: pseudoZ} by summing, $ w_iN$, across all $K$ of the GMMs in the MDN. 

The main difference between the Evidence, $\mathcal{Z}$ and pseudo-Evidence, $\hat{\mathcal{Z}}$ emerges because the Evidence is calculated by integrating the likelihood, which summarises the samples accross the prior directly.  
In contrast, the pseudo-Evidence still integrates samples throughout the prior space, but it does so indirectly through a secondary set of summaries created from the sampled data vector (represented by the joint density in Equation \ref{eq: pseudoP}). 
Instead of fitting the NDE joint density to the data directly (i.e. the set of $D(\theta)$ in Equation \ref{eq: pseudoP}), \textsc{pyDelfi} fits to summaries of the data by implementing score compression. 
Since we are already compressing the lightcones with the 3D CNN to produce $t[D(\theta)]$, the score function is the second layers of data compression, denoted, $s\{t[D(\theta)]\}$, which is then fed into \textsc{pyDelfi}. 
This reduces the data dimensionality significantly, from the order of the data vector to the order of the simulation parameters, enabling \textsc{pyDelfi} to scale independently of large data vectors. 
Given that the data summary provided preserves sufficient information, we should still have access to a valid $\hat{\mathcal{P}}$ and $\hat{\mathcal{Z}}$. 

Whether or not $\mathcal{Z} = \hat{\mathcal{Z}}$ is accurate enough to use the Bayes factor in model comparison will depend on how effectively the data is compressed into summaries across each of the compared models. 
This dictates the constants of proportionality in Equation \ref{eq: pseudoP} and therefore whether these constants cancel when calculating the Bayes factor. 
We infer that a bias can be introduced depending on how well a particular model's information content is compressed. 
This bias may be independent of how well the compressed summaries represent the original data sets. 

In the context of parameter estimation analyses DELFI excels. 
\citetalias{2018MNRAS.477.2874A} find a difference of $< 0.05 \sigma$ when comparing density estimation likelihood-free inference to a well-converged MCMC sampled posterior when analysing a wCDM and a light-curve calibration model from type Ia supernova apparent magnitudes \citepalias{2018MNRAS.477.2874A}. 

To summarise \textsc{pyDelfi}, the joint density is learnt by assessing the distribution of model-dependent data summaries across the parameter prior. 
It does so by compressing the data summary distribution in a model-independent way, which is then fit by the joint density distribution. 

To implement \textsc{pyDelfi}, we need to take the distribution of data summaries, $t$, and compress these further into a concise set of summaries, $s$, representing the entire data vector. 
To optimise compression of the data summaries, we define an approximate likelihood based on the score function of a Wishart distribution (detailed in Section \ref{sec: score_comp} \& \ref{sec: approxLikelihoods}).  
When \textsc{pyDelfi} is trained, the properties of the joint density, $\rho$, are compared to the score compression of the simulated data vector, $s$. 
The learnt $\hat{\mathcal{P}}$ is proportional to the learnt joint density when evaluated with the best-fit parameters provided by maximising the approximate likelihood. 
Since we have access to generating data summaries, $t$, from the simulation parameters, $\theta$, the $\hat{\mathcal{P}}$ should become an excellent approximation of the true posterior upon iteration and therefore provide the fiducial parameters of the mock data set. 
Considering that we desire $\hat{\mathcal{P}} = \mathcal{P}$ which infers $\hat{\mathcal{Z}} = \mathcal{Z}$, the Bayes factor results should provide reliable model selection conclusions with consistent use of either $\hat{\mathcal{Z}}$ or $\mathcal{Z}$. 

\subsection{Score compression summary}\label{sec: score_comp}

Compression to the score function of the likelihood, $\nabla \mathcal{L}$, is optimal since it provides a 1:1 mapping of compressed statistics to parameters that preserve the Fisher information content of a given data set. 
If the Fisher information is a valid measure of the information in the data set, score compression is an optimal compression. 
\citetalias{2018MNRAS.477.2874A} use the phrase asymptotically optimal here to highlight that the Fisher information may not be the most optimal measure of information. 
Despite being a reliable choice for a wide range of data sets, there may be better ways of compressing the data. 

Here, we briefly summarise the score compression process; please see \citetalias{2018MNRAS.476L..60A} for an in-depth derivation. 
We can expand the score function with the Taylor expansion as,
\begin{equation}\label{eq: score expansion}
    s = \nabla \mathcal{L} \approx \nabla \mathcal{L}_{\theta_{\rm p}} + \nabla \left ( \delta \theta^T \nabla \mathcal{L}_{\theta_{\rm p}} \right )- \frac{1}{2}  \delta \theta^T  \left [ - \nabla \nabla^T \mathcal{L}_{\theta_{\rm p}} \right  ]  \delta \theta , 
\end{equation}
where the term in square brackets is often referred to as the observed information matrix and $\theta_{{\rm p}}$ refers to the parameter at which the Taylor expansion is performed. 
We assume the score compression summary is an unbiased estimator of the parameters and take the covariance, $\mathcal{C}$, of the score function to obtain, 
\begin{equation}
    \mathcal{C}\left ( \nabla \mathcal{L}_{\theta_{\rm p}},\nabla \mathcal{L}_{\theta_{\rm p}} \right) = E_{\theta_{\rm p}} \left [ \nabla \mathcal{L}_{\theta_{\rm p}}\nabla^T \mathcal{L}_{\theta_{\rm p}} \right ], 
\end{equation}
where the right-hand side of this equation is the Fisher information matrix. 
The covariance of our summary is therefore maximally informative because the lower limit of the Cram\'er-Rao bound becomes saturated, see, e.g. \citet{jaynes03}. 
We have shown that the information content of our compression is optimal given that our estimator is unbiased in obtaining the likelihood maximum from the compressed data summaries, i.e. we require $E_{\theta_{\rm p}} = \theta_{\rm p}$ \& $\nabla \mathcal{L}|_{E_{\theta_{\rm p}}} = 0$ to be true for the fiducial parameter set, $\theta_p = \theta_f$. 
If this is the case, Fisher information will be conserved between the original data and the compressed summaries. 
The only caveats are that our compression must provide an injective mapping between the parameter sets \& their summaries, and that the mapping will transform our Bayesian prior in a way that is difficult to evaluate. 
When the compression procedure in Equation \ref{eq: score expansion} is applied to a Gaussian likelihood function, and only the linear part of the result is considered, we obtain exactly the well-known MOPED algorithm or Massively Optimized Parameter Estimation \& Data compression \citep{2000MNRAS.317..965H}. 
Linear compression, however, limits the machinery to a case where only the mean is dependent on the parameters. 
Expanding Equation \ref{eq: score expansion} to the quadratic terms obtains an optimal quadratic estimator for the power spectrum as in \citet{1997PhRvD..55.5895T, 1998PhRvD..57.2117B, 2000ApJ...533...19B} however this is only true for Gaussian likelihoods. 
To obtain a compression scheme where both the mean and covariance depend on the parameters as desired, a Karhunen-Lo\'eve eigenvalue decomposition method may be used \citep{1997ApJ...480...22T}; however, this is lossy when measured against Fisher information. 
A lossless equivalent has been derived for CMB analyses \citep{2016PhRvD..93h3525Z}; but, the results are not general beyond Gaussian likelihoods. 
The score compression procedure derived in A18 and summarised here provides an approach that is optimal in all cases and is the building blocks of the \textsc{pyDelfi} architecture described in \citetalias{2018MNRAS.477.2874A} \& Section \ref{sec: pydelfi-nittygritty}. 

To apply score compression in likelihood-free inference an approximate likelihood must be used. 
Crucially, the choice of likelihood here is disentangled from the method's ability to perform parameter estimation. 
The error in this choice will reduce only the optimisation of the compression rather than enhance the precision of the statistical analyses (\citetalias{2018MNRAS.477.2874A}, \citetalias{2019MNRAS.488.4440A}). 
Since we have assumed $\nabla \mathcal{L}$ to be an unbiased estimator of the parameters, our summary will become lossy at worst. 
If this assumption is incorrect, our parameter pseudo-posteriors will remain unbiased and merely worsen in precision compared to the posterior. 
The same is true for any mean, covariance or derivatives estimated from reducing the input data vector with score compression. 

\subsection{Approximate likelihood}\label{sec: approxLikelihoods}

To evaluate score compression, we need a tractable likelihood approximation. 
In Equation \ref{eq: score expansion}, $s$ is approximated by the maximum-likelihood estimate of the covariance matrix so that the covariance obtained will match that of the input data summary vector. 
This distribution is known as a Wishart distribution and, for a single data point, simplifies down to a $\chi^2$ \citep{2013MNRAS.432.1928T, 2019MNRAS.483..189H}. 

To fit the network efficiently, we want to minimise the information difference between the network and the training data. 
We can do this conveniently via entropy using the Kullback-Leibler (KL) divergence, $\mathcal{D}_{\rm KL}$, defined as,
\begin{equation}\label{eq: KL-divergence}
    \mathcal{D}_{\rm KL}(p^*|\rho) = \int p^*(t |\theta ) {\rm ln} \frac{\rho(t|\theta, w_n)}{p^*(t|\theta)}  d t ,
\end{equation}
between an NDE estimator $\rho(t|\theta, w_n)$ and a target distribution $p^*(t| \theta )$. 
Since we do not know the target distribution, we use a loss function defined as,
\begin{equation}\label{eq: loss function}
    -{\rm ln} \mathcal{L} \equiv \sum_{\rm samples} -{\rm ln} \rho(t|\theta_i, w)
\end{equation}
where $\rho$ is defined in Section \ref{sec: pydelfi-nittygritty}, and $\mathcal{L}$ will behave as a likelihood (allowing the application of MCMC algorithms) in the limit $\theta \rightarrow \theta_f$. 
The networks are then trained by minimising Equation \ref{eq: loss function} with respect to the network weights, allowing the posterior density to be inferred from the training data (and priors). 
\textsc{pyDelfi} produces a joint density that contains the maximum Fisher information obtainable from the summary set \citep{2012arXiv1205.2629L}, validating the requirements for a Bayesian likelihood (Appendix \ref{sec: ABC}). 
With a successfully converged MCMC, $p^* = \rho(t| \theta = \theta_f )$, implies, $\hat{\mathcal{P}} = \mathcal{P}$, via Equation \ref{eq: pseudoP} as desired. 
This is where the \textsc{Emcee} algorithm is implemented within \textsc{pyDelfi}. 
For example, minimising the K-L divergence in Equation \ref{eq: KL-divergence} takes the place of maximising the likelihood in a traditional Bayesian parameter estimation exercise. 
In \citetalias{2018MNRAS.477.2874A}, a comparison between the integration values of \textsc{MultiNest} and summing the elements of Equation \ref{eq: GMMpseudoZ} are shown to have an excellent agreement. 
Unfortunately the latter method is intractable when more complex NDE architectures are used. 
\textsc{Delfi-Nest} provides quantitative model comparison in a wider variety of scenarios than the Evidence obtained directly from \textsc{pyDelfi}. 

\section{Model selection statistics}\label{sec: stats}
\subsection{Bayesian model selection}\label{ssec: BMS}

Within Bayesian inference, Bayes theorem is conventionally written, 
\begin{equation}\label{eq: Bayestheorem}
    \mathcal{P} (\mathbf{\theta} | D, M) = \frac{\mathcal{L}(D|\mathbf{\theta},M)\Pi(\mathbf{\theta}|M)}{\mathcal{Z}(D|M)},
\end{equation}
where $D$ is the (mock) observed data set and $M$ a model with parameters $\mathbf{\theta}$. 
The probability distributions are labelled: $\mathcal{P}$, the parameter posterior; $\mathcal{L}$ the likelihood; $\Pi$ the parameter prior distribution, and the Bayesian evidence is labelled $\mathcal{Z}$.
Throughout this work, all $\Pi$ will be a uniform distribution for each $\theta$. 
This simplifies Equation \ref{eq: Bayestheorem} to $\mathcal{P} \propto \mathcal{L}$, implying parameter estimation can be summarised by finding the parameters, $\theta_{\rm  MAP}$, that provide the maximum likelihood point,. 
Model selection is summarised by integrating $\mathcal{P}$ to marginalise $\mathcal{L} \Pi$.
The Bayesian Evidence is crucial in model selection as it provides a probability describing how likely a given model is to be the truth when confronted with data. 
For two competing models (labelled $M_1$ and $M_2$), we can analyse them by taking the ratio of the corresponding Evidence values to obtain the Bayes factor as,
\begin{equation}\label{eq: Bayesfactor}
    \mathcal{B}_{1,2} = \frac{\mathcal{Z}_1}{\mathcal{Z}_2}.
\end{equation}

The Bayes factor results in this paper should be interpreted as betting odds that dictate which model is more likely given the data, and therefore which of competing models should be favoured. 
We include the Jeffreys' scale\footnote{The Jeffreys' scale is identified on our Bayes factor figures as dark grey (representing odds of $\mathcal{B}_{1,2} < 10$), light grey ($10 < \mathcal{B}_{1,2} < 150$) and white regions ($\mathcal{B}_{1,2} > 150$) corresponding to \textit{weak}, \textit{moderate}, and \textit{strong} scoring respectively.} in grey on our Figures for ease of interpretation \textit{only}. 
We emphasise that the model selection results are the odds themselves. 
Model selection statistics, like $\mathcal{Z}$ and $\mathcal{B}$, are best used in the context of a broader model comparison analyses.  
Please see e.g. \citet{al2009bayesian} for more details on Bayesian analyses. 

\subsection{\textsc{MultiNest}}\label{ssec: multinest}

Conventionally, $\mathcal{Z}$ is tricky to calculate because it is an integral written as, 
\begin{equation}\label{eq: Evidence}
    \mathcal{Z} = \int \mathcal{L}\Pi d\theta  ,
\end{equation}
with as many dimensions as we have parameters, $\theta$. 
We simplify this problem with nested sampling \citep{2004AIPC..735..395S}, where the integration dimensions are reduced to one - a fraction of the original prior volume. 

To implement nested sampling, the algorithm must reliably produce iso-likelihood contours from each sampled point. 
\textsc{MultiNest} \citep{2008MNRAS.384..449F, 2009MNRAS.398.1601F} does this by combining ellipsoidal rejection sampling with a form of k-means clustering to stack ellipsoids that engulf the iso-likelihood region. 
Although there are more sophisticated algorithms, such as \textsc{Dynesty} \citep{2020MNRAS.493.3132S}, we found these can increase computational intensity with negligible benefit. 
We find \textsc{MultiNest} to be a good balance between accurate and efficient given our analysed models' low number of parameters. 

The posteriors shown in this work have been checked for convergence with \textsc{NestCheck} \citep{2019MNRAS.483.2044H}. 
To cross-check our Bayesian Evidence calculation, we recalculated several results using \textsc{PocoMC} \citep{2022JOSS....7.4634K, 2022ascl.soft07018K}, which combines nested sampling with Sequential Mont\'e Carlo. 
The differences in results between the two algorithms were negligible.

\section{\textsc{Delfi-Nest}: Tuning \textsc{pyDelfi} for model selection}\label{ssec: delfi_nest}

To adapt \textsc{pyDelfi} to a more flexible model selection framework, we replace the \textsc{Emcee} sampler with \textsc{MultiNest}. 
Likelihood-free methods (or simulation-based inference) learn a joint density of the parameter and the summaries of a desired data set.
The joint density is proportional to the parameter posterior when trained appropriately and evaluated at the fiducial parameter set. 
In \citetalias{2018MNRAS.477.2874A}, \citetalias{2019MNRAS.488.4440A} and \citetalias{2022ApJ...926..151Z}, \textsc{pyDelfi} is shown to learn the parameter posterior, allowing for successful Bayesian parameter estimation. 
We produce the joint density with \textsc{pyDelfi} as in these previous works, and then integrate with  \textsc{MultiNest} to obtain the Evidence. 
We do model selection between two competing models with the Bayes factor by taking the ratio between the integrated joint densities. 
We assume that any constant of proportionality between the joint density and Posterior obtained in Equation \ref{eq: pseudoP}  (which follows through into the Evidence in Equation \ref{eq: pseudoZ}) will be largely cancelled. 
The NDE architecture and 3D CNN architecture are consistent in each analysis unless stated otherwise. 

To train \textsc{pyDelfi} for the cosmic shear analysis, we use 200 initial samples with 39 populations each of batch size 200. 
For the 21~cm analyses, we increased the populations to 100. 
Our 3D CNN is trained separately for each model with 10000 lightcones with parameters distributed across the prior distributions as in Table \ref{tab: priorranges}. 
This set is separated into 8000 training lightcones, 1000 for validating the training process, and 1000 for predictions made after the network has finished training for cross-validation. 

\subsection{\textit{Cross-check}: Cosmic-shear posterior comparison with \textsc{Delfi-Nest}}\label{sec: Delfi-Nest crosscheck}

\begin{figure}
    \centering
    \includegraphics[scale=0.3]{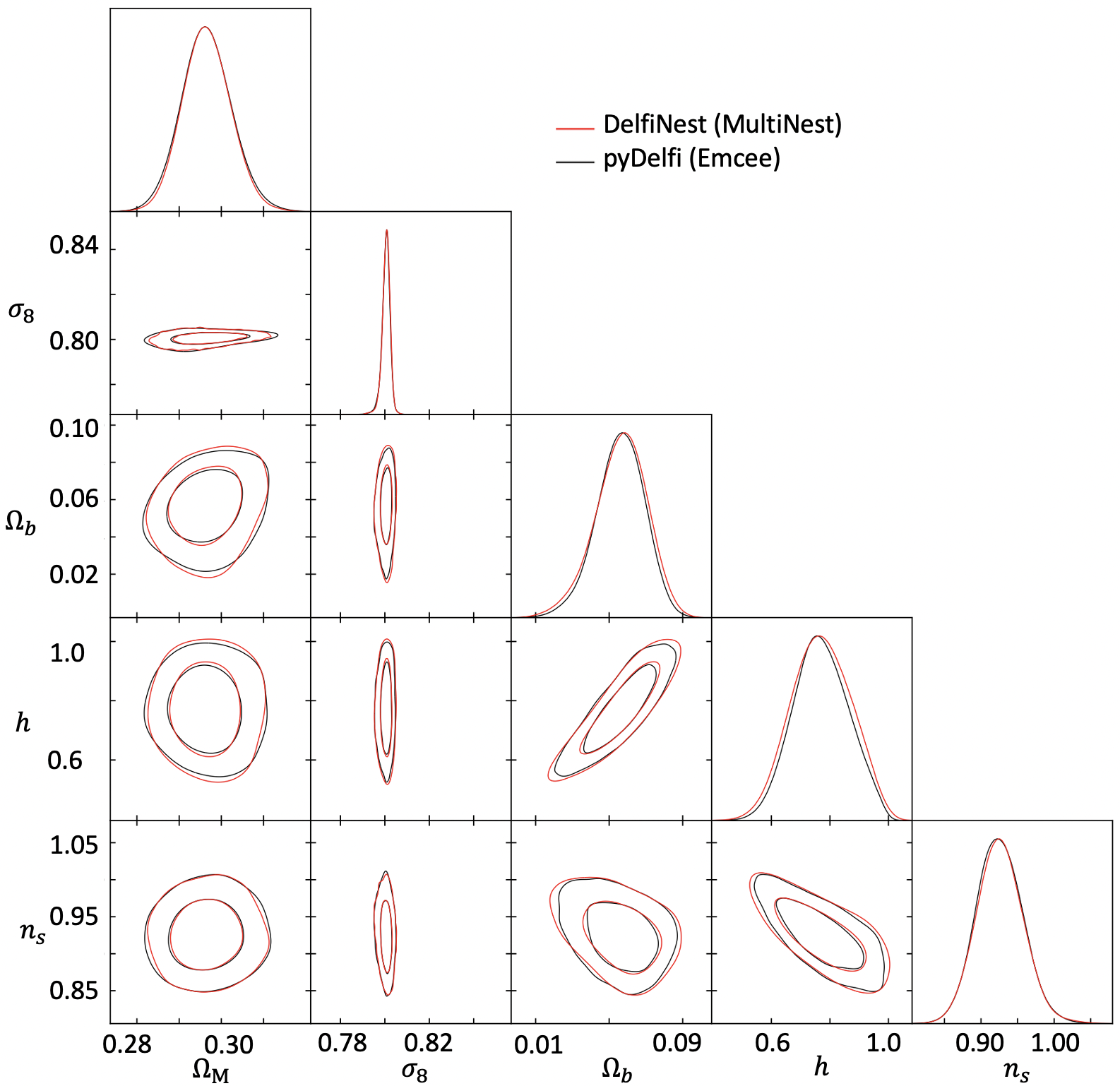}
    \caption{We compare the output of \textsc{Delfi-Nest} and \textsc{pyDelfi} on the cosmic shear model detailed in \citetalias{2019MNRAS.488.4440A}.
    In all cases, overlapping posterior distributions clearly show the agreement between the \textsc{Emcee} and \textsc{MultiNest} sampling techniques. 
    }
    \label{fig: Sampler_comparison}
\end{figure}

As a validation cross-check, we first show agreement between \textsc{pyDelfi} posteriors and our \textsc{Delfi-Nest} implementation. 
Figure \ref{fig: Sampler_comparison} shows overlapping posteriors produced by the two contrasting sampling mechanisms. 
We also test the model selection methodology by decisively rejecting a reduced parameter cosmic-shear model from the complete, well-established cosmic-shear model in Section \ref{sec: cosmic shear Model selection}. 
Both of these analyses use the cosmic shear model from\footnote{the cosmic shear model is freely available from the \textsc{pyDelfi} GitHub page \url{https://github.com/justinalsing/pydelfi}
} \citepalias{2019MNRAS.488.4440A}.
We found agreement between \textsc{MultiNest} and \textsc{pyDelfi}'s Evidence estimates when using the simple GMM network discussed in Section \ref{sec: pydelfi-nittygritty} (not shown). 
In Appendix \ref{sec: justify_INS}, we also perform further cross-checks of these results by showing that regular nested sampling agrees with importance nested sampling when calculating the Evidence from the joint-density posterior. 

\subsection{\textit{Cross-check}: Toy cosmic shear model comparison}\label{sec: cosmic shear Model selection}

In this section, we apply \textsc{MultiNest} to compare the full 5-parameter (5p) cosmic shear model against a reduced 2-parameter (2p) cosmic shear model. 
In this section, we aim to check the statistical machinery. 
The physics of cosmic shear and the details of these models are found in Appendix \ref{sec: Cosmic_shear}. 
We produce a mock data set each for the 5p and 2p cosmic shear models, labelled \textit{Good} and \textit{Bad}, respectively. 
The \textit{Good} mock data contains fiducial parameters that are in agreement with Planck 2018 observational data \citep{2020A&A...641A...6P} and the parameter posteriors obtained in the \textsc{pyDelfi} analysis in \citetalias{2019MNRAS.488.4440A}. 
By construction, the 2p model is equivalent to the 5p model but with three missing parameters, which are limited by $\delta-$function priors. 
The \textit{Bad} mock data contains a mean parameter shift of $2\sigma$ away\footnote{This $\sigma$ is taken from posteriors obtained when fitting the standard cosmic shear (5p) model to the corresponding \textit{Good} mock data (which agrees with observation).} from the Planck data. 
The fiducial parameters and specific prior details are shown in Table \ref{tab: cosmicShearParams}. 

As a hypothesis, the 2p model should be able to fit the \textit{Bad} mock data by construction but should struggle to fit the \textit{Good} data. 
The 5p should be able to fit both data sets, but it has a larger prior volume. 
We expect the 5p model to be favoured with \textit{strong} evidence on the \textit{Good} data and the 2p model to be favoured marginally on the \textit{Bad} data. 
This is reflected in the Bayes factors shown in Table \ref{tab: cosmicShearResults} where the 5p model is favoured $2566:1$ on the  \textit{Good} data and the 2p model is only favoured $26:1$ on the  \textit{Bad}. 
Although the 2p model falls into the \textit{moderate} category on the Jeffreys' scale, the change in prior volume between the two models is $\sim 20.8$. 
Since most of this result comes from prior penalisation, it is more appropriately associated with the \textit{weak} Evidence category. 

We consider the results of this section a success of Bayesian model selection in the context of likelihood-free posteriors and, therefore, proceed to analyse the toy EoR models. 

\begin{table}
    \centering
    \begin{tabular}{c|c c c c}
             & 5p Model & \textit{Good} Data  & 2p Model  &  \textit{Bad} Data \\ \hline 
         $\Omega_{\rm M}$ & [0,1] & 0.3  & [0,1] & 0.7  \\
         $\sigma_8$  & [0.4,1.2]  & 0.8 & $\delta(\sigma_8 - 0.9)$&  0.9  \\  
         $\Omega_{b}$ & [0,0.1] & 0.05  & $\delta(\Omega_{b} - 0.03)$ & 0.03  \\  
         $h$ & [0.4,1] & 0.67 & $\delta(h - 0.74)$ & 0.74 \\
        $n_{\rm s}$  & [0.7,1.3] &0.96 &  [0.7,1.3]  & 1.04 \\ 
     \end{tabular}
    \caption{The cosmic shear model parameter priors and the fiducial parameters used to calculate the cosmic shear mock data sets. 
    5p refers to the full (5-parameter) model used in \citetalias{2019MNRAS.488.4440A}, while 2p is a subset of this model that we aim to distinguish using the statistical machinery. 
    Square brackets indicate uniform distributions across the ranges above. 
    \textit{Good} and \textit{Bad} represent mock data sets calculated with the 5p and 2p models, respectively. 
    In the 2p model, there are two free parameters ($\Omega_{\rm M}$ \& $n_{\rm s}$). The other three parameters have $\delta$ functions placed on the fiducial parameter values for the \textit{Bad} mock data set. 
    The \textit{Good} data matches measured results while the \textit{Bad} data does not. }
    \label{tab: cosmicShearParams}
\end{table}

\begin{table}
    \centering
    \begin{tabular}{c|c}
     ${\rm ln}\mathcal{Z}(Good|5p)$ & -17.1 \\ 
      ${\rm ln}\mathcal{Z}(Bad|5p)$ & -15.7 \\ 
      ${\rm ln}\mathcal{Z}(Good|2p)$ & -24.9 \\ 
      ${\rm ln}\mathcal{Z}(Bad|2p)$  & -12.7 \\ \hline
      $\mathcal{B}(Good)_{5p,2p}$ & 2566 : 1 \\
      $\mathcal{B}(Bad)_{5p,2p}$ & 1 : 26 \\ 
    \end{tabular}
    \caption{
    The Evidence values (top half) calculated by \textsc{MultiNest} on the \textsc{pyDelfi} posteriors for the 5p and 2p cosmic shear models discussed in Section \ref{sec: cosmic shear Model selection}. 
    Each model has a corresponding mock data set where \textit{Good} and \textit{Bad} are associated with the 5p and 2p models, respectively. 
    On the bottom half of the table, we show the Bayes factors for each model recovering its own mock data set. 
    The \textit{Good} mock data set favours the 5p model, easily obtaining \textit{strong} Evidence on the Jeffreys' scale, while the \textit{Bad} data set favours the 2p model nearly ten times less favourably. 
    }
    \label{tab: cosmicShearResults}
\end{table}

\section{EoR Results}\label{sec: results}

\subsection{\textit{Cross-check}: Reproducing the \textsc{Delfi}-{\rm 3DCNN} results}\label{sec: XZcrosscheck}

\begin{figure}
    \centering
    \includegraphics[scale=0.4]{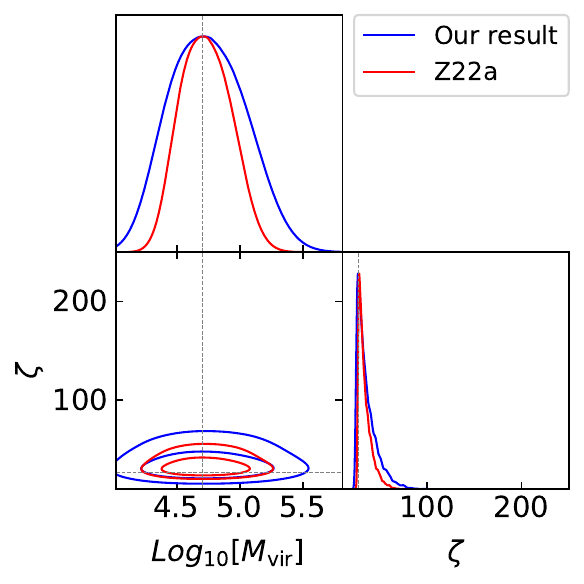}
    \caption{
    Both posteriors have been created with \textsc{Delfi-Nest}, the only difference being the box size used when creating the training \& mock data ($128~{\rm Mpc}^2$ in blue \& $66~{\rm Mpc}^2$ in red). 
    Both show that \textsc{Delfi-Nest} can recover the fiducial parameters used to create the mock data.
    In red, we show that \textsc{Delfi-Nest} recreates exactly the posteriors obtained with the \textsc{Delfi}-3DCNN in \citetalias{2022ApJ...926..151Z}. 
    In blue, the quality of the posterior fit degrades when only the box size is increased. 
    We discuss this further in Section \ref{sec: XZcrosscheck}. 
    }
    \label{fig: boxsize_posterior}
\end{figure}

The first results we obtain are fitting the FZH model to its own mock data set to recover the fiducial parameters used for the mock data. 
Parameter posteriors for a two-parameter \textsc{21cmFAST} faint-galaxy model from \textsc{21cmMC} are produced by compressing the lightcone into summaries with the 3D CNN and learning the posterior from the summaries with \textsc{Delfi-Nest}. 
In Figure \ref{fig: boxsize_posterior} we show (in blue) that the posterior MAP parameters agree with the fiducial parameters indicated by the grey dotted line. 

We also reproduce the procedure in \citetalias{2022ApJ...926..151Z} (shown in red) except: we do not rotate the density field box throughout the lightcone; the 3D CNN architecture has been updated since the \citetalias{2022ApJ...926..151Z} publication; and we use \textsc{MultiNest} for the sampling.
The only difference between our result and the \citetalias{2022ApJ...926..151Z} replication in Figure \ref{fig: boxsize_posterior} are from \textsc{21cmFAST}'s lightcone simulation parameters. 
We use a redshift range of 7.5 to 12 and boxes of 250 Mpc per side (128 pixels); \citetalias{2022ApJ...926..151Z} uses 7.51 to 11.67 and 66 Mpc (with 66 pixels), respectively. 
Both use the same redshift interpolation step size ($z_{\rm step} = 1.03$). 
Both recover the fiducial parameters used to produce the FZH mock data set; however, we obtain less precise posteriors than in \citetalias{2022ApJ...926..151Z}. 
We are able to reproduce the \citetalias{2022ApJ...926..151Z} posteriors (with exactly the same variances) when using the same simulation parameters. 

The broadening of the two posteriors therefore comes only from this change (and not from the seed rotation, sampler, or network updates). 
The posterior variances are, $\sigma_{\zeta} = 1.6$ \& $\sigma_{\rm logT_{\rm vir}} = 0.029$ (from \citetalias{2022ApJ...926..151Z} and re-simulated here) to $1.9$ \& $0.09$ (in Table \ref{tab: MAP}). 
The smaller box size used in \citetalias{2022ApJ...926..151Z} has a smaller $\sigma$ because there is a decrease in Poisson variance in the small-scale structure. 
Smaller boxes underestimate the frequency of small-scale structures in the lightcone \citep{2020arXiv200406709D}. 
Increasing the relative number of large structures will provide a signal that the 3D CNN can identify more clearly and therefore provide a more precise result. 
Since we do not include telescope noise or X-ray heating in these simulations, the choice of a smaller box in \citetalias{2022ApJ...926..151Z} is still valid; however, it should not be carried forward in future work. 

\subsection{Likelihood free model selection of EoR morphologies with \textsc{Delfi-Nest}}\label{sec: EoRresults}

\begin{figure*}
    \centering
    \subfigure[\label{fig: Bayesfactor FZH}]{\includegraphics[scale=0.24]{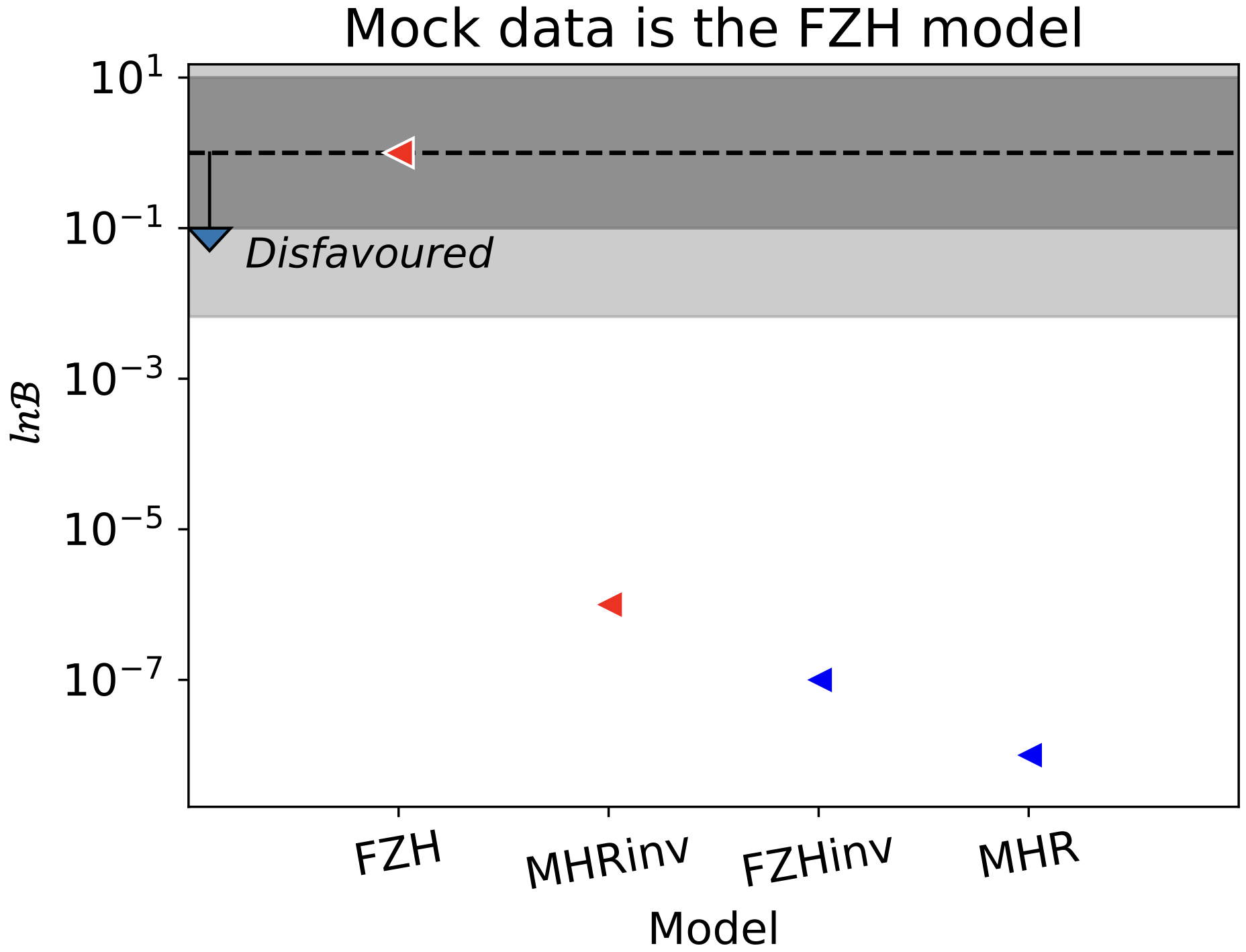}\hfill}
    \subfigure[\label{fig: Bayesfactor MHRinv}]{\includegraphics[scale=0.24]{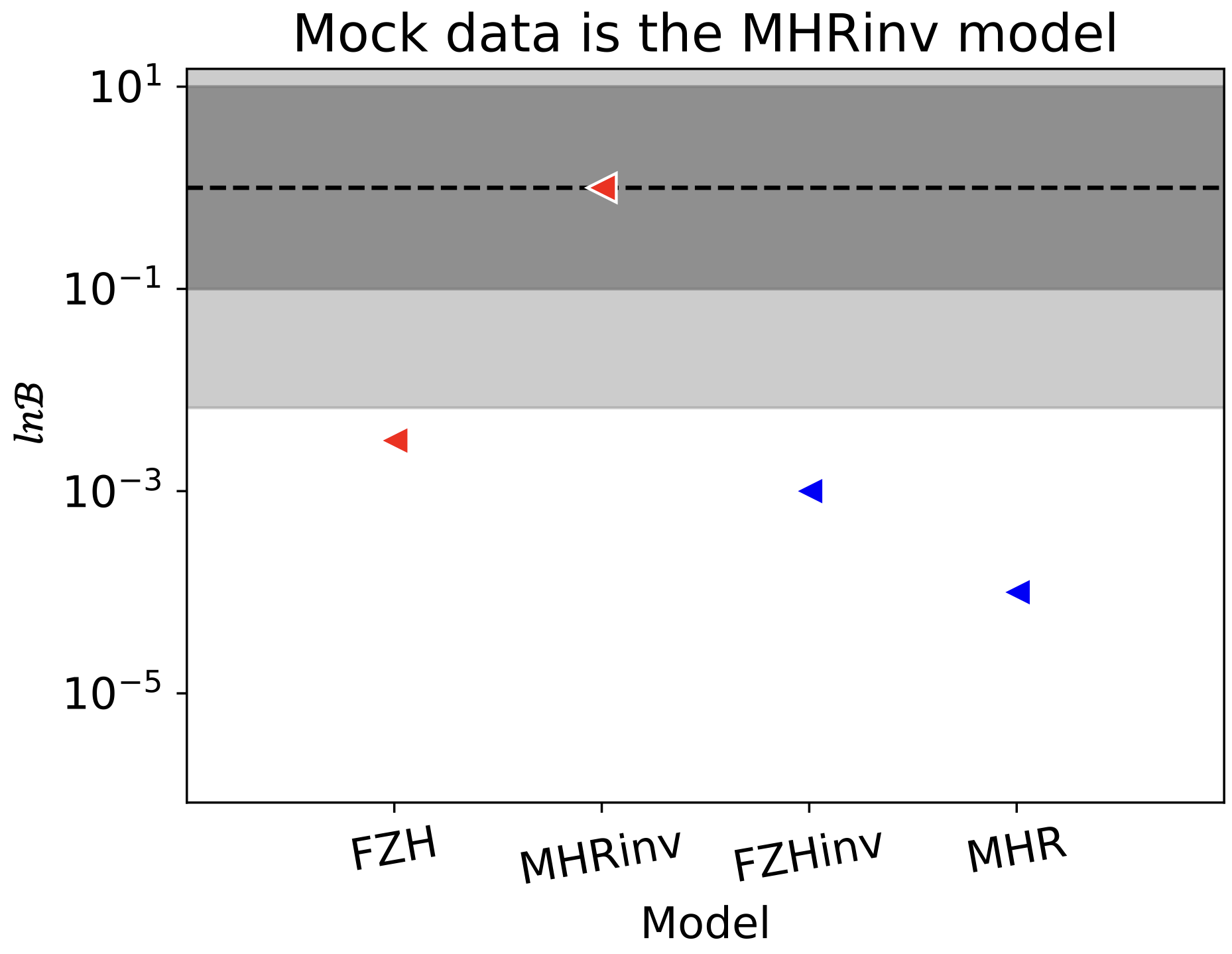}}
    \subfigure[\label{fig: Bayesfactor FZHinv}]{\includegraphics[scale=0.24]{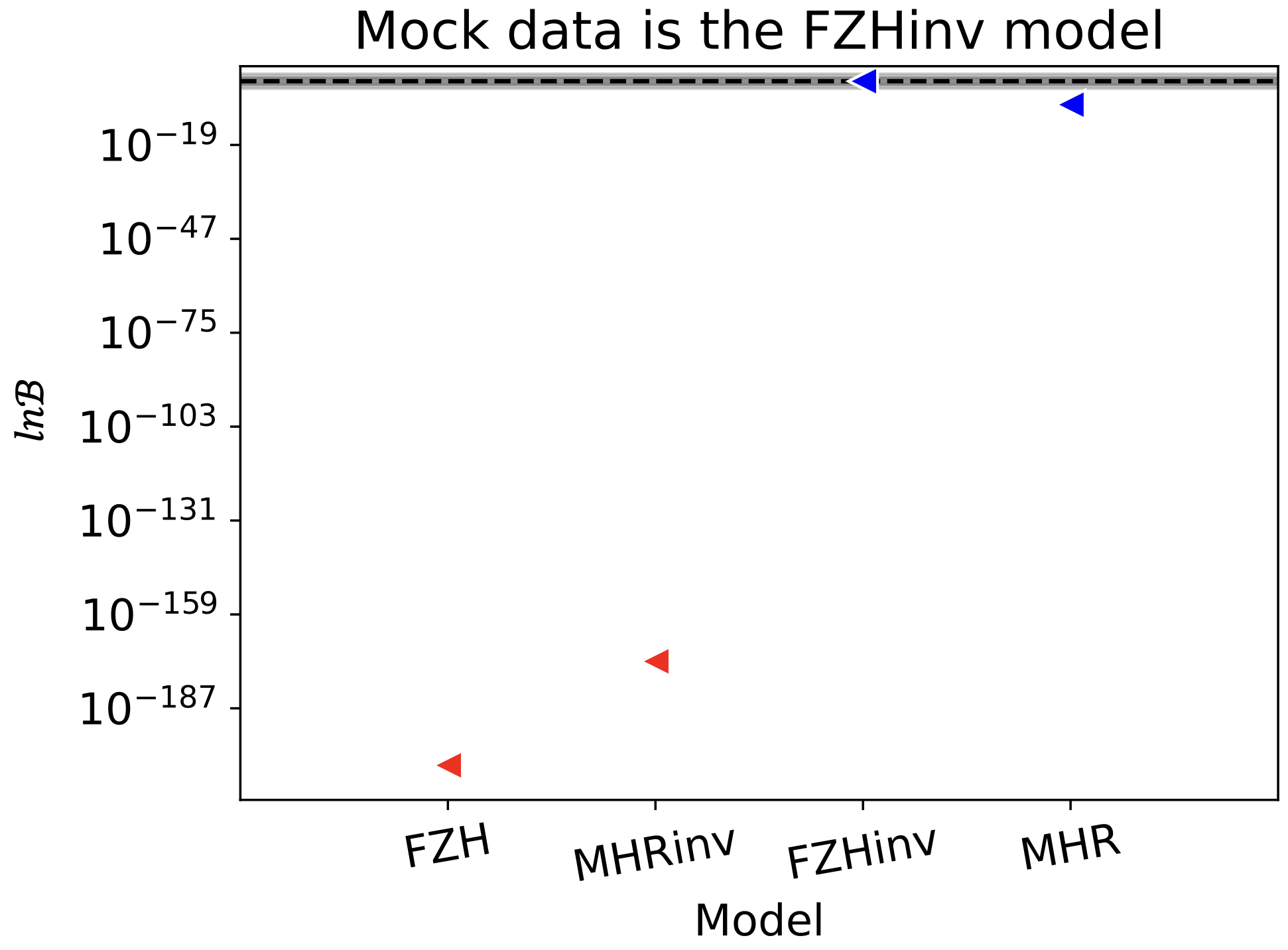}}
    \subfigure[\label{fig: Bayesfactor MHR}]{\includegraphics[scale=0.24]{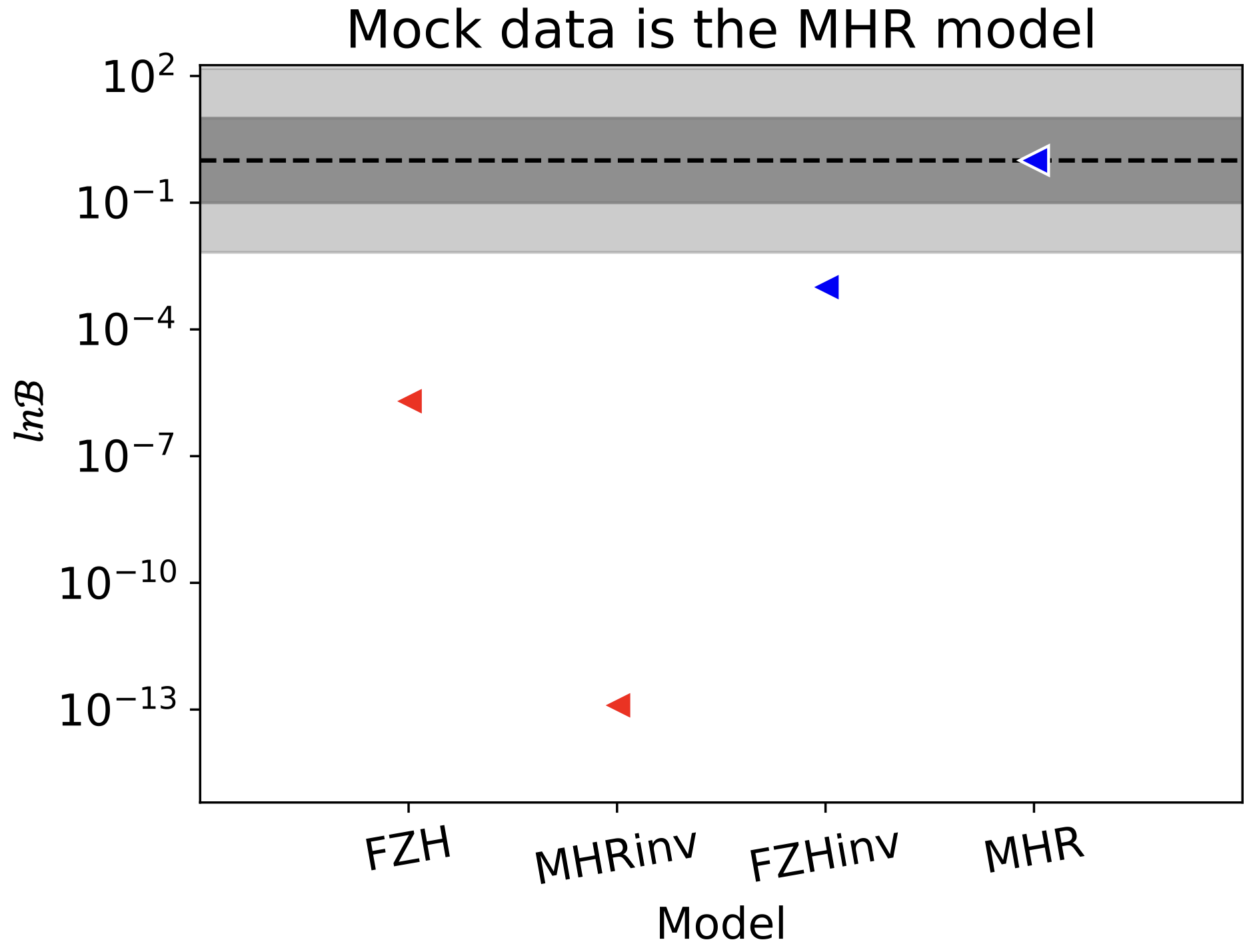}}
    \caption{
    The Bayes factor results when reproducing each of the four mock data sets created by the four toy EoR models. 
    Colour coding identifies the differing EoR morphologies as red for \textit{inside-out} and blue for \textit{outside-in}. 
    In each plot, the model which produced the mock data is favoured (outlined in white), with every other model being ruled out with \textit{strong} Evidence on the Jeffreys' scale (shown fading grey to white). 
    For each mock data set, the second best choice of model shares morphology with the mock data. 
    Please see Section \ref{sec: EoRresults} for more detail. 
    }
    \label{fig: BayesFactorresults}
\end{figure*}

In this section, we aim to distinguish the four EoR models introduced in Section \ref{sec: 21cm}, reproducing the main results from our previous work. 
We present the key Bayes factor results in this paper in Figure 
\ref{fig: BayesFactorresults}. 
The figure clearly shows that \textsc{Delfi-Nest} can favour the correct toy model when faced with each respective mock data set. 
In Figure \ref{fig: Bayesfactor FZH} the FZH mock data is used. 
The Bayes factors range $\mathcal{B} \sim 10^6:1$, $10^7:1$ to $10^8:1$ respectively, with MHR being most strongly rejected because it differs from MHR in morphology and uses a different ionisation definition. 
As FZHinv has an inverted ionisation threshold, it becomes somewhat unphysical and, therefore, fares worse than MHRinv, which shares \textit{inside-out} morphology with FZH. 
The next best model is the one that shares morphology, as expected. 
Figure \ref{fig: Bayesfactor MHRinv} uses the MHRinv mock data. 
FZH and FZHinv are rejected similarly with odds of $\mathcal{B} = 318:1$ and $1012:1$ respectively, again the model sharing morphology with the mock data is favoured. 
For the FZHinv mock data in Figure \ref{fig: Bayesfactor FZHinv}, the \textit{inside-out} models are rejected with \textit{strong} Evidence. 
MHR is also \textit{strongly} rejected with odds of $\mathcal{B} \sim 10^9:1$; these large Bayes factor values reflect that FZHinv is not a particularly physical model. 
Therefore, it is good that the machinery rejects the other models confidently when the FZHinv mock data set is used and that the other \textit{outside-in} model fares relatively well. 
Lastly, the MHR mock data set is used in Figure \ref{fig: Bayesfactor MHR} furthers the arguments above with the \textit{outside-in} FZHinv model being the closest contender with $\mathcal{B} = 998:1$. 
Of the \textit{inside-out} models, FZH is favoured over MHRinv with odds $\mathcal{B} \sim 10^6:1$ compared to $10^{13}:1$ respectively because it shares ionisation definition. 

\begin{figure}
    \centering
    \subfigure[\label{fig: FZH_post}FZH$|$FZH]{\includegraphics[scale=0.4]{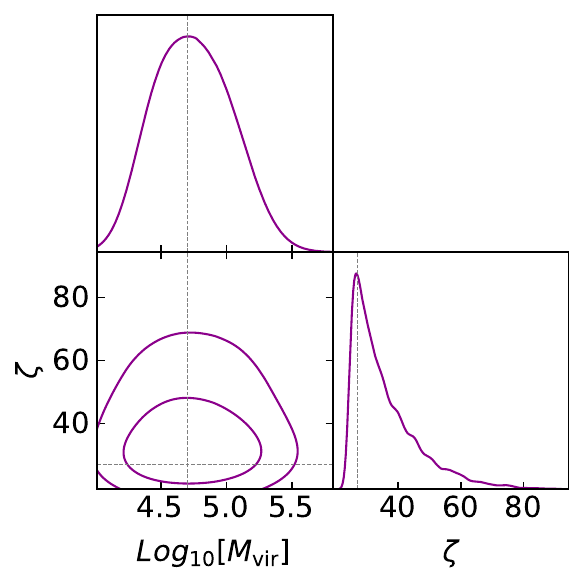}\hfill}
    \subfigure[\label{fig: MHRinv_post}MHRinv$|$MHRinv]{\includegraphics[scale=0.4]{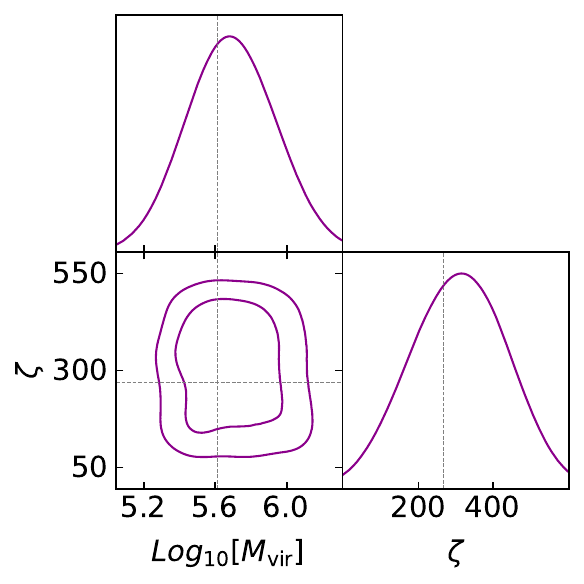}\hfill}
    \subfigure[\label{fig: FZHinv_post}FZHinv$|$FZHinv]{\includegraphics[scale=0.4]{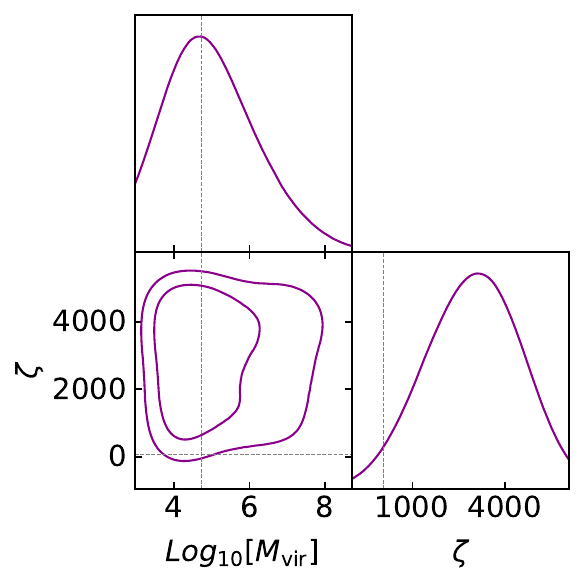}\hfill}
    \subfigure[\label{fig: MHR_post}MHR$|$MHR]{\includegraphics[scale=0.4]{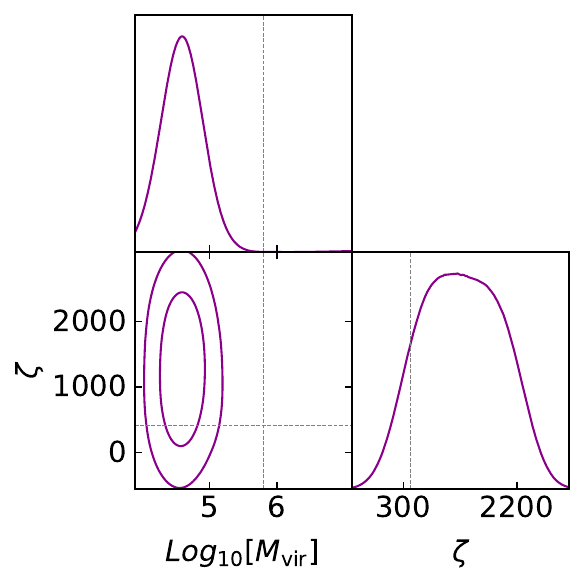}\hfill}
    \caption{
    The parameter posteriors obtained when each model is fit against its own mock data set. 
    These posteriors are a subset of the complete analysis shown in Table \ref{tab: MAP}. 
    Notice that the parameter posteriors for the \textit{inside-out} models (FZH and MHRinv) agree with their mock data set's fiducial parameters, while the  \textit{outside-in} models (MHR and FZHinv) are unable to obtain the correct input parameters. 
    We discuss this in the context of 3D CNN weightings in Section \ref{sec: discussion}. 
    }
    \label{fig: posteriors}
\end{figure}

\begin{table}
    \begin{flushleft}
    \begin{tabular}{l|c c c}
        \textbf{Data$|$Model} & ${\rm log}[T_{\rm vir}]_{\rm MAP }$ & $\zeta_{\rm MAP}$  & ${\rm ln}\mathcal{Z}$ \\ \hline 
        FZH$|$FZH& 4.67 $\pm$ 0.09 & 29 $\pm$ 1.9 & 14.7 $\pm$ 0.02 \\
        FZH$|$Finv& 6.00 $\pm$ 0.10 & 71.3 $\pm$ 4.9 & -453 $\pm$ 0.1 \\
        FZH$|$Minv& 5.54 $\pm$ 0.11 & 159 $\pm$ 18 & -0.71 $\pm$ 0.04 \\
        FZH$|$MHR& 5.28 $\pm$ 0.04 & 42.8 $\pm$ 25.1 & -3.65 $\pm$ 0.02 \\ 
        & & & \\
        Finv$|$FZH& 6.50 $\pm$ 0.41 & 1710 $\pm$ 1300 & -0.369 $\pm$ 0.023 \\
        Finv$|$Finv & 4.89 $\pm$ 0.67 & 2120 $\pm$ 1290 & 17.3 $\pm$ 0.1 \\
        Finv$|$Minv & 5.36 $\pm$ 0.19 & 3110 $\pm$ 550 & -2.94 $\pm$ 0.02 \\
        Finv$|$MHR& 5.54 $\pm$ 0.01 & 1250 $\pm$ 10 & 3.20 $\pm$ 0.01 \\ 
        & & & \\
        Minv$|$FZH& 5.64 $\pm$ 0.19 & 377 $\pm$ 211 & -0.26 $\pm$ 0.04 \\
        Minv$|$Finv & 6.00 $\pm$ 0.02 & 11.6 $\pm$ 1.6 & -417 $\pm$ 0.6 \\
        Minv$|$Minv & 5.61 $\pm$ 0.01 & 286 $\pm$ 141 & 5.03 $\pm$ 0.02 \\ 
        Minv$|$MHR& 4.71 $\pm$ 0.13 & 220 $\pm$ 208 & -21.3 $\pm$ 0.4 \\
        & & & \\
        MHR$|$FZH& 6.26 $\pm$ 0.10 & 805 $\pm$ 477 & -3.32 $\pm$ 0.03 \\
        MHR$|$Finv& 5.20 $\pm$ 0.07 & 425 $\pm$ 512 & 2.56 $\pm$ 0.03 \\
        MHR$|$Minv& 5.36 $\pm$ 0.05 & 3850 $\pm$ 1130 & -4.58 $\pm$ 0.30 \\
        MHR$|$MHR& 5.39 $\pm$ 0.07 & 1905 $\pm$ 682 & 9.42 $\pm$ 0.03 \\ 
    \end{tabular}
    \end{flushleft}
    \caption{The MAP parameters and Evidence values from each of our main EoR analyses. 
    MHRinv and FZHinv have been abbreviated to Minv and Finv, respectively. 
    Only posteriors using the same model and mock data set are plotted in Figure \ref{fig: posteriors}. 
    The Evidence values here are used to calculate the Bayes factor results in Figure \ref{fig: BayesFactorresults}.}
    \label{tab: MAP}
\end{table}

Our full analysis, which has been edited for brevity, includes four posteriors per each of the four models. 
In the posteriors shown we aim to recover the fiducial parameters each model used to create their respective mock data set. 
Table \ref{tab: MAP} shows the Evidence values used in the Bayes factor calculations as well as the maximum a posterior (MAP) parameters from the posteriors (a subset of the posteriors are shown in Figure \ref{fig: posteriors} while all possible Bayes factor combinations are plotted in Figure \ref{fig: BayesFactorresults}). 
In Figure \ref{fig: posteriors}, we present the parameter posteriors for each model only when fitting the same model's mock data set. 
The models with \textit{inside-out} morphology produce parameter posteriors that agree with the mock data fiducial parameters while the \textit{outside-in} models do not which is not what we expected. 
For FHZinv$|$ FZHinv (\ref{fig: FZHinv_post}), ${\rm log[T_{\rm vir}]}$ is recovered but $\zeta$ is $\sim3 \sigma$ too large despite a variance of 1290, which covers $25\%$ of the parameter prior\footnote{Variances that are a significant fraction of the prior range reflect a poor relationship between the parameter and its ability to influence the posterior. For example, if a posterior variance is $\sim 100\%$ of the prior range it implies there is nothing the analysis can infer about the properties of that parameter.}. 
In the MHR$|$MHR posterior (\ref{fig: MHR_post}), both variances are slightly better; however, ${\rm log[T_{\rm vir}]}$ is tightly constrained but to a value that is nearly $6\sigma$ higher than the fiducial parameter. 
For $\zeta$, the MAP value is $\sim 2.5 \sigma$ too large, with a variance $\sim12\%$ of the prior range. 
\ref{fig: FZH_post} (as in Section \ref{sec: XZcrosscheck}) shows the FZH$|$FZH posterior and reflects the 3D CNN performing well. 
As we have discussed earlier, the 3D CNN was designed in \citetalias{2022ApJ...926..151Z} to reproduce the MAP parameters of an FZH$|$FZH analysis. 
Figure \ref{fig: MHRinv_post} shows the MHRinv$|$MHRinv posterior for and shows a recovery of the fiducial parameters within $1\sigma$ for both parameters (although not as precisely as for FZH$|$FZH). 

We have validated Bayesian model selection while using likelihood-free inference in the context of reionisation, and we are confident that both networks are behaving as expected since we can reproduce both previous works, \citetalias{2019MNRAS.488.4440A} (Section \ref{sec: Delfi-Nest crosscheck}) and \citetalias{2022ApJ...926..151Z} (Section \ref{sec: XZcrosscheck}). 
We are also able to recover the input parameters for the \textit{inside-out} EoR models and reject models with \textit{strong} Evidence that have an \textit{outside-in} morphology. 
But in Figure \ref{fig: posteriors}, we are unable to reproduce the fiducial parameters for the MHR \& FZHinv mock data sets with those models respectively (in Figures \ref{fig: MHR_post} \& \ref{fig: FZHinv_post}), both of which are \textit{outside-in}, differing in morphology to FZH. 
In our previous work, we used the 21~cm PS likelihood to obtain the appropriate model's fiducial parameters in all cases. 
Since we are unable to do so in this work, we need to discuss the validity of the obtained Evidence values and Bayes factors. 
The following discussion aims suggest where assumptions in our methodology may have failed. 
We suggest that the issue is likely the flexibility of the 3D CNN since it was designed in \citetalias{2022ApJ...926..151Z} only for use on the FZH model within \textsc{21cmFAST}. 

\section{Discussion}\label{sec: discussion}

We separate this discussion into three major themes based on assumptions in the method (including information loss), artefacts in the simulation that may bias results, and our future outlook before distilling our conclusions and summarising in the final section. 

From the previous section we deduce that the \textit{outside-in} joint density distributions do not correctly map to posteriors because the MAP parameters for these models do not agree with the mock data parameters in Figure \ref{fig: posteriors}. 
The \textsc{Delfi-Nest} framework must have at least one incorrectly weighted network (whether that is in the 3D CNN summary space, or from the \textsc{Delfi} score compression used in the approximate likelihood, or both). 

Score compression is used in both parts of the cosmic-shear cross-check presented in Section \ref{sec: cosmic shear Model selection}, where the machinery successfully produces posteriors and distinguishes models. 
We therefore suggest that CNN architecture robustness needs to be scrutinised because score compression is lossless while the CNN compression is not. 

Summaries of the lightcone provided by the 3D CNN may not be sufficient to allow tractable mapping of the joint density to the posterior for all our morphological models. 
However our model selection results are promising. 
We conclude that these issues do not hinder our ability to provide usable Bayes factors but care must be taken with different models of EoR morphologies. 

Comparing to our previous work, both the 21~cm PS likelihood and \textsc{Delfi-Nest} can distinguish the four models of toy EoR morphology.
The 21~cm PS likelihood also allowed retrieval of the mock data parameters in all cases. 
In \citetalias{2022ApJ...926..151Z}, the 3D CNN compression can produce tighter parameter posterior constraints than when a likelihood is defined by summarising the lightcone with the 21~cm PS. 

If a CNN is going to be used on real data in the near future, we need to know if it can be trusted to distinguish morphologies. 
Physics suggests reionization began \textit{inside-out}: because structure initially developed in over-densities, where galaxy formation starts; and ended \textit{outside-in}: as an accumulating UV background ionised the remaining dense clumps of \hi\ that could not ignite star formation \citep{2019MNRAS.483.5301G}. 
As a result, observational data should be a mix of morphologies at all times. 
We think it is feasible that this network will have some comfort zone within which it performs well, but we are surprised to find that this comfort zone has been outstretched with \textit{outside-in} toy EoR models. This motivates concern in the inflexibility we have found when applying \textsc{Delfi-Nest} to our extreme morphologies. 

$\bullet$ \textbf{Assumptions}: 
%
Firstly, it is unclear if the proportionality in Equation \ref{eq: pseudoP} contains a dependence on the mock data parameter set (introduced in Section \ref{sec: pyDelfi}). 
Compression algorithms preserve information differently between models and depending on where the lightcones are simulated in the parameter space. 
The region of the parameter space used for the mock data in each case could cause the summary vector ($t$) to insufficiently describe the lightcone's information content. 
Regions of the parameter space may be suitable e.g. Figures \ref{fig: FZH_post} \& \ref{fig: MHRinv_post}, while for other regions it might fail e.g. \ref{fig: FZHinv_post} \& \ref{fig: MHR_post}. 
Therefore, \textit{outside-in} models likely have information losses that are sufficient to remove a tractable relationship between the posterior ($\mathcal{P}$) and joint density ($\rho$). 
While for \textit{inside-out} models, the 3D CNN conserves sufficient information so that \textsc{pyDelfi} can provide a reliable $\mathcal{P} \approx \rho$.
It is worth noting that if the summaries fed to \textsc{pyDelfi} are lossless then the proportionality in Equation \ref{eq: pseudoP} will become an equal sign giving, $\mathcal{P} = \rho$. 

Secondly, we do not know how the prior space transforms under data compression (and whether or not the summary space bound by the parameter prior edge is relevant when integrating the joint density in Equation \ref{eq: pseudoP}). 
The 3D CNN causes an extreme reduction in data volume, decreasing the Fisher information of the data. 
We have assumed that the summaries contain enough relevant information to proceed with our analysis. 
If too much information has been lost from the data set this will follow on to the accuracy of the posteriors. 
Posteriors are learnt with \textsc{pyDelfi} based on a cross-entropy loss function (and the data losses are not equal throughout each training set). 
Bias is therefore introduced depending on which lightcones can be compressed by the 3D CNN most efficiently. 
If Fisher information is fully preserved from the lightcone, a fraction of the samples required for an MCMC can provide useful posterior distributions \citep{2023MNRAS.524.4711M}. 
But if the information loss from compression is large, increasing the size of the training set is the most viable option to compensate so that enough relevant information can be salvaged by \textsc{pyDelfi} to produce meaningful posteriors. 
It might be the case that the $\textit{outside-in}$ models are compressed inefficiently in comparison to the $\textit{inside-out}$ models in which case our discrepancy in Section \ref{sec: EoRresults} could be solved by increasing the training data set size for the $\textit{outside-in}$ models (which is beyond the scope of this work). 

In \citetalias{2022ApJ...926..151Z} and \cite{2019MNRAS.484..282G}, the networks relate signal structure to the galaxy parameters by identifying the edges of bubbles. 
This must mean that bubble morphology contains at least similar information to the parameter posteriors. 
If the model is \textit{outside-in} however, the link between parameter posterior and the description of bubble topology is non-trivial. 
Parameters still relate to the galaxies, but as the network is focusing on the edges of structure these will be voids and the relationship between voids and the parameter posteriors is not as well understood (and in the case of FZHinv is non-physical). 
Part of this pitfall could be resolved if the \textit{outside-in} parameters were better chosen\footnote{An example of this would be re-parameterising, $\zeta$, as the instantaneous emissivity of the background radiation discussed in \citet{2000ApJ...530....1M}; or using, $f_{\rm coll}$, directly instead of the virial temperature.}.  
Because we are using a neural network, it is non-trivial to check parameter dependencies. 

To illustrate this with an example, For a Gaussian likelihood parameter dependency in the covariance matrix causes a bias in the results which can be measured and solved by marginalising over the parameters \citep{2016MNRAS.456L.132S}. 
A similar parameter dependence in the covariance of the network weights would feed into the joint-density (and the posterior) but we do not know how to measure it. 
It is also possible the score compression within \textsc{pyDelfi}  is parameter dependent. 
If so the precision of the posterior would worsen, but should remain accurate as using score compression in a Bayesian model selection context has been verified in \citet{2023JCAP...11..048H}. 
If parameter bias is present in the 3D CNN compression, the entire results could be inaccurate, potentially explaining the poorly fit posteriors when $\textit{outside-in}$ models are used in Figure \ref{fig: posteriors}. 

Regardless of these concerns, we have proceeded to integrate the posteriors obtained with the same prior ranges used to calculate the training data. 
Since the joint density is learnt from the data summaries, it is non-trivial to write down how the effect of varying parameter priors affects the learnt Evidence. 
As a preliminary check, we plotted the parameter distribution of the training data and how that distribution transformed when calculating summaries from the 3D CNN. 
However, we were unable to discern any usable inference from this process. 
The summary space has been transformed away from the prior in a non-negligible way in each case, as can be seen in Appendix \ref{sec: alternate_priors}. 
In this appendix, we found the summary space to be unchanged to $<1\%$ when increasing or decreasing the parameter prior by $10\%$ of the original range, suggesting the summary space coverage is converged for each model. 

$\bullet$ \textbf{Simulation artefacts}: semi-numerical techniques contain various non-physical mechanisms for increasing efficiency while producing accurate results (e.g. the \citet{1970A&A.....5...84Z} approximation). 
We are concerned that the 3D CNN will identify non-physical simulation features that might bias the results. 
To try and cross-check our results against this behaviour, we present two alternate analyses in Appendix \ref{sec: remedies}. 

Broadly, we assess this by altering the network architecture to a 2D-CNN (Appendix \ref{sec: alt_cnn}). 
We found the results produced by this network to be identical (within $0.1\%$) to those made by the 3D CNN. 
We also alter the NDE structure within \textsc{pyDelfi} to gain negligible differences (stated in Appendix \ref{sec: NDEs}). 

The second analysis looks at a specific problem where structure is correlated along the \textsc{21cmFAST} lightcone due to the wrapping of a single density field along the line-of-sight. 
We present an alternate simulation that draws a new density field for each coeval cube that is wrapped into the lightcone in Appendix \ref{sec: no_recurring_seeds}. 
Our appendix concludes that these correlations are insignificant. 
It is well documented (within the machine learning field) that CNNs contain a quantifiable translation-invariance \citep{DBLP:journals/corr/abs-1801-01450}, particularly when applied without knowledge of the desired solution. 
This is remedied by adapting the kernels with filters specific to the problem, however knowledge of the solution is required \citep{8578193, 9156444}. 

Unknown artefacts in 21~cm data will worsen when noise (and noise cleaning) is included in the data. 
If CNNs are used without adequate cross-checks, observational artefacts may lead to a bias toward specific simulation techniques. 
In this work all our EoR models are using the same underlying simulation assumptions which may explain why we have been unable in Appendix \ref{sec: remedies} to obtain results that differ from Section \ref{sec: results}. 

$\bullet$ \textbf{Outlook}: future work requires training sets that use models with richer astrophysics. 
We also defer analysis of the flexibility of CNN data compression with more subtle simulation parameters (such as box size, pixel size, redshift step length and the maximum heating redshift) to future work. 
In \citet{2020MNRAS.498..373P}, the correlation between the density and ionisation fields is parameterised allowing access to a range of mixed morphologies. 
A model based on this work could be used in future work to pinpoint the comfort zone for reliable posteriors produced with \textsc{Delfi-Nest}. 

An alternative to using an approximate likelihood is introduced in \citep{2018PhRvD..97h3004C}, where an optimal compression scheme is learnt by maximising the Fisher information instead. 
This information maximising neural network (IMNN) technique will improve on the approximate likelihoods used here, but will suffer from being asymptotically optimal given the choice of data summary (as does \textsc{pyDelfi}). 
Other possible alternatives include Recurrent Neural Networks \citep{2022MNRAS.509.3852P} as well as the wavelet scattering transform \citep{zhao2023simulationbased} which have both been shown to outperform the standard CNNs used here.
The latter concluding that standard CNN's have a limited flexibility for adequately training on varied data sets because they can only adapt as far as the complexity of the network architecture allows (which is constant after the network has been designed). 
In \citet{2022MNRAS.511.3446N}, 3D CNN 1st layer kernels are elongated along the line-of-sight to help the network adapt to structure evolution. 
We defer a further analyses with different data summaries, non-standard CNN's, larger training sets and other joint density estimation techniques to future work. 

\section{Conclusion \& Summary}\label{sec: conc+summ}

We have achieved likelihood-free Bayesian Model selection in the context of Reionisation for the first time by combining multiple frameworks into the publicly available\footnote{\url{https://github.com/binnietom/Delfi-Nest}}  \textsc{Delfi-Nest}. 
We learnt parameter posteriors with \textsc{pyDelphi} (\citetalias{2018MNRAS.476L..60A}, \citetalias{2018MNRAS.477.2874A}, \citetalias{2019MNRAS.488.4440A}) from distributions of summaries produced by the 3D CNN \citepalias{2022ApJ...926..151Z} which compressed 21~cm lightcones simulated by \textsc{21cmFAST} \citep{2011MNRAS.411..955M, 2020JOSS....5.2582M}. 

To summarise the method, \textsc{pyDelfi} uses a model to fit the joint density, $\rho(t, \theta)$, which is then used to learn the posterior by recursively evaluating $\rho$ at the mock observed data, $D(\theta_f)$. 
In this work, observed data refers to a mock data set simulated with fiducial parameters, $\theta_f$. 
The posterior can therefore be obtained as, $\mathcal{P} \propto \rho[\theta | D(\theta_f)]$.
We also altered the ionisation model within \textsc{21cmFAST} to produce three alternate EoR models containing either \textit{inside-out} or \textit{outside-in} morphology based on \citet{2014MNRAS.443.3090W}. 
By training four 3D CNNs with four training sets, each simulated with a different toy EoR model, we have been able to decisively distinguish a mock data set which represents each of the four toy models by feeding each network into \textsc{Delfi-Nest}. 
To perform model selection, we integrated the learnt posterior distributions with \textsc{MultiNest} to produce the Bayes factor, an odds ratio describing which model is most likely to produce a data set. 
Each of the four EoR models produced a mock data set by simulating a lightcone with fiducial parameters. 
By applying Bayesian model selection, we comfortably distinguished which model was used to create the mock data set in each case. 
The next best model in all cases shared morphology with the mock data model. 
However, only \textit{inside-out} models produce posteriors which recover accurate fiducial parameters. 

We have discussed using CNNs outside their intended construction purpose and highlighted where concern should be applied. 
The 3D CNN compression is not lossless while \textsc{pyDelfi} uses score compression, which is lossless. 
Information losses in the 3D CNN are likely model and parameter dependent. 
In brief, the leeway in which CNNs can be reliably applied must be quantified to be trusted within any broader statistical analysis context, such as model comparison. 
We have shown that the Evidence obtained by integrating the joint density can provide a Bayes factor that is useful within model comparison. 

However, the use of a CNN obscures a reliable tracking between the results and the simulation. 
Optimisation of 3D CNNs is non-trivial when transferred between different model morphologies. 
Concerns of bias from the Bayesian prior are also unknown in the context of likelihood-free model selection as transformation to the CNNs summary space is highly non-linear. 
Our attempted remedies and cross-checks have not debunked our issues; however, the model comparison results agree with our previous work \citep{2019MNRAS.487.1160B}. 

In comparison to our previous work, there is a significant speed increase while using \textsc{Delfi-Nest} \& we see an improvement of $\sim 50\%$ on the parameter posterior variance for FZH$|$FZH. 
However, this improvement is similar to changes reported when simulating lightcones rather than coeval cubes as in \citet{2018MNRAS.477.3217G}. 

If we are going to take advantage of all the information available in the 21~cm lightcone, further work needs to be done on how information content differs between more detailed simulations and semi-numerical techniques.  
More detailed end-to-end analysis pipelines need to be developed and analysed. 
Despite having achieved our initial hypothesis of distinguishing EoR morphologies, using the 3D CNN for EoR model selection is significantly less flexible and less transparent than using a 21~cm PS likelihood as in our previous work. 

\section*{Acknowledgements}
This work is supported by the National SKA Program of China (grant No.~2020SKA0110401) and NSFC (grants No.~12350410353 and No.~11821303). We thank Richard Grumitt for the useful discussions. We acknowledge the Tsinghua Astrophysics High-Performance Computing platform at Tsinghua University for providing computational and data storage resources that have contributed to the research results reported within this paper.

\software{\textsc{21cmFAST} \citep{2011MNRAS.411..955M, 2020JOSS....5.2582M}, 
3D CNN \citep{2022ApJ...926..151Z}, 
\textsc{pyDelfi} \citep{2019MNRAS.488.4440A}, 
\textsc{MultiNest} \citep{2008MNRAS.384..449F, 2009MNRAS.398.1601F, 2019OJAp....2E..10F}, 
\textsc{Emcee} \citep{2013PASP..125..306F},
\textsc{PocoMC} \citep{2022JOSS....7.4634K},
\textsc{NestCheck} \citep{2019MNRAS.483.2044H}, 
scipy \citep{DBLP:journals/corr/abs-1907-10121}, 
numpy \citep{DBLP:journals/corr/abs-2006-10256}, 
GetDist \citep{lewis2019getdist},
matplotlib \citep{4160265}, 
Python3 \citep{10.5555/1593511}, 
Keras \citep{chollet2015keras}, 
scikit-learn \citep{Pedregosa2011scikit-learn},  
draw.io (\url{https://www.drawio.com})
}

\appendix

\section{Approximate Bayesian Computation (ABC)}\label{sec: ABC}

In most applied analyses, a direct formulation of the Bayesian likelihood is intractable. 
As an alternate way to gain inference from a forward model, ABC methods can provide reliable parameter estimation and model selection results in various contexts \citep{2008ConPh..49...71T, al2009bayesian,2015JCAP...08..043A}. 
To implement ABC we rely on a distance metric, $x$, being smaller than a user-defined threshold, $\epsilon$ so that when,
\begin{equation}\label{eq: ABC}
    x(D_{\rm true}, D_{\rm sample})  \leq \epsilon ,
\end{equation}
an approximate posterior is produced by assembling the parameters that satisfy Equation \ref{eq: ABC}. 
The ABC posterior agrees with the true posterior defined in Equation \ref{eq: Bayestheorem} if, $x$ captures the desired properties and, $\epsilon$ is suitably small. 
Within \textsc{PyDelfi} for example, minimising the cross-entropy between networks derived from summaries of the mock data and sampled data in Equation \ref{eq: KL-divergence} represents $x$,  providing the joint density distribution which can be sampled with an MCMC algorithm (with convergence criteria, $\epsilon$).
when enough measurements have been iterated, ABC techniques can produce reliable posterior distributions. 

Analytic likelihoods are not necessarily better than ABC. 
Bayesian parameter estimation is well established with both analytic likelihood and likelihood-free methods. 
The latter is possible because conditional density estimation can be free of a traditional likelihood and still be Bayesian through various neural density estimation techniques \citep{2016arXiv160506376P}, like those used in \citetalias{2018MNRAS.476L..60A}, \citetalias{2018MNRAS.477.2874A}, and \citetalias{2019MNRAS.488.4440A}. 
It is important to note that to be suitably Bayesian, the summary \& likelihood (or distance metric) must capture all of the information about how the data is conditional on the parameters for a specific model \citep{jaynes03}. 
In this work, we are concerned with the validity of integrating posteriors created by ABC through \textsc{pyDelfi} with \textsc{MultiNest} in the same way as our previous work does for the analytic likelihoods defined in \textsc{21cmMC}  \citep{2015MNRAS.449.4246G, 2018MNRAS.477.3217G, 2017MNRAS.472.2651G}. 

\section{Alternate Analyses for robustness}\label{sec: remedies}

\subsection{\textit{Cross-check}: using a 2D-CNN network architecture}\label{sec: alt_cnn}

In Section \ref{sec: discussion} we discuss the flexibility of applying the 3D CNN to differing toy EoR models. 
To validate our discussion, we repeated the results of this paper with a 2D-CNN architecture for comparison, motivated by \citet{2019MNRAS.484..282G}.  
Table \ref{tab: NN architectures} shows the original 3D CNN and the alternative 2D-CNN architectures in full. 
Both \citetalias{2022ApJ...926..151Z} and \citet{2019MNRAS.484..282G} acknowledge that the networks are primarily sensitive to the edges of bubbles throughout the 21~cm lightcone structure. 
Therefore, we expect both networks to be similarly successful in model selection, given the minor improvements in the precision of parameter posterior distributions reported in \citetalias{2022ApJ...926..151Z}. 

Our model selection results change insignificantly between the two networks. 
The results of this section are, therefore, as presented in Section \ref{sec: EoRresults}. 
We do not report the same improvement on the precision of $\zeta$ parameter posterior when using the 3D CNN in \citetalias{2022ApJ...926..151Z} and the 2D-CNN in \citet{2019MNRAS.484..282G}. 
The change in precision reported is likely due to the added degeneracy of the reionisation parameters when comparing the simplified FZH model against the more complex model with X-ray heating \citet{2007MNRAS.376.1680P, 2017MNRAS.472.2651G}. 
We therefore deem the improvement of precision between these two architectures insignificant. 
Differences are also likely to arise from changing the prior on $\zeta$ from uniform to log (given the fiducial parameter value for $\zeta$ is low relative to the prior width); however we measured a 1\% change on MAP parameters during the development of our previous work and this agrees with the analyses shown in the appendices of \citetalias{2022ApJ...926..151Z} (their Table 4). 

\begin{table*}
    \begin{tabular}{l l l|l l l}
            & \textbf{3D CNN} & & & \textbf{2D-CNN} & \\ \hline \hline 
        Layer & Output Shape & Parameter No. & Layer & Output Shape & Parameter No. \\ \hline
        Input & [128, 128, 534, 1] & 0 & Input & [128, 128, 534]  & 0 \\
        3D Convolution & [62, 62, 265, 32] & 4032 & 2D Convolution & [97, 126, 1] & 51265 \\
        Batch Normalisation& [62, 62, 265, 32] & 248 & Batch Normalisation & [97, 126, 1] & 4 \\
        Activation & [62, 62, 265, 32] & 0 & Activation & [97, 126, 1] & 0 \\
        3D Max Pooling & [31, 31, 132, 32] & 0 & Batch Normalisation & [97, 126, 1] & 4 \\
        Zero Padding 3D & [33, 33, 134, 32] & 0 & Activation & [97, 126, 1] & 0\\
        3D Convolution & [29, 29, 130, 64] & 256064 & 2D Max Pooling & [48, 63, 1] & 0\\
        Batch Normalisation & [29, 29, 130, 64]  & 116& 2D Convolution & [17, 32, 3] & 3075\\
        Activation & [29, 29, 130, 64]  & 0& Batch Normalisation & [17, 32, 3] & 12\\
        3D Max Pooling & [14, 14, 65, 64]  & 0& Activation & [17, 32, 3] & 0\\
        3D Spatial Dropout & [14, 14, 65, 64]  & 0& 2D Convolution & [2, 2, 1] & 289\\
        3D Zero Padding & [16, 16, 67, 64]  & 0& Batch Normalisation & [2, 2, 1] & 4\\
        3D Convolution & [12, 12, 63, 128] & 1024128 & Activation & [2, 2, 1] & 0\\
        Batch Normalisation & [12, 12, 63, 128] & 48 & 2D Max Pooling & [1, 1, 1] & 0\\
        Activation & [12, 12, 63, 128] & 0 & Layer Reshape & 1 & 0\\
        3D Max Pooling & [6, 6, 31, 128] & 0 &  Dense & 512 & 1024\\
        3D Spatial Dropout & [6, 6, 31, 128] & 0 & Batch Normalisation & 512 & 2048\\
        Flatten & 142848 & 0 & Activation & 512 & 0\\
        Dense & 64 & 9142336 & Dropout & 512 & 0\\
        Batch Normalisation & 64 & 256 & $[{\rm Dense, Dense}]$ & [1,1] & [513,513]\\
        Activation & 64 & 0 & & & \\
        Dropout & 64 & 0 & & & \\
        Dense & 16 & 1040 & & & \\
        Batch Normalisation & 16 & 64 & & & \\
        Activation & 16 & 0 & & & \\
        Dense & 4 & 68 & & & \\
        Activation & 4 & 0 & & & \\
        $[{\rm Dense, Dense}]$ & [1,1] & [5,5]& & & \\
    \end{tabular} 
  \caption{Detailed descriptions of the 3D CNN architecture (left) and the alternate 2D-CNN architecture we used for a supplementary analysis. 
  Each layer is connected to the previous row in the table. 
  The final output summaries in both cases are a pair of dense layers connected to an activation layer and a dropout layer for the 3D CNN and 2D-CNN, respectively.
  Both networks use a ReLU activation function.
  The $(n+1){\rm th}$ quantity in each table represents the number of channel divisions produced throughout the procedure.} 
  \label{tab: NN architectures}
\end{table*}

\subsection{\textit{Cross-check}: no recurring seeds}\label{sec: no_recurring_seeds}

\begin{figure*}
    \centering
    {\includegraphics[scale=0.38]{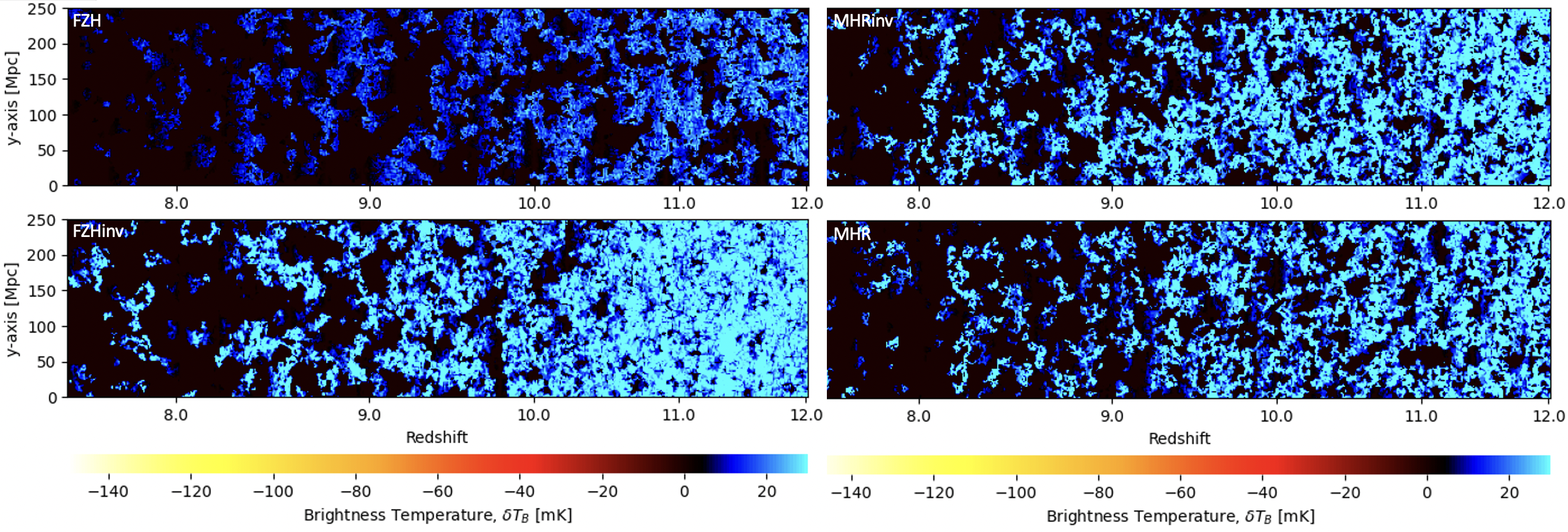}\hfill}
    \caption{\label{fig: LCs}
    Brightness temperature lightcones for the four toy EoR scenarios plotted $z \sim [7.5,12]$ similar to Figure \ref{fig: Toy_model_light-cones} but without any recurring density field seeds. 
    This unfruitful procedural change to \textsc{21cmFAST} is detailed in Section \ref{sec: no_recurring_seeds}. 
    }
    \label{fig: LCs_nowrapping}
\end{figure*}

In \citetalias{2022ApJ...926..151Z}, rotating the density field boxes before inputting them into a 21~cm lightcone is said to help reduce the correlation of the structure that can be measured in the brightness temperature co-variance. 
Repeated structures can be seen by eye in the lightcones in Figure \ref{fig: Toy_model_light-cones}. 
When using these lightcones to train the 3D CNN, as we do in Section \ref{sec: 3D-CNN}, we can reproduce the posteriors obtained by \citetalias{2022ApJ...926..151Z} (Figure \ref{fig: boxsize_posterior}). 
In this Section, we take a different approach to addressing the correlation of structure caused by wrapping coeval cubes into the lightcone.
Here, we draw a different random seed for every coeval cube used. 
Although the repeated structures no longer occur in the lightcone, we find that different structures fade in and out of the lightcone as the coeval cubes are interpolated. 
The resulting lightcone signal is noticeably not as smooth as before, as can be seen as we re-plot the toy model lightcones (Figure \ref{fig: LCs_nowrapping}). 

Training the 3D CNN with lightcones that have no recurring seeds causes negligible changes to the behaviour of the 3D CNNs when re-trained and fed into \textsc{Delfi-Nest}. 
We find this result somewhat surprising and defer an in-depth analysis into the continuity of large-scale line-of-sight structure modes when simulating 21~cm lightcones to future work. 
The following subsection contains a more detailed description of how the \textsc{21cmFAST} lightcones are produced and discusses why this is important. 

\subsection{Stitching coeval cubes into the lightcone}\label{sec: light-cone-detail}

The standard \textsc{21cmFAST} lightcone is constructed by multiple coeval cubes simulated at different redshifts before being stitched together to emulate the desired structure evolution along the line of sight \citep{2018MNRAS.477.3217G}. 
Redshift space sampling is done with a linear interpolated across cosmic time (we use a redshift interpolation step size, $\Delta z_{\rm step}=1.03$ which provides coeval cubes at redshifts 7.5, 7.76, 8.02, 8.29, 8.57, 8.85, 9.15, 9.45, 9.77, 10.09, 10.42, 10.77, 11.12, 11.48, 11.86, \& 12.24, with structures repeating every $\sim 1.03$ redshifts). 
The lightcone's properties along the line-of-sight are constructed by interpolating the properties of the two nearest coeval boxes.
For example, if the desired redshift matches one coeval box, the lightcone properties will match that coeval cube. 
For redshifts in between two boxes, the properties of the next coeval box will gradually fade in (with a weighted mean) until the next sampled redshift is reached and the process repeats. 
The structure modes will, therefore, repeat along the line-of-sight every redshift interpolation step. 
Since only one coeval density field needs to be simulated per lightcone, the computational burden is significantly reduced; however, a similar structure is visibly repeated periodically along the lightcone.  

When analysing reionisation, the spherically averaged 21~cm PS is the most typical summary statistic for the brightness temperature. 
It is worth noting that the cosmological principle should cause the 21~cm PS to be spherically symmetrical; interferometers lend themselves to producing it; and a hearty proportion of EoR physics can be constrained with it, see e.g. \citet{2012RPPh...75h6901P}. 
Not including any evolutionary precision along the line-of-sight can cause the parameter posteriors to be skewed by up to $\sim 10 \sigma$ when the 21~cm signal evolves rapidly, e.g. \citet{2015MNRAS.449.4246G, 2018MNRAS.477.3217G}. 
For implementation in the \textsc{21cmMC} likelihood, equal comoving distance chunks of the lightcone are used to stack 21~cm PS into the likelihood. 
This is not dissimilar to observing discrete frequency bands dictated by radio interferometer bandwidths (like running a non-overlapping top-hat filter across the lightcone). 
However, box-car sampling causes the lightcone effect, where the evolution of structure across the line-of-sight is not accounted for within each chunk and will bias the results \citep{2012MNRAS.424.1877D, 2014MNRAS.442.1491D}. 
When measuring the 21~cm PS from lightcone chunks, any periodicity-related bias in the simulation method ends up being outweighed by the non-ergodicity of the 21~cm PS because the lightcone is sliced into chunks. 

The need for statistics beyond the power spectrum is well documented, e.g. \citet{2016MNRAS.461..126T, 2017MNRAS.472.2436W, zhao2023simulationbased}. 
Arguments typically involve the demand for increased sensitivity on small angular scales and the quantification of 21~cm non-Gaussianity (as well as deviation from the cosmological principle). 
We wish to emphasise that the properties of large-scale line-of-sight structure modes within simulation should be more prominently included in this context. 
It is unclear whether structure modes of larger lengths will alter the behaviour of summaries that use the entire lightcone length. 
Our attempts to remedy correlated structure formation along the line-of-sight have failed. 
But we have found that the 3D CNN summaries are just as usable with a different large-scale line-of-sight structure. 
We therefore deem that correlation of structure in the lightcone is insignificant in this analysis. 
However, to faithfully asses the behaviour of lightcone summaries with lightcone-length line-of-sight structure modes, a more continuous lightcone simulation is needed. 
We defer this further analysis to future work. 

\subsection{\textit{Cross-check}: Changing prior ranges}\label{sec: alternate_priors}

\begin{figure*}
    \centering
    \subfigure[\label{}]{\includegraphics[scale=0.2]{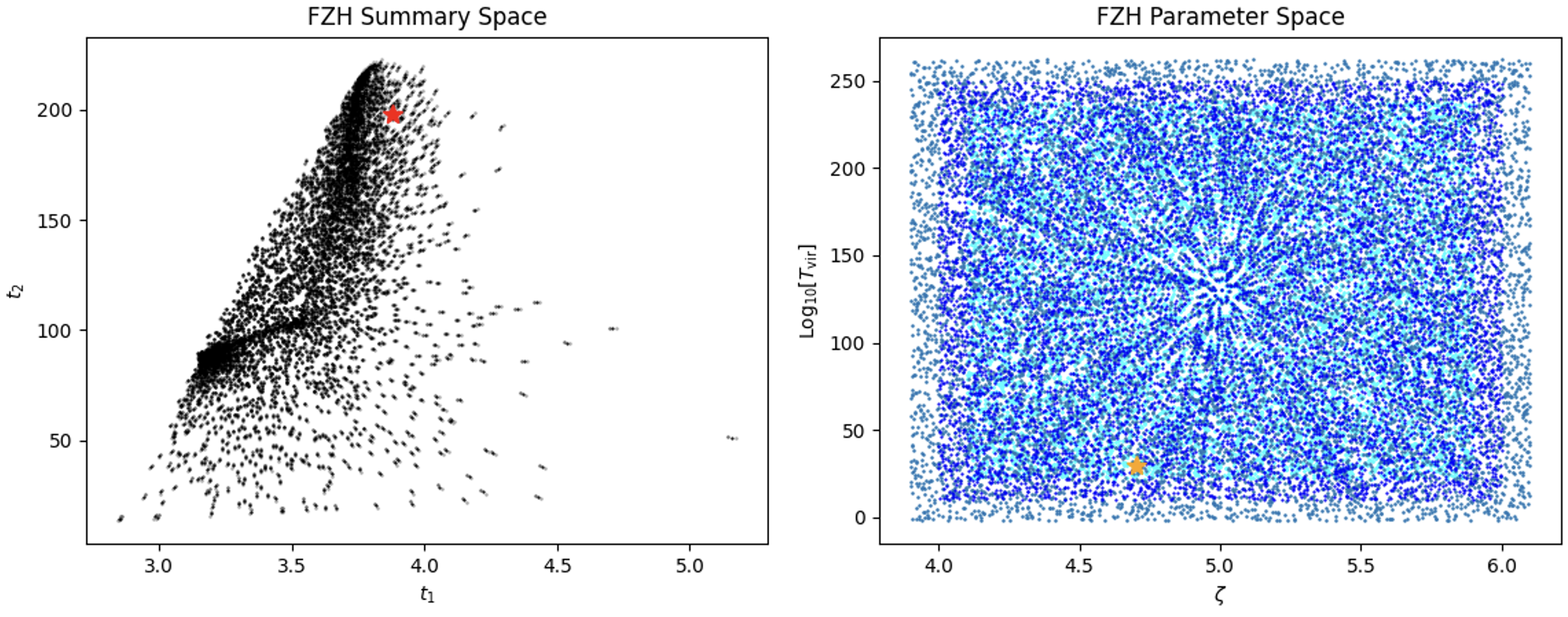}\hfill}
    \subfigure[\label{}]{\includegraphics[scale=0.2]{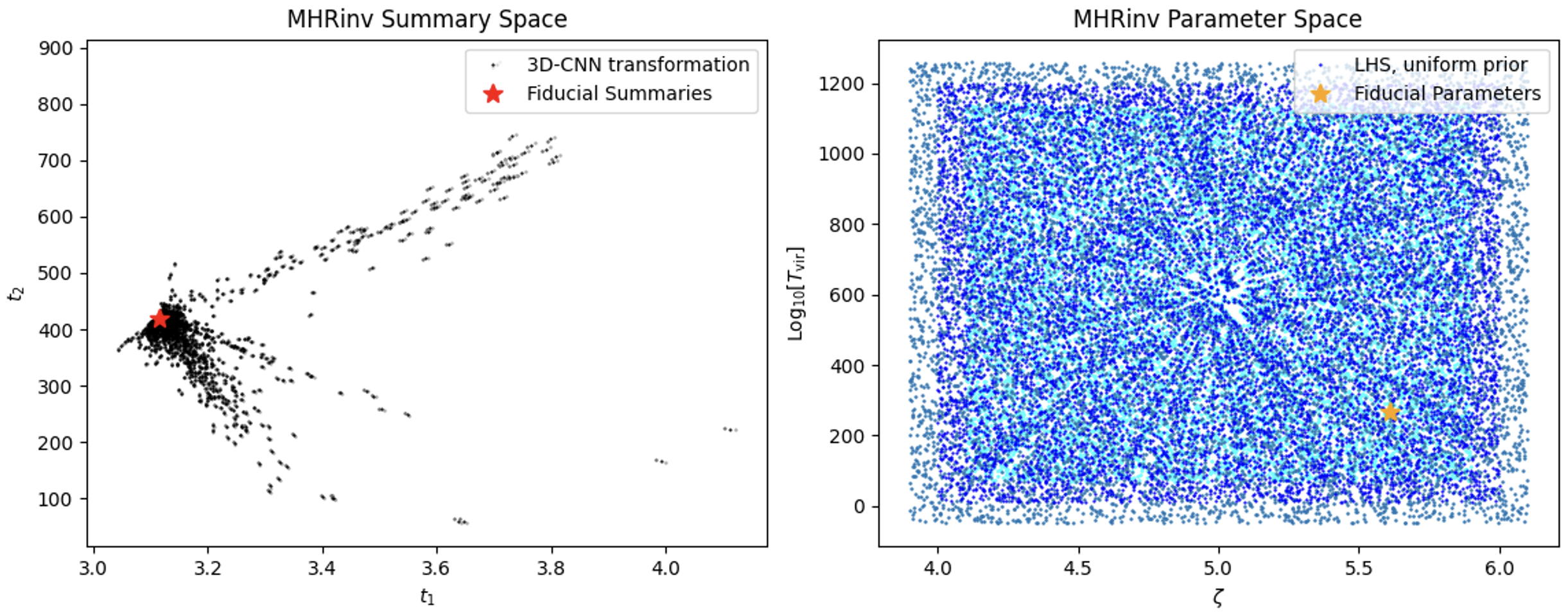}\hfill}
    \subfigure[\label{}]{\includegraphics[scale=0.2]{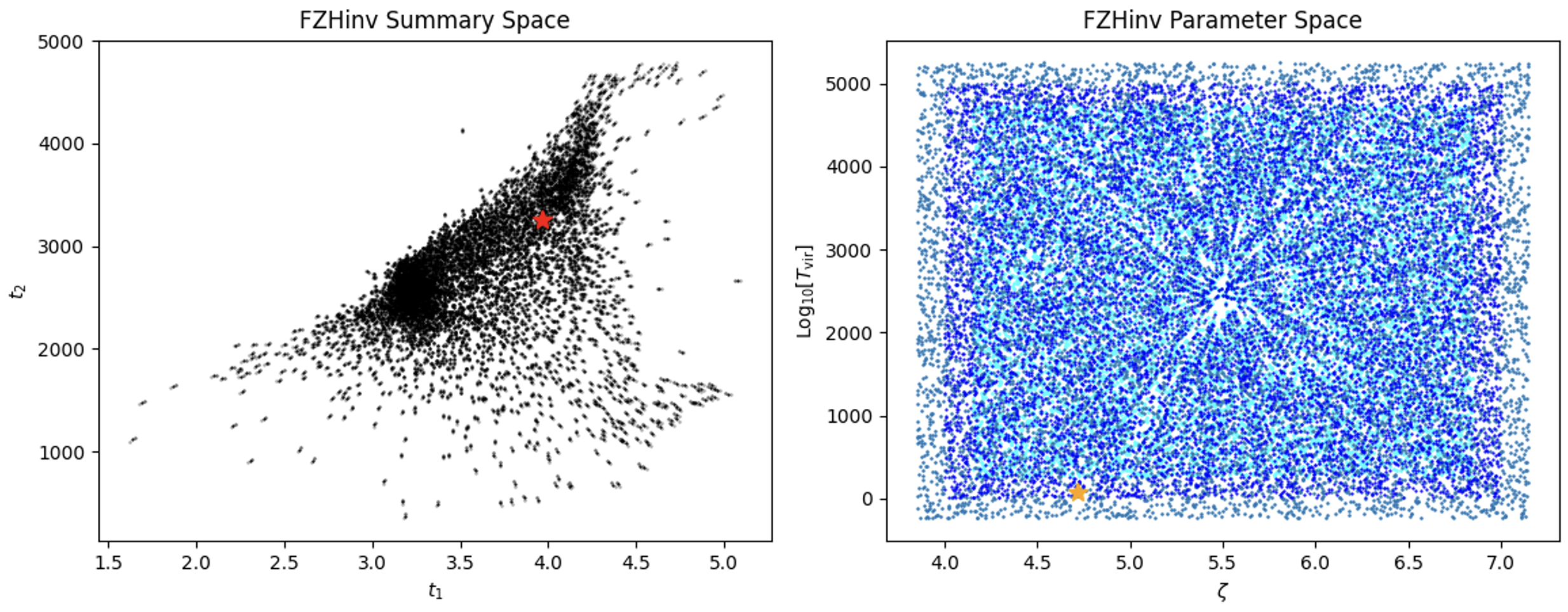}\hfill}
    \subfigure[\label{}]{\includegraphics[scale=0.2]{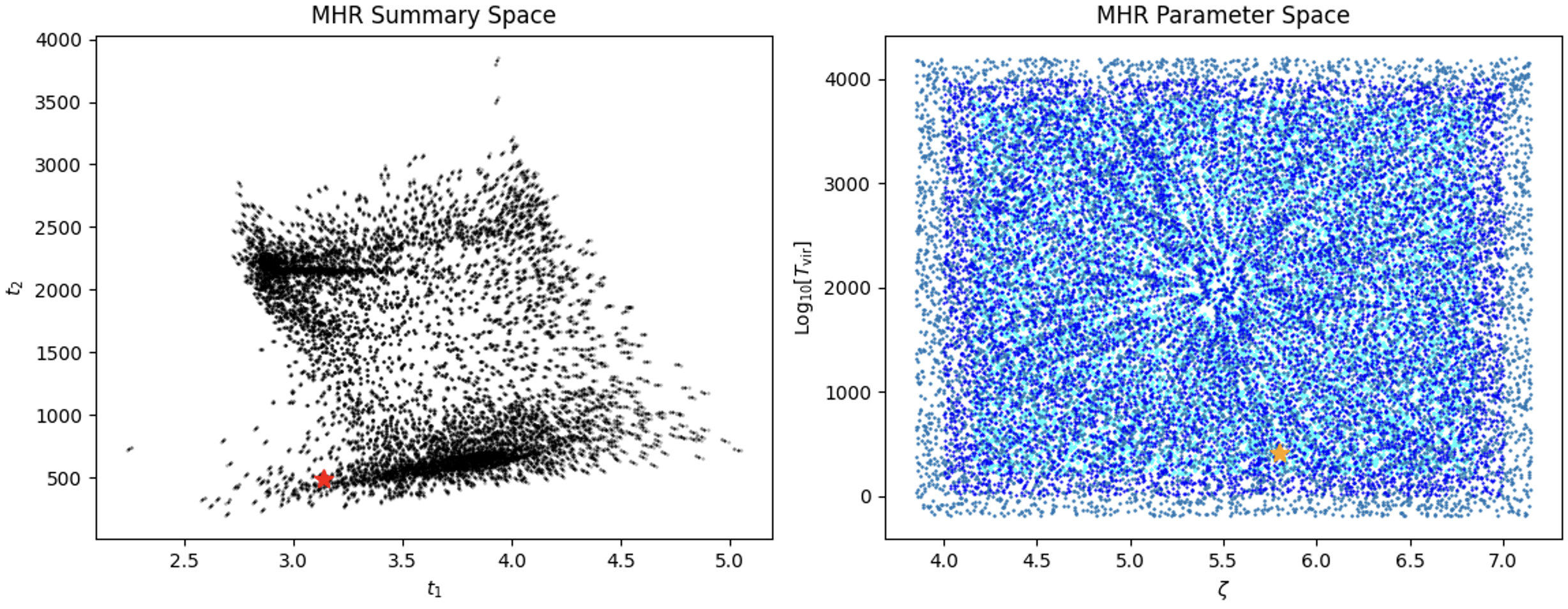}\hfill}
    \caption{
    Each pair of plots shows the distribution of samples used to train each model's 3D CNN (right) and the summary values obtained when summarising these lightcones with the 3D CNN (left). 
    In Section \ref{sec: alternate_priors} we discuss increasing and decreasing the prior range by $10\%$, illustrated by the different shades of blue on the right-hand side of each pair of plots. 
    Parameter values are Latin-hypercube sampled within the parameter prior ranges. 
    We obtained the same summary space distribution for each of the 3 prior range distributions within $<1\%$ highlighting that the \textsc{Delfi}-3DCNN is robust to prior choice within the ranges used in this work (plotted in grey but masked by the black points). 
    The fiducial values for each mock data set are highlighted with an orange star, this becomes the red star when transformed into the summary space. 
    }
    \label{fig: priorpoints}
\end{figure*}

To check our results for robustness against prior volume, we recalculated the results for each data set with the priors increased to $110\%$ and decreased to $90\%$ of the original. 
To simplify this we do not consider changes to the fiducial parameters for the mock-data here. 
For each prior we simulated 10000 lightcones parameters of the same LHS distribution scaled throughout the new prior. 
In both cases, we found negligible changes to the results obtained by \textsc{Delfi-Nest}. 
These results were identical by eye, which was somewhat surprising; however, we show in Figure \ref{fig: priorpoints} that despite changing the range of samples used to train the networks, the network summary space distribution remains unchanged to $<1\%$. 
This suggests that the summary space has converged on the presented distribution of summary values. 
Because the summary space is unchanged, it makes some sense that the posterior emulated by \textsc{pyDelfi} should contain the same volume (and therefore will integrate to the same Bayesian Evidence with \textsc{Delfi-Nest}). 

\section{A Summary of the Neural Density Estimators (NDEs)}\label{sec: NDEs}

The \textsc{pyDelfi} architecture uses 5 sets of Mixture Density Networks (MDNs) and 1 Masked Auto-regressive Flow (MAF). 
These are stacked into a single ensemble. 
The posterior obtained results from stacking the weights dictated by minimising training loss in each network. 
Each network uses a $\rm{tanh}$ activation function. 
Within \textsc{pyDelfi}, the MDN network weights will either be initially randomised or a Fisher pre-training analysis is done using 250 simulations per epoch. 
We use random weights for the initial conditions of the 21~cm analysis and use the default Fisher pre-training to guide the weights in the cosmic-shear analysis. 
To train the NDEs the desired estimator distribution; $p(t|\theta, w_n)$, where $t$ is the neural density-weight vector (aimed at reproducing the 3D CNN summaries, $t$), $w_n$ are the network weights and $\theta$ parameters; is fit to a target distribution, $p^*(t|\theta)$.
Traditionally, this is estimated with MCMC algorithms by minimising the cross-entropy, or Kullback-Leibler (KL) divergence (Equation \ref{eq: KL-divergence}) between the two distributions. 
For an in-depth, comprehensive background on machine learning and neural networks, please see \citet{2007JEI....16d9901B}.
We also direct the reader to \citetalias{2018MNRAS.477.2874A}, \citetalias{2019MNRAS.488.4440A}, \citetalias{2022ApJ...926..151Z} and \citealt{2022ApJ...933..236Z} for more formalism on choosing network architectures and varying tuning parameters. 
The original \textsc{pyDelfi} work constructs the NDE with 5 MDN (with 1, 2, 4, 6 \& 8 GMMs each) and 1 MAF (with 5 MADEs); in the 3D CNN implementation, there are 4 MAF (with 8 MADEs each). 
We experimented with both these combinations and a few alternatives for the 3D CNN. 
Namely, we tried a 4 MDN architecture (with 1, 4, 6 \& 8 GMMs each) and a 5 MAF and 1 MDN (with 8 GMMs) architecture combinations. 
Altering architectures results in insignificant changes; we found the architectures in the original work obtain the best results (and are therefore presented in this work). 
We iterate \textsc{pyDelfi} with a batch size of 200 and epochs set to a maximum of 1000. 
The training iterations are stopped if the loss function stops decreasing between the maximum and minimum number of epochs. 

\subsection{Mixture Density Network (MDN)}\label{sec: MDN}

In the context of \textsc{pyDelfi}, MDNs refer specifically to Gaussian MDNs. 
These 5 MDNs have a different number of Gaussian components, $n_c=1,~2,~3,~4,$ and $5$. 
Every MDN contains 2 hidden layers of 50 hidden units each. 
The density estimators are defined as, 
\begin{equation}\label{eq: MDNs}
    \rho( t | \theta, w_n) = \sum^{n_c}_{k=1} ~w_k \times  \mathcal{N} \left [~ t ~|~ \mu_k, \mathcal{C}_k \mathcal{C}_k^{T} ~ \right ],
\end{equation}
where $\mu_k$, $\mathcal{C}_k$, and $w_k$ are the mean, covariance, and component weights respectively.
All the $\mu_k$, $\mathcal{C}_k$, and $w_k$ are functions of $(\theta, w_n)$ and $\mathcal{N}$ is the multivariate Gaussian distribution. 
This is the full description of the MDNs used in \textsc{pyDelfi} and expands on the example in the main text in Equation \ref{eq: GMMpseudoZ}.
Please see e.g. \citet{Bishop94mixturedensity,2016arXiv160506376P, 2017arXiv170507057P, 2018A&C....25..183M} for a more detailed review of MDNs. 

\subsection{Masked Auto-regressive Flow (MAF)}\label{sec: MAF}

Here, the output density $\rho( t | \theta)$ is the function of a conserved flow of probability from a set of points $z$ sampled from a base density $\eta$ spanning the parameter space.
By defining this flow as the transform, $T$, we write, 
\begin{equation}\label{eq: flow}
    t = T(z)~,  ~~~~{\rm and}~~~~~~   z \propto \rho(z|\theta) .
\end{equation}
We express the desired output density as,
\begin{equation}\label{eq: theoryMAF}
    \rho( t | \theta) = \eta(z|\theta) \left | \frac{\partial z}{\partial t} \right | ,
\end{equation}
where $z$ is obtained through the inverse transform of $T$. 
For flow-based models, the model density is determined from $\eta$, the Jacobian, $ \left| \frac{\partial z}{\partial t} \right |$, and by evaluating the model density with the inverse transform, $T^{-1}$. 
\textsc{pyDelfi} uses a Masked Auto-encoder for Density Estimation (MADE) as $T^{-1}$ to ensure the flow is auto-regressive within a connected auto-encoder. 

We can now decompose the Jacobian into individual transforms as $T = T_k \circ T_{k-1} \circ ... \circ T_1$.
Using a MADE here also ensures that the $T$ is differentiable, can be inverted, and that the Jacobian in Equation \ref{eq: theoryMAF} is easy to handle. 
The auto-regressive conditional density can now be expressed as, 
\begin{equation}\label{eq: MADE}
    \rho(t|\theta, w_n) = \Pi^{d_{t}}_{i=1} \rho(t_i|t_{1:i-1},\theta, w_n),
\end{equation}
where each $\rho(t_i|t_{1:i-1},\theta, w_n)$ is a 1-D Gaussian with mean, $\mu_i(t_{1:i-1},\theta, w_n)$, and variance, $\sigma_i(t_{1:i-1},\theta, w_n)$, and $d_{t}$ is the dimensionality of $t$. 
It is worth mentioning that a MADE is defined by the properties of the parameterisation used for $\rho(t_i|t_{1:i-1},\theta, w_n)$ in Equation \ref{eq: MADE} (which is a 1D-Gaussian in this work). 
The sampled $z$ points can then be related to $t$ via the affine transformation, $z_i = (t_i-\mu_i)/\sigma_i$. 
Hence, we can rewrite the determinant of the Jacobian from Equation \ref{eq: theoryMAF} as,
\begin{equation}\label{eq: MAFJacobian}
     \left | \frac{\partial T^{-1}(t, \theta, w_n)}{\partial t} \right |  =  \Pi^{{\rm dim}(t)}_{i=1} \frac{1}{\sigma_i(t, \theta, w_n)}. 
\end{equation}
By stacking the transformations hierarchically, the next layer's base density, $z_n$, becomes $z_{n-1}$. 
Taking our initial layer, $z_0$, to be the input Gaussian (with null mean and identity variance). 
We can finally write the conditional density as,
\begin{equation}\label{eq: analyticMAF}
    \rho( t | \theta, w_n) = \mathcal{N} \left [~ z_0( t, \theta, w_n) ~|~ 0, {\rm I} ~ \right ]  \Pi^{N_{\rm m}}_{j=1} \Pi^{d_{t}}_{i=1} \frac{1}{\sigma^j_i(t, \theta, w_n)},
\end{equation}
where $N_{\rm m}$ is the number of MADEs. 
For a comprehensive description of MAFs, please see \citep{2017arXiv170507057P}. 

\section{Supplementary Cosmic Shear material}\label{sec: Cosmic_shear}

Cosmic shear refers to the application of gravitational lensing to the light from distant galaxies due to the universe's large-scale structure. 
The two key features of the universe here are its matter distribution and geometry.
Hence, the parameters of this model are contained within the $\Lambda{\rm CDM}$ cosmology.
However, we use a flat universe in this analysis (to agree with \citetalias{2019MNRAS.488.4440A}), deviating from the Planck results used (in \textsc{21cmFAST}) throughout the rest of this work. 

\subsection{The cosmic shear tomographic $C_{l}$ model}\label{sec: cosmic_shear_model}

Given the weak lensing limit, the applied coherent distortion to the galaxy images observed on the field of the sky with ellipticity $e$ is, 
\begin{equation}\label{eq: cosmic_shear-def}
    e = e_0 + \delta,
\end{equation}
where $e_0$ is the unlensed galactic ellipticity per field and $\delta$ is the ellipticity contribution from cosmic shear across the field. 
Note that it is the statistical properties of the cosmic shear fields $\delta_i$ that provide valuable insight into cosmology; please see \citetalias{2019MNRAS.488.4440A} for more context or \citet{2015RPPh...78h6901K} for a detailed review. 
These fields of cosmic shear are combined per redshift, $z_i$, into one of $n$ tomographic redshift bins to create the full data set, $\rm d$ = $(\delta_1,...,\delta_i,...,\delta_n)$, which will be compressed later in Section \ref{sec: cosmic_shear_comp}. 
Since the images are pixelated, we estimate the shear in the $i$'th bin per pixel, $p$ as,
\begin{equation}\label{eq: comsic_shear-data}
    \delta_{p,i} = \sum_{p,i} \frac{\langle e  \rangle_{p,i} }{ N_{p,i}},
\end{equation}
where the angular brackets denote a spatial average across each pixel and $N_{p, i}$ is the number of galaxies (within a given pixel in a given redshift bin). 
Each cosmic shear measurement is averaged per galaxy across the space defined by each pixel's boundary. 
By assuming that there are many galaxies per pixel, the intrinsic variation between each cosmic shear contribution is approximated as zero, given a Gaussian random noise fluctuation (i.e. a Gaussian with mean $\mu_{p, i} = 0$, and variance $\sigma_{p, i}^2 = \frac{\sigma^2_e}{N_{p, i}}$ per bin, per pixel). 

Our mock cosmic shear data set is analysed by measuring the tomographic power spectra of each cosmic shear field. 
Assuming a flat $\Lambda$CDM cosmology and using Limber's Equation \citep{1954ApJ...119..655L}, this can be written between redshift bins $a$ and $b$ as the spherical harmonics,
\begin{equation}\label{eq: cosmic shear Cl}
    C_{l, ab} =  \int \frac{ d \chi }{\chi^2} \omega_a \omega_b [1 + z(\chi)] P_{\delta} (k, \chi) .
\end{equation}
which contains, $P_{\delta}$, as the matter power spectrum;  $\chi(z)$, the comoving distance-redshift relationship. 
The lensing weight functions are then written, 
\begin{equation}\label{eq: lensing weights}
    \omega_a = \frac{3 \Omega_{\rm M} H_{0}^2}{ 2} \chi (z) \int_{\chi (z)}^{\chi (z_h)} d\chi(z')  n_a(z') \frac{\chi(z')-\chi(z)}{\chi(z')} ,
\end{equation}
where $n_a$ is the redshift distribution for galaxies in bin $a$, and the integration is performed to the comoving distance of the horizon. 
For more detail please see e.g. \citet{1992ApJ...388..272K, 1998ApJ...498...26K, Hu_1999, PhysRevD.65.023003, 10.1093/mnrasl/slt045, 2004MNRAS.348..897T}. 
The $\Lambda$CDM cosmological parameters change the cosmic shear power spectrum via the comoving distance-redshift relation and the matter power spectrum. 
Measurement of the $C_{l, ab}$ therefore constrains the cosmology. 

\subsection{The cosmic shear mock observation}\label{sec: cosmic_shear_obs}

Initially, we simulate tomographic shear maps with a Gaussian random field that agrees with a power spectrum corresponding to the sampled cosmology with parameters, $\theta~=~(\Omega_{\rm M}$, $\sigma_8$, $\Omega_b$, $h$, $n_s)$. 
The simulation of the data exists within \textsc{HEALPix} \citep{2005ApJ...622..759G} pixelisations ($n_{\rm side} = 128$) following \citetalias{2019MNRAS.488.4440A} (for more information about the Hierarchical Equal Area isoLatitude Pixelation of a sphere please see \url{https://healpix.sourceforge.io}). 
An isotropic-shaped noise is added to the maps (as a Gaussian random field), followed by the Euclid mask.
Equation \ref{eq: cosmic shear Cl} is then used to calculate the $C_l$ for auto- and cross-band powers from these noisy tomographic maps. 
The mock data set used here is similar to the ESA Euclid survey \citep{2011arXiv1110.3193L}. 
Maps are masked appropriately to show incomplete sky coverage. 
The mean galaxy number density is $30~{\rm arcmin}^{-2}$ and the galaxy redshift distribution is written,
\begin{equation}\
    n(z) \propto z^2 \exp{ \left [- \left ( \frac{1.41}{0.9} z \right )^{-1.5} \right]},
\end{equation}
with photo-z errors being modelled as Gaussian with standard deviation $\sigma_z = 0.05(1+z)$.
The observation is performed across 15,000 square degrees with five tomographic bins, each with equal mean galaxy number density. 
Between $l \in [10,~383]$ modes are binned into ten log-space bands. 
For the shape noise, galaxies are Poisson distributed throughout the pixels per tomographic slice, and the pixel variance is calculated as in \ref{sec: cosmic_shear_model} with $\sigma_e=0.3$. 

\subsection{Compressing the cosmic shear data object}\label{sec: cosmic_shear_comp}

First, we organise the cosmic shear maps into the format which they are compressed from, $\tilde{d}$. 
The data object $\rm d$ is combined with noise and masked before being used to calculate auto- and cross-band powers for the angular $C_l$ within $K$ $l$-bands and $n_z$ tomographic bins. 
From the noisy, masked, tomographic cosmic shear maps, the E-mode spherical harmonic coefficients are combined as,
\begin{equation}
    \tilde{C}_{B_{k}, ij} = \sum_{l\in B_k} \sum^l_{m=-l} \hat{a}^i_{lm} \hat{a}^{j*}_{lm} .
\end{equation}
These then construct our tomographic band powers (the desired data object) as, 
\begin{equation}\label{eq: cosmic_shear_dataObj}
    \tilde{d} = \left ( \tilde{C}_{B_{1}, 11}, \tilde{C}_{B_{1}, 12}, ...,\tilde{C}_{B_{1}, n_z n_z},\tilde{C}_{B_{2}, 11},...,\tilde{C}_{B_{K}, n_z n_z} \right )  ,
\end{equation}
which are then compressed to provide the data summary, $t$. 
Our compressed data object is constructed around a mean, $\mu$ and covariance, $\mathcal{C}$ calculated from $10^3$ forward simulations of the cosmic shear model. 
We then apply score compression (Section \ref{sec: score_comp}) to obtain the compressed summary, 
\begin{equation}
    t \equiv \nabla_\theta \mathcal{L} = \nabla^T_\theta \mu ~\mathcal{C}^{-1}(\tilde{d} - \mu) ,
\end{equation}
where the gradients are estimated per parameter using 100 pairs of simulations with matched random seeds and a step size of $5\%$. 
The $C_l$s are assumed to be Gaussian distributed to perform this compression step. 
Although this is not strictly true (particularly at low $l$), it has proven to be an adequate assumption in the context of the \textsc{pyDelfi} analyses in \citetalias{2019MNRAS.488.4440A}. 
Throughout this work, $t$, represents data summaries, while $s$ is the compression used within \textsc{pyDelfi}. 
The cosmic shear analysis uses score compression twice. 

\subsection{\textit{Cross-check:} comparing importance nested sampling results}\label{sec: justify_INS}

We compare Evidence values obtained with and without importance nested sampling (INS) in the context of the cosmic shear analysis in Section \ref{sec: cosmic shear Model selection}. 
Nested sampling discards a significant proportion of points failing to meet the likelihood threshold. 
INS \citep{2013arXiv1301.6450C, 2019OJAp....2E..10F} uses the discarded points and compares their density to the density of accepted points to estimate the Evidence. 
The convergence of INS on the marginal likelihood is better constrained than regular nested sampling; however, they should agree. 
Table \ref{tab: importance} shows a close agreement between the nested sampling $\mathcal{Z}$ and the INS $\mathcal{Z}$ when using the cosmic shear data discussed in Section \ref{sec: cosmic shear Model selection}. 

\begin{table}
    \centering
    \begin{tabular}{c|c c}
        Model & ${\rm ln}\mathcal{Z}$ & INS ${\rm ln}\mathcal{Z}$ \\ \hline
       $Good|5p$ & -17.1 $\pm$ 0.03 & -17.4 $\pm$ 0.02 \\
       $Bad|5p$ & -16.0 $\pm$ 0.07 & -16.0 $\pm$ 0.04 \\
       $Good|2p$ & -24.9 $\pm$ 0.01 & -25.0 $\pm$ 0.01 \\
       $Bad|2p$ & -12.7 $\pm$ 0.05 & -12.8 $\pm$ 0.03 \\
    \end{tabular}
    \caption{This table shows agreement between the Evidence values obtained with nested sampling and importance nested sampling within \textsc{MultiNest} on the toy cosmic shear model analysis. 
    As both algorithms agree, this is a promising cross-check of the results in Section \ref{sec: cosmic shear Model selection}. 
    Please see Appendix \ref{sec: cosmic_shear_model} \& Section \ref{sec: Delfi-Nest crosscheck} for more detail about these models. }
    \label{tab: importance}
\end{table}

\bibliography{_ADS_postThesis}{}

\begin{thebibliography}{}
\expandafter\ifx\csname natexlab\endcsname\relax\def\natexlab#1{#1}\fi
\providecommand{\url}[1]{\href{#1}{#1}}
\providecommand{\dodoi}[1]{doi:~\href{http://doi.org/#1}{\nolinkurl{#1}}}
\providecommand{\doeprint}[1]{\href{http://ascl.net/#1}{\nolinkurl{http://ascl.net/#1}}}
\providecommand{\doarXiv}[1]{\href{https://arxiv.org/abs/#1}{\nolinkurl{https://arxiv.org/abs/#1}}}

\bibitem[{{Abdurashidova} {et~al.}(2022){Abdurashidova}, {Aguirre},
  {Alexander}, {Ali}, {Balfour}, {Beardsley}, {Bernardi}, {Billings}, {Bowman},
  {Bradley}, {Bull}, {Burba}, {Carey}, {Carilli}, {Cheng}, {DeBoer}, {Dexter},
  {de Lera Acedo}, {Dibblee-Barkman}, {Dillon}, {Ely}, {Ewall-Wice}, {Fagnoni},
  {Fritz}, {Furlanetto}, {Gale-Sides}, {Glendenning}, {Gorthi}, {Greig},
  {Grobbelaar}, {Halday}, {Hazelton}, {Hewitt}, {Hickish}, {Jacobs}, {Julius},
  {Kern}, {Kerrigan}, {Kittiwisit}, {Kohn}, {Kolopanis}, {Lanman}, {La Plante},
  {Lekalake}, {Lewis}, {Liu}, {MacMahon}, {Malan}, {Malgas}, {Maree},
  {Martinot}, {Matsetela}, {Mesinger}, {Molewa}, {Morales}, {Mosiane},
  {Murray}, {Neben}, {Nikolic}, {Nunhokee}, {Parsons}, {Patra}, {Pascua},
  {Pieterse}, {Pober}, {Razavi-Ghods}, {Ringuette}, {Robnett}, {Rosie}, {Sims},
  {Singh}, {Smith}, {Syce}, {Thyagarajan}, {Williams}, {Zheng}, \& {HERA
  Collaboration}}]{2022ApJ...925..221A}
{Abdurashidova}, Z., {Aguirre}, J.~E., {Alexander}, P., {et~al.} 2022, \apj,
  925, 221, \dodoi{10.3847/1538-4357/ac1c78}

\bibitem[{{Akeret} {et~al.}(2015){Akeret}, {Refregier}, {Amara}, {Seehars}, \&
  {Hasner}}]{2015JCAP...08..043A}
{Akeret}, J., {Refregier}, A., {Amara}, A., {Seehars}, S., \& {Hasner}, C.
  2015, \jcap, 2015, 043, \dodoi{10.1088/1475-7516/2015/08/043}

\bibitem[{{Alsing} {et~al.}(2019){Alsing}, {Charnock}, {Feeney}, \&
  {Wandelt}}]{2019MNRAS.488.4440A}
{Alsing}, J., {Charnock}, T., {Feeney}, S., \& {Wandelt}, B. 2019, \mnras, 488,
  4440, \dodoi{10.1093/mnras/stz1960}

\bibitem[{{Alsing} \& {Wandelt}(2018)}]{2018MNRAS.476L..60A}
{Alsing}, J., \& {Wandelt}, B. 2018, \mnras, 476, L60,
  \dodoi{10.1093/mnrasl/sly029}

\bibitem[{{Alsing} {et~al.}(2018){Alsing}, {Wandelt}, \&
  {Feeney}}]{2018MNRAS.477.2874A}
{Alsing}, J., {Wandelt}, B., \& {Feeney}, S. 2018, \mnras, 477, 2874,
  \dodoi{10.1093/mnras/sty819}

\bibitem[{Baldi \& Sadowski(2013)}]{NIPS2013_71f6278d}
Baldi, P., \& Sadowski, P.~J. 2013, NeurIPS, 26,
  \dodoi{https://proceedings.neurips.cc/paper_files/paper/2013/file/71f6278d140af599e06ad9bf1ba03cb0-Paper.pdf}

\bibitem[{{Barkana} \& {Loeb}(2001)}]{2001PhR...349..125B}
{Barkana}, R., \& {Loeb}, A. 2001, \physrep, 349, 125,
  \dodoi{10.1016/S0370-1573(01)00019-9}

\bibitem[{{Binnie} \& {Pritchard}(2019)}]{2019MNRAS.487.1160B}
{Binnie}, T., \& {Pritchard}, J.~R. 2019, \mnras, 487, 1160,
  \dodoi{10.1093/mnras/stz1297}

\bibitem[{{Bishop}(1994)}]{Bishop94mixturedensity}
{Bishop}, C.~M. 1994, Technical Report, Mixture density networks, CiteSeer

\bibitem[{{Bishop} \& {Nasrabadi}(2007)}]{2007JEI....16d9901B}
{Bishop}, C.~M., \& {Nasrabadi}, N.~M. 2007, JEI, 16, 049901,
  \dodoi{10.1117/1.2819119}

\bibitem[{Blum {et~al.}(2013)Blum, Nunes, Prangle, \&
  Sisson}]{10.1214/12-STS406}
Blum, M. G.~B., Nunes, M.~A., Prangle, D., \& Sisson, S.~A. 2013, Stat. Sci.,
  28, 189 , \dodoi{10.1214/12-STS406}

\bibitem[{{Bond} {et~al.}(1998){Bond}, {Jaffe}, \&
  {Knox}}]{1998PhRvD..57.2117B}
{Bond}, J.~R., {Jaffe}, A.~H., \& {Knox}, L. 1998, \prd, 57, 2117,
  \dodoi{10.1103/PhysRevD.57.2117}

\bibitem[{{Bond} {et~al.}(2000){Bond}, {Jaffe}, \&
  {Knox}}]{2000ApJ...533...19B}
---. 2000, \apj, 533, 19, \dodoi{10.1086/308625}

\bibitem[{{Bowman} {et~al.}(2018{\natexlab{a}}){Bowman}, {Rogers}, {Monsalve},
  {Mozdzen}, \& {Mahesh}}]{2018Natur.555...67B}
{Bowman}, J.~D., {Rogers}, A. E.~E., {Monsalve}, R.~A., {Mozdzen}, T.~J., \&
  {Mahesh}, N. 2018{\natexlab{a}}, \nat, 555, 67, \dodoi{10.1038/nature25792}

\bibitem[{{Bowman} {et~al.}(2018{\natexlab{b}}){Bowman}, {Rogers}, {Monsalve},
  {Mozdzen}, \& {Mahesh}}]{2018Natur.564E..35B}
---. 2018{\natexlab{b}}, \nat, 564, E35, \dodoi{10.1038/s41586-018-0797-4}

\bibitem[{{Cameron} \& {Pettitt}(2013)}]{2013arXiv1301.6450C}
{Cameron}, E., \& {Pettitt}, A. 2013, arXiv e-prints, 1301.6450,
  \dodoi{10.48550/arXiv.1301.6450}

\bibitem[{{Charnock} {et~al.}(2018){Charnock}, {Lavaux}, \&
  {Wandelt}}]{2018PhRvD..97h3004C}
{Charnock}, T., {Lavaux}, G., \& {Wandelt}, B.~D. 2018, \prd, 97, 083004,
  \dodoi{10.1103/PhysRevD.97.083004}

\bibitem[{Chollet {et~al.}(2015)}]{chollet2015keras}
Chollet, F., {et~al.} 2015, Keras, \url{https://github.com/fchollet/keras},
  GitHub

\bibitem[{{Cohen} {et~al.}(2019){Cohen}, {Fialkov}, {Barkana}, \&
  {Monsalve}}]{2019arXiv191006274C}
{Cohen}, A., {Fialkov}, A., {Barkana}, R., \& {Monsalve}, R. 2019, arXiv
  e-prints, 1910.06274, \dodoi{1910.06274}

\bibitem[{Dahl {et~al.}(2013)Dahl, Sainath, \& Hinton}]{6639346}
Dahl, G.~E., Sainath, T.~N., \& Hinton, G.~E. 2013, ICASSP, 8609,
  \dodoi{10.1109/ICASSP.2013.6639346}

\bibitem[{{Datta} {et~al.}(2014){Datta}, {Jensen}, {Majumdar}, {Mellema},
  {Iliev}, {Mao}, {Shapiro}, \& {Ahn}}]{2014MNRAS.442.1491D}
{Datta}, K.~K., {Jensen}, H., {Majumdar}, S., {et~al.} 2014, \mnras, 442, 1491,
  \dodoi{10.1093/mnras/stu927}

\bibitem[{{Datta} {et~al.}(2012){Datta}, {Mellema}, {Mao}, {Iliev}, {Shapiro},
  \& {Ahn}}]{2012MNRAS.424.1877D}
{Datta}, K.~K., {Mellema}, G., {Mao}, Y., {et~al.} 2012, \mnras, 424, 1877,
  \dodoi{10.1111/j.1365-2966.2012.21293.x}

\bibitem[{{Davies} {et~al.}(2018){Davies}, {Hennawi}, {Eilers}, \&
  {Luki{\'c}}}]{2018ApJ...855..106D}
{Davies}, F.~B., {Hennawi}, J.~F., {Eilers}, A.-C., \& {Luki{\'c}}, Z. 2018,
  \apj, 855, 106, \dodoi{10.3847/1538-4357/aaaf70}

\bibitem[{{Dayal} {et~al.}(2014){Dayal}, {Ferrara}, {Dunlop}, \&
  {Pacucci}}]{2014MNRAS.445.2545D}
{Dayal}, P., {Ferrara}, A., {Dunlop}, J.~S., \& {Pacucci}, F. 2014, \mnras,
  445, 2545, \dodoi{10.1093/mnras/stu1848}

\bibitem[{{Deep Kaur} {et~al.}(2020){Deep Kaur}, {Gillet}, \&
  {Mesinger}}]{2020arXiv200406709D}
{Deep Kaur}, H., {Gillet}, N., \& {Mesinger}, A. 2020, arXiv e-prints,
  arXiv:2004.06709.
\newblock \doarXiv{2004.06709}

\bibitem[{{Ellis}(2014)}]{2014arXiv1411.3330E}
{Ellis}, R.~S. 2014, arXiv e-prints, 1411.3330, \dodoi{1411.3330}

\bibitem[{{Feng} {et~al.}(2016){Feng}, {Di-Matteo}, {Croft}, {Bird},
  {Battaglia}, \& {Wilkins}}]{2016MNRAS.455.2778F}
{Feng}, Y., {Di-Matteo}, T., {Croft}, R.~A., {et~al.} 2016, \mnras, 455, 2778,
  \dodoi{10.1093/mnras/stv2484}

\bibitem[{{Feroz} \& {Hobson}(2008)}]{2008MNRAS.384..449F}
{Feroz}, F., \& {Hobson}, M.~P. 2008, \mnras, 384, 449,
  \dodoi{10.1111/j.1365-2966.2007.12353.x}

\bibitem[{{Feroz} {et~al.}(2009){Feroz}, {Hobson}, \&
  {Bridges}}]{2009MNRAS.398.1601F}
{Feroz}, F., {Hobson}, M.~P., \& {Bridges}, M. 2009, \mnras, 398, 1601,
  \dodoi{10.1111/j.1365-2966.2009.14548.x}

\bibitem[{{Feroz} {et~al.}(2019){Feroz}, {Hobson}, {Cameron}, \&
  {Pettitt}}]{2019OJAp....2E..10F}
{Feroz}, F., {Hobson}, M.~P., {Cameron}, E., \& {Pettitt}, A.~N. 2019, DOAJ, 2,
  10, \dodoi{10.21105/astro.1306.2144}

\bibitem[{{Fialkov} {et~al.}(2014){Fialkov}, {Barkana}, \&
  {Visbal}}]{2014Natur.506..197F}
{Fialkov}, A., {Barkana}, R., \& {Visbal}, E. 2014, \nat, 506, 197,
  \dodoi{10.1038/nature12999}

\bibitem[{{Field}(1958)}]{1958PIRE...46..240F}
{Field}, G.~B. 1958, Proceedings of the IRE, 46, 240,
  \dodoi{10.1109/JRPROC.1958.286741}

\bibitem[{{Finlator} {et~al.}(2018){Finlator}, {Keating}, {Oppenheimer},
  {Dav{\'e}}, \& {Zackrisson}}]{2018MNRAS.480.2628F}
{Finlator}, K., {Keating}, L., {Oppenheimer}, B.~D., {Dav{\'e}}, R., \&
  {Zackrisson}, E. 2018, \mnras, 480, 2628, \dodoi{10.1093/mnras/sty1949}

\bibitem[{{Foreman-Mackey} {et~al.}(2013){Foreman-Mackey}, {Hogg}, {Lang}, \&
  {Goodman}}]{2013PASP..125..306F}
{Foreman-Mackey}, D., {Hogg}, D.~W., {Lang}, D., \& {Goodman}, J. 2013, \pasp,
  125, 306, \dodoi{10.1086/670067}

\bibitem[{{Furlanetto} \& {Briggs}(2004)}]{2004NewAR..48.1039F}
{Furlanetto}, S.~R., \& {Briggs}, F.~H. 2004, \nar, 48, 1039,
  \dodoi{10.1016/j.newar.2004.09.034}

\bibitem[{{Furlanetto} {et~al.}(2006){Furlanetto}, {Oh}, \&
  {Briggs}}]{2006PhR...433..181F}
{Furlanetto}, S.~R., {Oh}, S.~P., \& {Briggs}, F.~H. 2006, \physrep, 433, 181,
  \dodoi{10.1016/j.physrep.2006.08.002}

\bibitem[{{Furlanetto} {et~al.}(2004){Furlanetto}, {Zaldarriaga}, \&
  {Hernquist}}]{2004ApJ...613....1F}
{Furlanetto}, S.~R., {Zaldarriaga}, M., \& {Hernquist}, L. 2004, \apj, 613, 1,
  \dodoi{10.1086/423025}

\bibitem[{{Garaldi} {et~al.}(2019){Garaldi}, {Compostella}, \&
  {Porciani}}]{2019MNRAS.483.5301G}
{Garaldi}, E., {Compostella}, M., \& {Porciani}, C. 2019, \mnras, 483, 5301,
  \dodoi{10.1093/mnras/sty3414}

\bibitem[{{Gillet} {et~al.}(2019){Gillet}, {Mesinger}, {Greig}, {Liu}, \&
  {Ucci}}]{2019MNRAS.484..282G}
{Gillet}, N., {Mesinger}, A., {Greig}, B., {Liu}, A., \& {Ucci}, G. 2019,
  \mnras, 484, 282, \dodoi{10.1093/mnras/stz010}

\bibitem[{{Giri} {et~al.}(2020){Giri}, {Mellema}, \&
  {Jensen}}]{2020JOSS....5.2363G}
{Giri}, S., {Mellema}, G., \& {Jensen}, H. 2020, The Journal of Open Source
  Software, 5, 2363, \dodoi{10.21105/joss.02363}

\bibitem[{{Giri} {et~al.}(2023){Giri}, {Schneider}, {Maion}, \&
  {Angulo}}]{2023A&A...669A...6G}
{Giri}, S.~K., {Schneider}, A., {Maion}, F., \& {Angulo}, R.~E. 2023, \aap,
  669, A6, \dodoi{10.1051/0004-6361/202244986}

\bibitem[{{G{\'o}rski} {et~al.}(2005){G{\'o}rski}, {Hivon}, {Banday},
  {Wandelt}, {Hansen}, {Reinecke}, \& {Bartelmann}}]{2005ApJ...622..759G}
{G{\'o}rski}, K.~M., {Hivon}, E., {Banday}, A.~J., {et~al.} 2005, \apj, 622,
  759, \dodoi{10.1086/427976}

\bibitem[{{Greig} \& {Mesinger}(2015)}]{2015MNRAS.449.4246G}
{Greig}, B., \& {Mesinger}, A. 2015, \mnras, 449, 4246,
  \dodoi{10.1093/mnras/stv571}

\bibitem[{{Greig} \& {Mesinger}(2017)}]{2017MNRAS.472.2651G}
---. 2017, \mnras, 472, 2651, \dodoi{10.1093/mnras/stx2118}

\bibitem[{{Greig} \& {Mesinger}(2018)}]{2018MNRAS.477.3217G}
---. 2018, \mnras, 477, 3217, \dodoi{10.1093/mnras/sty796}

\bibitem[{{Hall} \& {Taylor}(2019)}]{2019MNRAS.483..189H}
{Hall}, A., \& {Taylor}, A. 2019, \mnras, 483, 189,
  \dodoi{10.1093/mnras/sty3102}

\bibitem[{Harris {et~al.}(2020)Harris, Millman, van~der Walt, Gommers,
  Virtanen, Cournapeau, Wieser, Taylor, Berg, Smith, Kern, Picus, Hoyer, van
  Kerkwijk, Brett, Haldane, del R{\'{\i}}o, Wiebe, Peterson,
  G{\'{e}}rard{-}Marchant, Sheppard, Reddy, Weckesser, Abbasi, Gohlke, \&
  Oliphant}]{DBLP:journals/corr/abs-2006-10256}
Harris, C.~R., Millman, K.~J., van~der Walt, S., {et~al.} 2020, CoRR,
  abs/2006.10256

\bibitem[{{Hassan} {et~al.}(2019){Hassan}, {Liu}, {Kohn}, \& {La
  Plante}}]{2019MNRAS.483.2524H}
{Hassan}, S., {Liu}, A., {Kohn}, S., \& {La Plante}, P. 2019, \mnras, 483,
  2524, \dodoi{10.1093/mnras/sty3282}

\bibitem[{Heavens {et~al.}(2013)Heavens, Alsing, \&
  Jaffe}]{10.1093/mnrasl/slt045}
Heavens, A., Alsing, J., \& Jaffe, A.~H. 2013, \mnras, 433, L6,
  \dodoi{10.1093/mnrasl/slt045}

\bibitem[{{Heavens} {et~al.}(2000){Heavens}, {Jimenez}, \&
  {Lahav}}]{2000MNRAS.317..965H}
{Heavens}, A.~F., {Jimenez}, R., \& {Lahav}, O. 2000, \mnras, 317, 965,
  \dodoi{10.1046/j.1365-8711.2000.03692.x}

\bibitem[{{Heavens} {et~al.}(2023){Heavens}, {Mootoovaloo}, {Trotta}, \&
  {Sellentin}}]{2023JCAP...11..048H}
{Heavens}, A.~F., {Mootoovaloo}, A., {Trotta}, R., \& {Sellentin}, E. 2023,
  \jcap, 2023, 048, \dodoi{10.1088/1475-7516/2023/11/048}

\bibitem[{{Higson} {et~al.}(2019){Higson}, {Handley}, {Hobson}, \&
  {Lasenby}}]{2019MNRAS.483.2044H}
{Higson}, E., {Handley}, W., {Hobson}, M., \& {Lasenby}, A. 2019, \mnras, 483,
  2044, \dodoi{10.1093/mnras/sty3090}

\bibitem[{{Hills} {et~al.}(2018){Hills}, {Kulkarni}, {Meerburg}, \&
  {Puchwein}}]{2018Natur.564E..32H}
{Hills}, R., {Kulkarni}, G., {Meerburg}, P.~D., \& {Puchwein}, E. 2018, \nat,
  564, E32, \dodoi{10.1038/s41586-018-0796-5}

\bibitem[{{Hobson} {et~al.}(2009){Hobson}, {Jaffe}, {Liddle}, {Mukherjee}, \&
  D.}]{al2009bayesian}
{Hobson}, H., {Jaffe}, A., {Liddle}, A., {Mukherjee}, P., \& D., P. 2009,
  Bayesian Methods in Cosmology (CUP)

\bibitem[{Hu(1999)}]{Hu_1999}
Hu, W. 1999, ApJ, 522, L21, \dodoi{10.1086/312210}

\bibitem[{Hu(2001)}]{PhysRevD.65.023003}
---. 2001, Phys. Rev. D, 65, 023003, \dodoi{10.1103/PhysRevD.65.023003}

\bibitem[{Hunter(2007)}]{4160265}
Hunter, J.~D. 2007, Computing in Science \& Engineering, 9, 90,
  \dodoi{10.1109/MCSE.2007.55}

\bibitem[{{Ioffe} \& {Szegedy}(2015)}]{2015arXiv150203167I}
{Ioffe}, S., \& {Szegedy}, C. 2015, arXiv e-prints, 1502.03167,
  \dodoi{10.48550/arXiv.1502.03167}

\bibitem[{{Jaynes}(2003)}]{jaynes03}
{Jaynes}, E.~T. 2003, Probability theory: The logic of science (CUP)

\bibitem[{{Kaiser}(1992)}]{1992ApJ...388..272K}
{Kaiser}, N. 1992, \apj, 388, 272, \dodoi{10.1086/171151}

\bibitem[{{Kaiser}(1998)}]{1998ApJ...498...26K}
---. 1998, \apj, 498, 26, \dodoi{10.1086/305515}

\bibitem[{{Karamanis} {et~al.}(2022{\natexlab{a}}){Karamanis}, {Beutler},
  {Peacock}, {Nabergoj}, \& {Seljak}}]{2022ascl.soft07018K}
{Karamanis}, M., {Beutler}, F., {Peacock}, J.~A., {Nabergoj}, D., \& {Seljak},
  U. 2022{\natexlab{a}}, ASCL, \dodoi{ascl:2207.018}

\bibitem[{{Karamanis} {et~al.}(2022{\natexlab{b}}){Karamanis}, {Nabergoj},
  {Beutler}, {Peacock}, \& {Seljak}}]{2022JOSS....7.4634K}
{Karamanis}, M., {Nabergoj}, D., {Beutler}, F., {Peacock}, J., \& {Seljak}, U.
  2022{\natexlab{b}}, JOSS, 7, 4634, \dodoi{10.21105/joss.04634}

\bibitem[{Kauderer{-}Abrams(2018)}]{DBLP:journals/corr/abs-1801-01450}
Kauderer{-}Abrams, E. 2018, CoRR, abs/1801.01450

\bibitem[{{Kern} {et~al.}(2017){Kern}, {Liu}, {Parsons}, {Mesinger}, \&
  {Greig}}]{2017ApJ...848...23K}
{Kern}, N.~S., {Liu}, A., {Parsons}, A.~R., {Mesinger}, A., \& {Greig}, B.
  2017, \apj, 848, 23, \dodoi{10.3847/1538-4357/aa8bb4}

\bibitem[{{Kilbinger}(2015)}]{2015RPPh...78h6901K}
{Kilbinger}, M. 2015, Rep. Prog. in Phys., 78, 086901,
  \dodoi{10.1088/0034-4885/78/8/086901}

\bibitem[{Koopmans {et~al.}(2015)Koopmans, Pritchard, Mellema, Aguirre, Ahn,
  Barkana, van Bemmel, Bernardi, Bonaldi, Briggs, de~Bruyn, Chang, Chapman,
  Chen, Courty, Dayal, Ferrara, Fialkov, Fiore, Ichiki, Illiev, Inoue, Jelic,
  Jones, Lazio, Maio, Majumdar, Mack, Mesinger, Morales, Parsons, Pen, Santos,
  Schneider, Semelin, de~Souza, Subrahmanyan, Takeuchi, Vedantham, Wagg,
  Webster, Wyithe, Datta, \& Trott}]{Koopmans:2015K0}
Koopmans, L., Pritchard, J., Mellema, G., {et~al.} 2015, PoS, AASKA14, 001,
  \dodoi{10.22323/1.215.0001}

\bibitem[{{Laureijs} {et~al.}(2011){Laureijs}, {Amiaux}, {Arduini},
  {Augu{\`e}res}, {Brinchmann}, {Cole}, {Cropper}, {Dabin}, {Duvet}, {Ealet},
  {Garilli}, {Gondoin}, {Guzzo}, {Hoar}, {Hoekstra}, {Holmes}, {Kitching},
  {Maciaszek}, {Mellier}, {Pasian}, {Percival}, {Rhodes}, {Saavedra Criado},
  {Sauvage}, {Scaramella}, {Valenziano}, {Warren}, {Bender}, {Castander},
  {Cimatti}, {Le F{\`e}vre}, {Kurki-Suonio}, {Levi}, {Lilje}, {Meylan},
  {Nichol}, {Pedersen}, {Popa}, {Rebolo Lopez}, {Rix}, {Rottgering},
  {Zeilinger}, {Grupp}, {Hudelot}, {Massey}, {Meneghetti}, {Miller}, {Paltani},
  {Paulin-Henriksson}, {Pires}, {Saxton}, {Schrabback}, {Seidel}, {Walsh},
  {Aghanim}, {Amendola}, {Bartlett}, {Baccigalupi}, {Beaulieu}, {Benabed},
  {Cuby}, {Elbaz}, {Fosalba}, {Gavazzi}, {Helmi}, {Hook}, {Irwin}, {Kneib},
  {Kunz}, {Mannucci}, {Moscardini}, {Tao}, {Teyssier}, {Weller}, {Zamorani},
  {Zapatero Osorio}, {Boulade}, {Foumond}, {Di Giorgio}, {Guttridge}, {James},
  {Kemp}, {Martignac}, {Spencer}, {Walton}, {Bl{\"u}mchen}, {Bonoli},
  {Bortoletto}, {Cerna}, {Corcione}, {Fabron}, {Jahnke}, {Ligori}, {Madrid},
  {Martin}, {Morgante}, {Pamplona}, {Prieto}, {Riva}, {Toledo}, {Trifoglio},
  {Zerbi}, {Abdalla}, {Douspis}, {Grenet}, {Borgani}, {Bouwens}, {Courbin},
  {Delouis}, {Dubath}, {Fontana}, {Frailis}, {Grazian}, {Koppenh{\"o}fer},
  {Mansutti}, {Melchior}, {Mignoli}, {Mohr}, {Neissner}, {Noddle}, {Poncet},
  {Scodeggio}, {Serrano}, {Shane}, {Starck}, {Surace}, {Taylor},
  {Verdoes-Kleijn}, {Vuerli}, {Williams}, {Zacchei}, {Altieri}, {Escudero
  Sanz}, {Kohley}, {Oosterbroek}, {Astier}, {Bacon}, {Bardelli}, {Baugh},
  {Bellagamba}, {Benoist}, {Bianchi}, {Biviano}, {Branchini}, {Carbone},
  {Cardone}, {Clements}, {Colombi}, {Conselice}, {Cresci}, {Deacon}, {Dunlop},
  {Fedeli}, {Fontanot}, {Franzetti}, {Giocoli}, {Garcia-Bellido}, {Gow},
  {Heavens}, {Hewett}, {Heymans}, {Holland}, {Huang}, {Ilbert}, {Joachimi},
  {Jennins}, {Kerins}, {Kiessling}, {Kirk}, {Kotak}, {Krause}, {Lahav}, {van
  Leeuwen}, {Lesgourgues}, {Lombardi}, {Magliocchetti}, {Maguire}, {Majerotto},
  {Maoli}, {Marulli}, {Maurogordato}, {McCracken}, {McLure}, {Melchiorri},
  {Merson}, {Moresco}, {Nonino}, {Norberg}, {Peacock}, {Pello}, {Penny},
  {Pettorino}, {Di Porto}, {Pozzetti}, {Quercellini}, {Radovich}, {Rassat},
  {Roche}, {Ronayette}, {Rossetti}, {Sartoris}, {Schneider}, {Semboloni},
  {Serjeant}, {Simpson}, {Skordis}, {Smadja}, {Smartt}, {Spano}, {Spiro},
  {Sullivan}, {Tilquin}, {Trotta}, {Verde}, {Wang}, {Williger}, {Zhao},
  {Zoubian}, \& {Zucca}}]{2011arXiv1110.3193L}
{Laureijs}, R., {Amiaux}, J., {Arduini}, S., {et~al.} 2011, arXiv e-prints,
  1110.3193, \dodoi{arXiv:1110.3193}

\bibitem[{Lewis(2019)}]{lewis2019getdist}
Lewis, A. 2019, GetDist: a Python package for analysing Monte Carlo samples.
\newblock \doarXiv{1910.13970}

\bibitem[{{Limber}(1954)}]{1954ApJ...119..655L}
{Limber}, D.~N. 1954, \apj, 119, 655, \dodoi{10.1086/145870}

\bibitem[{Lintusaari {et~al.}(2016)Lintusaari, Gutmann, Dutta, Kaski, \&
  Corander}]{10.1093/sysbio/syw077}
Lintusaari, J., Gutmann, M.~U., Dutta, R., Kaski, S., \& Corander, J. 2016,
  Syst. Biol., 66, e66, \dodoi{10.1093/sysbio/syw077}

\bibitem[{{Loeb} \& {Furlanetto}(2013)}]{FGU}
{Loeb}, A., \& {Furlanetto}, S. 2013, The First Galaxies in the Universe (PUP)

\bibitem[{{Lueckmann} {et~al.}(2018){Lueckmann}, {Bassetto}, {Karaletsos}, \&
  {Macke}}]{2018arXiv180509294L}
{Lueckmann}, J.-M., {Bassetto}, G., {Karaletsos}, T., \& {Macke}, J.~H. 2018,
  arXiv e-prints, 1805.09294, \dodoi{10.48550/arXiv.1805.09294}

\bibitem[{{Lueckmann} {et~al.}(2017){Lueckmann}, {Goncalves}, {Bassetto},
  {{\"O}cal}, {Nonnenmacher}, \& {Macke}}]{2017arXiv171101861L}
{Lueckmann}, J.-M., {Goncalves}, P.~J., {Bassetto}, G., {et~al.} 2017, arXiv
  e-prints, 1711.01861, \dodoi{10.48550/arXiv.1711.01861}

\bibitem[{{Lyu}(2012)}]{2012arXiv1205.2629L}
{Lyu}, S. 2012, arXiv e-prints, 1205.2629, \dodoi{10.48550/arXiv.1205.2629}

\bibitem[{Magdon-Ismail \& Atiya(1998)}]{NIPS1998_93279690}
Magdon-Ismail, M., \& Atiya, A. 1998, NeurIPS, 11,
  \dodoi{https://proceedings.neurips.cc/paper_files/paper/1998/file/9327969053c0068dd9e07c529866b94d-Paper.pdf}

\bibitem[{{Mason} {et~al.}(2023){Mason}, {Mu{\~n}oz}, {Greig}, {Mesinger}, \&
  {Park}}]{2023MNRAS.524.4711M}
{Mason}, C.~A., {Mu{\~n}oz}, J.~B., {Greig}, B., {Mesinger}, A., \& {Park}, J.
  2023, \mnras, 524, 4711, \dodoi{10.1093/mnras/stad2145}

\bibitem[{McKay {et~al.}(1979)McKay, Beckman, \&
  Conover}]{ef76b040-2f28-37ba-b0c4-02ed99573416}
McKay, M.~D., Beckman, R.~J., \& Conover, W.~J. 1979, Technometrics, 21, 239,
  \dodoi{http://www.jstor.org/stable/1268522}

\bibitem[{{Melchior} \& {Goulding}(2018)}]{2018A&C....25..183M}
{Melchior}, P., \& {Goulding}, A.~D. 2018, A\& C, 25, 183,
  \dodoi{10.1016/j.ascom.2018.09.013}

\bibitem[{{Mellema} {et~al.}(2006){Mellema}, {Iliev}, {Alvarez}, \&
  {Shapiro}}]{2006NewA...11..374M}
{Mellema}, G., {Iliev}, I.~T., {Alvarez}, M.~A., \& {Shapiro}, P.~R. 2006, \na,
  11, 374, \dodoi{10.1016/j.newast.2005.09.004}

\bibitem[{{Mellema} {et~al.}(2013){Mellema}, {Koopmans}, {Abdalla}, {Bernardi},
  {Ciardi}, {Daiboo}, {de Bruyn}, {Datta}, {Falcke}, {Ferrara}, {Iliev},
  {Iocco}, {Jeli{\'c}}, {Jensen}, {Joseph}, {Labroupoulos}, {Meiksin},
  {Mesinger}, {Offringa}, {Pandey}, {Pritchard}, {Santos}, {Schwarz},
  {Semelin}, {Vedantham}, {Yatawatta}, \& {Zaroubi}}]{2013ExA....36..235M}
{Mellema}, G., {Koopmans}, L. V.~E., {Abdalla}, F.~A., {et~al.} 2013, Exp.
  Astron., 36, 235, \dodoi{10.1007/s10686-013-9334-5}

\bibitem[{{Mertens} {et~al.}(2020){Mertens}, {Mevius}, {Koopmans}, {Offringa},
  {Mellema}, {Zaroubi}, {Brentjens}, {Gan}, {Gehlot}, {Pand ey}, {Sardarabadi},
  {Vedantham}, {Yatawatta}, {Asad}, {Ciardi}, {Chapman}, {Gazagnes}, {Ghara},
  {Ghosh}, {Giri}, {Iliev}, {Jeli{\'c}}, {Kooistra}, {Mondal}, {Schaye}, \&
  {Silva}}]{2020MNRAS.493.1662M}
{Mertens}, F.~G., {Mevius}, M., {Koopmans}, L.~V.~E., {et~al.} 2020, \mnras,
  493, 1662, \dodoi{10.1093/mnras/staa327}

\bibitem[{{Mesinger} \& {Furlanetto}(2007)}]{2007ApJ...669..663M}
{Mesinger}, A., \& {Furlanetto}, S. 2007, \apj, 669, 663,
  \dodoi{10.1086/521806}

\bibitem[{{Mesinger} {et~al.}(2011){Mesinger}, {Furlanetto}, \&
  {Cen}}]{2011MNRAS.411..955M}
{Mesinger}, A., {Furlanetto}, S., \& {Cen}, R. 2011, \mnras, 411, 955,
  \dodoi{10.1111/j.1365-2966.2010.17731.x}

\bibitem[{{Miralda-Escud{\'e}} {et~al.}(2000){Miralda-Escud{\'e}}, {Haehnelt},
  \& {Rees}}]{2000ApJ...530....1M}
{Miralda-Escud{\'e}}, J., {Haehnelt}, M., \& {Rees}, M.~J. 2000, \apj, 530, 1,
  \dodoi{10.1086/308330}

\bibitem[{{Mondal} {et~al.}(2015){Mondal}, {Bharadwaj}, {Majumdar}, {Bera}, \&
  {Acharyya}}]{2015MNRAS.449L..41M}
{Mondal}, R., {Bharadwaj}, S., {Majumdar}, S., {Bera}, A., \& {Acharyya}, A.
  2015, \mnras, 449, L41, \dodoi{10.1093/mnrasl/slv015}

\bibitem[{{Monsalve} {et~al.}(2023){Monsalve}, {Bye}, {Sievers}, {Bidula},
  {Bustos}, {Chiang}, {Guo}, {Hendricksen}, {McGee}, {Patricio Mena},
  {Prabhakar}, {Restrepo}, \& {Thyagarajan}}]{2023arXiv231007741M}
{Monsalve}, R.~A., {Bye}, C.~H., {Sievers}, J.~L., {et~al.} 2023, arXiv
  e-prints, 2310.07741, \dodoi{10.48550/arXiv.2310.07741}

\bibitem[{{Murray} \& et~al.(2020)}]{2020JOSS....5.2582M}
{Murray}, S., \& et~al. 2020, JOSS, 5, 2582, \dodoi{10.21105/joss.02582}

\bibitem[{{Neutsch} {et~al.}(2022){Neutsch}, {Heneka}, \&
  {Br{\"u}ggen}}]{2022MNRAS.511.3446N}
{Neutsch}, S., {Heneka}, C., \& {Br{\"u}ggen}, M. 2022, \mnras, 511, 3446,
  \dodoi{10.1093/mnras/stac218}

\bibitem[{{Ocvirk} {et~al.}(2016){Ocvirk}, {Gillet}, {Shapiro}, {Aubert},
  {Iliev}, {Teyssier}, {Yepes}, {Choi}, {Sullivan}, {Knebe}, {Gottl{\"o}ber},
  {D'Aloisio}, {Park}, {Hoffman}, \& {Stranex}}]{2016MNRAS.463.1462O}
{Ocvirk}, P., {Gillet}, N., {Shapiro}, P.~R., {et~al.} 2016, \mnras, 463, 1462,
  \dodoi{10.1093/mnras/stw2036}

\bibitem[{{Ocvirk} {et~al.}(2018){Ocvirk}, {Aubert}, {Sorce}, {Shapiro},
  {Deparis}, {Dawoodbhoy}, {Lewis}, {Teyssier}, {Yepes}, {Gottl{\"o}ber},
  {Ahn}, {Iliev}, \& {Hoffman}}]{2018arXiv181111192O}
{Ocvirk}, P., {Aubert}, D., {Sorce}, J.~G., {et~al.} 2018, arXiv e-prints,
  1811.11192, \dodoi{1811.11192}

\bibitem[{{Pagano} \& {Liu}(2020)}]{2020MNRAS.498..373P}
{Pagano}, M., \& {Liu}, A. 2020, \mnras, 498, 373,
  \dodoi{10.1093/mnras/staa2118}

\bibitem[{{Papamakarios}(2019)}]{2019arXiv191013233P}
{Papamakarios}, G. 2019, arXiv e-prints, 1910.13233,
  \dodoi{10.48550/arXiv.1910.13233}

\bibitem[{{Papamakarios} \& {Murray}(2016)}]{2016arXiv160506376P}
{Papamakarios}, G., \& {Murray}, I. 2016, arXiv e-prints, 1605.06376,
  \dodoi{1605.06376}

\bibitem[{{Papamakarios} {et~al.}(2017){Papamakarios}, {Pavlakou}, \&
  {Murray}}]{2017arXiv170507057P}
{Papamakarios}, G., {Pavlakou}, T., \& {Murray}, I. 2017, arxiv e-prints,
  1705.07057, \dodoi{1705.07057}

\bibitem[{{Park} {et~al.}(2019){Park}, {Mesinger}, {Greig}, \&
  {Gillet}}]{2019MNRAS.484..933P}
{Park}, J., {Mesinger}, A., {Greig}, B., \& {Gillet}, N. 2019, \mnras, 484,
  933, \dodoi{10.1093/mnras/stz032}

\bibitem[{{Pattison} {et~al.}(2023){Pattison}, {Anstey}, \& {de Lera
  Acedo}}]{2023arXiv230702908P}
{Pattison}, J. H.~N., {Anstey}, D.~J., \& {de Lera Acedo}, E. 2023, arXiv
  e-prints, 2307.02908, \dodoi{10.48550/arXiv.2307.02908}

\bibitem[{Pedregosa {et~al.}(2011)Pedregosa, Varoquaux, Gramfort, Michel,
  Thirion, Grisel, Blondel, Prettenhofer, Weiss, Dubourg,
  {et~al.}}]{Pedregosa2011scikit-learn}
Pedregosa, F., Varoquaux, G., Gramfort, A., {et~al.} 2011, Journal of Machine
  Learning Research, 12, 2825

\bibitem[{{Planck Collaboration} {et~al.}(2016){Planck Collaboration}, {Ade},
  {Aghanim}, {Arnaud}, {Ashdown}, {Aumont}, {Baccigalupi}, {Banday},
  {Barreiro}, {Bartlett}, {Bartolo}, {Battaner}, {Battye}, {Benabed},
  {Beno{\^\i}t}, {Benoit-L{\'e}vy}, {Bernard}, {Bersanelli}, {Bielewicz},
  {Bock}, {Bonaldi}, {Bonavera}, {Bond}, {Borrill}, {Bouchet}, {Boulanger},
  {Bucher}, {Burigana}, {Butler}, {Calabrese}, {Cardoso}, {Catalano},
  {Challinor}, {Chamballu}, {Chary}, {Chiang}, {Chluba}, {Christensen},
  {Church}, {Clements}, {Colombi}, {Colombo}, {Combet}, {Coulais}, {Crill},
  {Curto}, {Cuttaia}, {Danese}, {Davies}, {Davis}, {de Bernardis}, {de Rosa},
  {de Zotti}, {Delabrouille}, {D{\'e}sert}, {Di Valentino}, {Dickinson},
  {Diego}, {Dolag}, {Dole}, {Donzelli}, {Dor{\'e}}, {Douspis}, {Ducout},
  {Dunkley}, {Dupac}, {Efstathiou}, {Elsner}, {En{\ss}lin}, {Eriksen},
  {Farhang}, {Fergusson}, {Finelli}, {Forni}, {Frailis}, {Fraisse},
  {Franceschi}, {Frejsel}, {Galeotta}, {Galli}, {Ganga}, {Gauthier}, {Gerbino},
  {Ghosh}, {Giard}, {Giraud-H{\'e}raud}, {Giusarma}, {Gjerl{\o}w},
  {Gonz{\'a}lez-Nuevo}, {G{\'o}rski}, {Gratton}, {Gregorio}, {Gruppuso},
  {Gudmundsson}, {Hamann}, {Hansen}, {Hanson}, {Harrison}, {Helou},
  {Henrot-Versill{\'e}}, {Hern{\'a}ndez-Monteagudo}, {Herranz}, {Hildebrand t},
  {Hivon}, {Hobson}, {Holmes}, {Hornstrup}, {Hovest}, {Huang}, {Huffenberger},
  {Hurier}, {Jaffe}, {Jaffe}, {Jones}, {Juvela}, {Keih{\"a}nen}, {Keskitalo},
  {Kisner}, {Kneissl}, {Knoche}, {Knox}, {Kunz}, {Kurki-Suonio}, {Lagache},
  {L{\"a}hteenm{\"a}ki}, {Lamarre}, {Lasenby}, {Lattanzi}, {Lawrence}, {Leahy},
  {Leonardi}, {Lesgourgues}, {Levrier}, {Lewis}, {Liguori}, {Lilje},
  {Linden-V{\o}rnle}, {L{\'o}pez-Caniego}, {Lubin}, {Mac{\'\i}as-P{\'e}rez},
  {Maggio}, {Maino}, {Mandolesi}, {Mangilli}, {Marchini}, {Maris}, {Martin},
  {Martinelli}, {Mart{\'\i}nez-Gonz{\'a}lez}, {Masi}, {Matarrese}, {McGehee},
  {Meinhold}, {Melchiorri}, {Melin}, {Mendes}, {Mennella}, {Migliaccio},
  {Millea}, {Mitra}, {Miville-Desch{\^e}nes}, {Moneti}, {Montier}, {Morgante},
  {Mortlock}, {Moss}, {Munshi}, {Murphy}, {Naselsky}, {Nati}, {Natoli},
  {Netterfield}, {N{\o}rgaard-Nielsen}, {Noviello}, {Novikov}, {Novikov},
  {Oxborrow}, {Paci}, {Pagano}, {Pajot}, {Paladini}, {Paoletti}, {Partridge},
  {Pasian}, {Patanchon}, {Pearson}, {Perdereau}, {Perotto}, {Perrotta},
  {Pettorino}, {Piacentini}, {Piat}, {Pierpaoli}, {Pietrobon}, {Plaszczynski},
  {Pointecouteau}, {Polenta}, {Popa}, {Pratt}, {Pr{\'e}zeau}, {Prunet},
  {Puget}, {Rachen}, {Reach}, {Rebolo}, {Reinecke}, {Remazeilles}, {Renault},
  {Renzi}, {Ristorcelli}, {Rocha}, {Rosset}, {Rossetti}, {Roudier},
  {Rouill{\'e} d'Orfeuil}, {Rowan-Robinson}, {Rubi{\~n}o-Mart{\'\i}n},
  {Rusholme}, {Said}, {Salvatelli}, {Salvati}, {Sandri}, {Santos},
  {Savelainen}, {Savini}, {Scott}, {Seiffert}, {Serra}, {Shellard}, {Spencer},
  {Spinelli}, {Stolyarov}, {Stompor}, {Sudiwala}, {Sunyaev}, {Sutton},
  {Suur-Uski}, {Sygnet}, {Tauber}, {Terenzi}, {Toffolatti}, {Tomasi},
  {Tristram}, {Trombetti}, {Tucci}, {Tuovinen}, {T{\"u}rler}, {Umana},
  {Valenziano}, {Valiviita}, {Van Tent}, {Vielva}, {Villa}, {Wade}, {Wandelt},
  {Wehus}, {White}, {White}, {Wilkinson}, {Yvon}, {Zacchei}, \&
  {Zonca}}]{2016A&A...594A..13P}
{Planck Collaboration}, {Ade}, P.~A.~R., {Aghanim}, N., {et~al.} 2016, \aap,
  594, A13, \dodoi{10.1051/0004-6361/201525830}

\bibitem[{{Planck Collaboration} {et~al.}(2020){Planck Collaboration},
  {Aghanim}, {Akrami}, {Ashdown}, {Aumont}, {Baccigalupi}, {Ballardini},
  {Banday}, {Barreiro}, {Bartolo}, {Basak}, {Battye}, {Benabed}, {Bernard},
  {Bersanelli}, {Bielewicz}, {Bock}, {Bond}, {Borrill}, {Bouchet}, {Boulanger},
  {Bucher}, {Burigana}, {Butler}, {Calabrese}, {Cardoso}, {Carron},
  {Challinor}, {Chiang}, {Chluba}, {Colombo}, {Combet}, {Contreras}, {Crill},
  {Cuttaia}, {de Bernardis}, {de Zotti}, {Delabrouille}, {Delouis}, {Di
  Valentino}, {Diego}, {Dor{\'e}}, {Douspis}, {Ducout}, {Dupac}, {Dusini},
  {Efstathiou}, {Elsner}, {En{\ss}lin}, {Eriksen}, {Fantaye}, {Farhang},
  {Fergusson}, {Fernandez-Cobos}, {Finelli}, {Forastieri}, {Frailis},
  {Fraisse}, {Franceschi}, {Frolov}, {Galeotta}, {Galli}, {Ganga},
  {G{\'e}nova-Santos}, {Gerbino}, {Ghosh}, {Gonz{\'a}lez-Nuevo}, {G{\'o}rski},
  {Gratton}, {Gruppuso}, {Gudmundsson}, {Hamann}, {Handley}, {Hansen},
  {Herranz}, {Hildebrandt}, {Hivon}, {Huang}, {Jaffe}, {Jones}, {Karakci},
  {Keih{\"a}nen}, {Keskitalo}, {Kiiveri}, {Kim}, {Kisner}, {Knox},
  {Krachmalnicoff}, {Kunz}, {Kurki-Suonio}, {Lagache}, {Lamarre}, {Lasenby},
  {Lattanzi}, {Lawrence}, {Le Jeune}, {Lemos}, {Lesgourgues}, {Levrier},
  {Lewis}, {Liguori}, {Lilje}, {Lilley}, {Lindholm}, {L{\'o}pez-Caniego},
  {Lubin}, {Ma}, {Mac{\'\i}as-P{\'e}rez}, {Maggio}, {Maino}, {Mandolesi},
  {Mangilli}, {Marcos-Caballero}, {Maris}, {Martin}, {Martinelli},
  {Mart{\'\i}nez-Gonz{\'a}lez}, {Matarrese}, {Mauri}, {McEwen}, {Meinhold},
  {Melchiorri}, {Mennella}, {Migliaccio}, {Millea}, {Mitra},
  {Miville-Desch{\^e}nes}, {Molinari}, {Montier}, {Morgante}, {Moss}, {Natoli},
  {N{\o}rgaard-Nielsen}, {Pagano}, {Paoletti}, {Partridge}, {Patanchon},
  {Peiris}, {Perrotta}, {Pettorino}, {Piacentini}, {Polastri}, {Polenta},
  {Puget}, {Rachen}, {Reinecke}, {Remazeilles}, {Renzi}, {Rocha}, {Rosset},
  {Roudier}, {Rubi{\~n}o-Mart{\'\i}n}, {Ruiz-Granados}, {Salvati}, {Sandri},
  {Savelainen}, {Scott}, {Shellard}, {Sirignano}, {Sirri}, {Spencer},
  {Sunyaev}, {Suur-Uski}, {Tauber}, {Tavagnacco}, {Tenti}, {Toffolatti},
  {Tomasi}, {Trombetti}, {Valenziano}, {Valiviita}, {Van Tent}, {Vibert},
  {Vielva}, {Villa}, {Vittorio}, {Wandelt}, {Wehus}, {White}, {White},
  {Zacchei}, \& {Zonca}}]{2020A&A...641A...6P}
{Planck Collaboration}, {Aghanim}, N., {Akrami}, Y., {et~al.} 2020, \aap, 641,
  A6, \dodoi{10.1051/0004-6361/201833910}

\bibitem[{{Prelogovi{\'c}} {et~al.}(2022){Prelogovi{\'c}}, {Mesinger},
  {Murray}, {Fiameni}, \& {Gillet}}]{2022MNRAS.509.3852P}
{Prelogovi{\'c}}, D., {Mesinger}, A., {Murray}, S., {Fiameni}, G., \& {Gillet},
  N. 2022, \mnras, 509, 3852, \dodoi{10.1093/mnras/stab3215}

\bibitem[{{Pritchard} \& {Furlanetto}(2006)}]{2006MNRAS.367.1057P}
{Pritchard}, J.~R., \& {Furlanetto}, S.~R. 2006, \mnras, 367, 1057,
  \dodoi{10.1111/j.1365-2966.2006.10028.x}

\bibitem[{{Pritchard} \& {Furlanetto}(2007)}]{2007MNRAS.376.1680P}
---. 2007, \mnras, 376, 1680, \dodoi{10.1111/j.1365-2966.2007.11519.x}

\bibitem[{{Pritchard} \& {Loeb}(2012)}]{2012RPPh...75h6901P}
{Pritchard}, J.~R., \& {Loeb}, A. 2012, Rep. Prog. in Phys., 75, 086901,
  \dodoi{10.1088/0034-4885/75/8/086901}

\bibitem[{{Santos} {et~al.}(2010){Santos}, {Ferramacho}, {Silva}, {Amblard}, \&
  {Cooray}}]{2010MNRAS.406.2421S}
{Santos}, M.~G., {Ferramacho}, L., {Silva}, M.~B., {Amblard}, A., \& {Cooray},
  A. 2010, \mnras, 406, 2421, \dodoi{10.1111/j.1365-2966.2010.16898.x}

\bibitem[{Santurkar {et~al.}(2018)Santurkar, Tsipras, Ilyas, \&
  Madry}]{conf/nips/SanturkarTIM18}
Santurkar, S., Tsipras, D., Ilyas, A., \& Madry, A. 2018, NeurIPS, 2488.
\newblock
  \url{http://dblp.uni-trier.de/db/conf/nips/nips2018.html#SanturkarTIM18}

\bibitem[{{Schmit} \& {Pritchard}(2018)}]{2018MNRAS.475.1213S}
{Schmit}, C.~J., \& {Pritchard}, J.~R. 2018, \mnras, 475, 1213,
  \dodoi{10.1093/mnras/stx3292}

\bibitem[{{Sellentin} \& {Heavens}(2016)}]{2016MNRAS.456L.132S}
{Sellentin}, E., \& {Heavens}, A.~F. 2016, \mnras, 456, L132,
  \dodoi{10.1093/mnrasl/slv190}

\bibitem[{Semih~Kayhan \& van Gemert(2020)}]{9156444}
Semih~Kayhan, O., \& van Gemert, J.~C. 2020, in 2020 IEEE/CVF Conference on
  Computer Vision and Pattern Recognition (CVPR), 14262--14273,
  \dodoi{10.1109/CVPR42600.2020.01428}

\bibitem[{{Shimabukuro} \& {Semelin}(2017)}]{2017MNRAS.468.3869S}
{Shimabukuro}, H., \& {Semelin}, B. 2017, \mnras, 468, 3869,
  \dodoi{10.1093/mnras/stx734}

\bibitem[{{Singh} {et~al.}(2018){Singh}, {Subrahmanyan}, {Udaya Shankar},
  {Sathyanarayana Rao}, {Fialkov}, {Cohen}, {Barkana}, {Girish}, {Raghunathan},
  {Somashekar}, \& {Srivani}}]{2018ApJ...858...54S}
{Singh}, S., {Subrahmanyan}, R., {Udaya Shankar}, N., {et~al.} 2018, \apj, 858,
  54, \dodoi{10.3847/1538-4357/aabae1}

\bibitem[{{Skilling}(2004)}]{2004AIPC..735..395S}
{Skilling}, J. 2004, AIP Conf Proc, 735, 395, \dodoi{10.1063/1.1835238}

\bibitem[{{Sobacchi} \& {Mesinger}(2014)}]{2014MNRAS.440.1662S}
{Sobacchi}, E., \& {Mesinger}, A. 2014, \mnras, 440, 1662,
  \dodoi{10.1093/mnras/stu377}

\bibitem[{{Speagle}(2020)}]{2020MNRAS.493.3132S}
{Speagle}, J.~S. 2020, \mnras, 493, 3132, \dodoi{10.1093/mnras/staa278}

\bibitem[{Srivastava {et~al.}(2014)Srivastava, Hinton, Krizhevsky, Sutskever,
  \& Salakhutdinov}]{JMLR:v15:srivastava14a}
Srivastava, N., Hinton, G., Krizhevsky, A., Sutskever, I., \& Salakhutdinov, R.
  2014, JMLR, 15, 1929, \dodoi{http://jmlr.org/papers/v15/srivastava14a.html}

\bibitem[{{Takada} \& {Jain}(2004)}]{2004MNRAS.348..897T}
{Takada}, M., \& {Jain}, B. 2004, \mnras, 348, 897,
  \dodoi{10.1111/j.1365-2966.2004.07410.x}

\bibitem[{{Taylor} {et~al.}(2013){Taylor}, {Joachimi}, \&
  {Kitching}}]{2013MNRAS.432.1928T}
{Taylor}, A., {Joachimi}, B., \& {Kitching}, T. 2013, \mnras, 432, 1928,
  \dodoi{10.1093/mnras/stt270}

\bibitem[{{Tegmark}(1997)}]{1997PhRvD..55.5895T}
{Tegmark}, M. 1997, \prd, 55, 5895, \dodoi{10.1103/PhysRevD.55.5895}

\bibitem[{{Tegmark} {et~al.}(1997){Tegmark}, {Taylor}, \&
  {Heavens}}]{1997ApJ...480...22T}
{Tegmark}, M., {Taylor}, A.~N., \& {Heavens}, A.~F. 1997, \apj, 480, 22,
  \dodoi{10.1086/303939}

\bibitem[{Tieleman \& Hinton(2012)}]{RMSProp}
Tieleman, T., \& Hinton, G. 2012, COURSERA: Neural networks for machine
  learning, 4, 26

\bibitem[{{Trott}(2016)}]{2016MNRAS.461..126T}
{Trott}, C.~M. 2016, \mnras, 461, 126, \dodoi{10.1093/mnras/stw1310}

\bibitem[{{Trotta}(2008)}]{2008ConPh..49...71T}
{Trotta}, R. 2008, Contemporary Physics, 49, 71,
  \dodoi{10.1080/00107510802066753}

\bibitem[{Van~Rossum \& Drake(2009)}]{10.5555/1593511}
Van~Rossum, G., \& Drake, F.~L. 2009, Python 3 Reference Manual (Scotts Valley,
  CA: CreateSpace)

\bibitem[{Virtanen {et~al.}(2019)Virtanen, Gommers, Oliphant, Haberland, Reddy,
  Cournapeau, Burovski, Peterson, Weckesser, Bright, van~der Walt, Brett,
  Wilson, Millman, Mayorov, Nelson, Jones, Kern, Larson, Carey, Polat, Feng,
  Moore, VanderPlas, Laxalde, Perktold, Cimrman, Henriksen, Quintero, Harris,
  Archibald, Ribeiro, Pedregosa, van Mulbregt, \&
  SciPy}]{DBLP:journals/corr/abs-1907-10121}
Virtanen, P., Gommers, R., Oliphant, T.~E., {et~al.} 2019, CoRR, abs/1907.10121

\bibitem[{{Watkinson} {et~al.}(2017){Watkinson}, {Majumdar}, {Pritchard}, \&
  {Mondal}}]{2017MNRAS.472.2436W}
{Watkinson}, C.~A., {Majumdar}, S., {Pritchard}, J.~R., \& {Mondal}, R. 2017,
  \mnras, 472, 2436, \dodoi{10.1093/mnras/stx2130}

\bibitem[{{Watkinson} \& {Pritchard}(2014)}]{2014MNRAS.443.3090W}
{Watkinson}, C.~A., \& {Pritchard}, J.~R. 2014, \mnras, 443, 3090,
  \dodoi{10.1093/mnras/stu1384}

\bibitem[{{Watkinson} {et~al.}(2020){Watkinson}, {Trott}, \&
  {Hothi}}]{2020arXiv200205992W}
{Watkinson}, C.~A., {Trott}, C.~M., \& {Hothi}, I. 2020, arXiv e-prints,
  2002.05992, \dodoi{2002.05992}

\bibitem[{Weiler {et~al.}(2018)Weiler, Hamprecht, \& Storath}]{8578193}
Weiler, M., Hamprecht, F.~A., \& Storath, M. 2018, in 2018 IEEE/CVF Conference
  on Computer Vision and Pattern Recognition, 849--858,
  \dodoi{10.1109/CVPR.2018.00095}

\bibitem[{Wilkinson(2013)}]{Wilkinson+2013+129+141}
Wilkinson, R.~D. 2013, Stat. Appl. Genet. Mol. Biol., 12, 129,
  \dodoi{doi:10.1515/sagmb-2013-0010}

\bibitem[{{Wise}(2019)}]{2019arXiv190706653W}
{Wise}, J.~H. 2019, arXiv e-prints, 1907.06653, \dodoi{arXiv:1907.06653}

\bibitem[{{Wise} {et~al.}(2012{\natexlab{a}}){Wise}, {Abel}, {Turk}, {Norman},
  \& {Smith}}]{2012MNRAS.427..311W}
{Wise}, J.~H., {Abel}, T., {Turk}, M.~J., {Norman}, M.~L., \& {Smith}, B.~D.
  2012{\natexlab{a}}, \mnras, 427, 311,
  \dodoi{10.1111/j.1365-2966.2012.21809.x}

\bibitem[{{Wise} {et~al.}(2014){Wise}, {Demchenko}, {Halicek}, {Norman},
  {Turk}, {Abel}, \& {Smith}}]{2014MNRAS.442.2560W}
{Wise}, J.~H., {Demchenko}, V.~G., {Halicek}, M.~T., {et~al.} 2014, \mnras,
  442, 2560, \dodoi{10.1093/mnras/stu979}

\bibitem[{{Wise} {et~al.}(2012{\natexlab{b}}){Wise}, {Turk}, {Norman}, \&
  {Abel}}]{2012ApJ...745...50W}
{Wise}, J.~H., {Turk}, M.~J., {Norman}, M.~L., \& {Abel}, T.
  2012{\natexlab{b}}, \apj, 745, 50, \dodoi{10.1088/0004-637X/745/1/50}

\bibitem[{{Yeh} {et~al.}(2023){Yeh}, {Smith}, {Kannan}, {Garaldi},
  {Vogelsberger}, {Borrow}, {Pakmor}, {Springel}, \&
  {Hernquist}}]{2023MNRAS.tmp..220Y}
{Yeh}, J. Y.~C., {Smith}, A., {Kannan}, R., {et~al.} 2023, \mnras,
  \dodoi{10.1093/mnras/stad210}

\bibitem[{{Yoshiura} {et~al.}(2021){Yoshiura}, {Pindor}, {Line}, {Barry},
  {Trott}, {Beardsley}, {Bowman}, {Byrne}, {Chokshi}, {Hazelton}, {Hasegawa},
  {Howard}, {Greig}, {Jacobs}, {Jordan}, {Joseph}, {Kolopanis}, {Lynch},
  {McKinley}, {Mitchell}, {Morales}, {Murray}, {Pober}, {Rahimi}, {Takahashi},
  {Tingay}, {Wayth}, {Webster}, {Wilensky}, {Wyithe}, {Zhang}, \&
  {Zheng}}]{2021MNRAS.505.4775Y}
{Yoshiura}, S., {Pindor}, B., {Line}, J.~L.~B., {et~al.} 2021, \mnras, 505,
  4775, \dodoi{10.1093/mnras/stab1560}

\bibitem[{{Zablocki} \& {Dodelson}(2016)}]{2016PhRvD..93h3525Z}
{Zablocki}, A., \& {Dodelson}, S. 2016, \prd, 93, 083525,
  \dodoi{10.1103/PhysRevD.93.083525}

\bibitem[{{Zahn} {et~al.}(2011){Zahn}, {Mesinger}, {McQuinn}, {Trac}, {Cen}, \&
  {Hernquist}}]{2011MNRAS.414..727Z}
{Zahn}, O., {Mesinger}, A., {McQuinn}, M., {et~al.} 2011, \mnras, 414, 727,
  \dodoi{10.1111/j.1365-2966.2011.18439.x}

\bibitem[{{Zel'Dovich}(1970)}]{1970A&A.....5...84Z}
{Zel'Dovich}, Y.~B. 1970, \aap, 500, 13

\bibitem[{{Zhao} {et~al.}(2022{\natexlab{a}}){Zhao}, {Mao}, {Cheng}, \&
  {Wandelt}}]{2022ApJ...926..151Z}
{Zhao}, X., {Mao}, Y., {Cheng}, C., \& {Wandelt}, B.~D. 2022{\natexlab{a}},
  \apj, 926, 151, \dodoi{10.3847/1538-4357/ac457d}

\bibitem[{{Zhao} {et~al.}(2022{\natexlab{b}}){Zhao}, {Mao}, \&
  {Wandelt}}]{2022ApJ...933..236Z}
{Zhao}, X., {Mao}, Y., \& {Wandelt}, B.~D. 2022{\natexlab{b}}, \apj, 933, 236,
  \dodoi{10.3847/1538-4357/ac778e}

\bibitem[{Zhao {et~al.}(2023)Zhao, Mao, Zuo, \&
  Wandelt}]{zhao2023simulationbased}
Zhao, X., Mao, Y., Zuo, S., \& Wandelt, B.~D. 2023, arXiv e-prints,
  2310.17602v1, \dodoi{2310.17602v1}

\end{thebibliography}
\bibliographystyle{aasjournal}

\end{document}